\documentclass[twocolumn,showpacs,aps,prd,floatfix]{revtex4}
\usepackage{graphicx}
\usepackage{dcolumn}
\usepackage{amsmath}
\usepackage{epsfig}
\usepackage{multirow}

\graphicspath{{figures/}}

\newcommand{\BABARPubYear}{09}
\newcommand{\BABARPubNumber}{001}
\newcommand{\SLACPubNumber}{13640}
\newcommand{\LANLNumber} {0905.3615}

\input {babarsym}
\newcommand{\bei}{\begin{itemize}}
\newcommand{\eei}{\end{itemize}}
\newcommand{\beq}{\begin{equation}}
\newcommand{\eeq}{\end{equation}}
\newcommand{\beqn}{\begin{eqnarray}}
\newcommand{\eeqn}{\end{eqnarray}}
\newcommand{\beqns}{\begin{eqnarray*}}
\newcommand{\eeqns}{\end{eqnarray*}}

\newcommand{\intl}{\int\limits}

\newcommand{\pvec}{{\bf p}}
\def\logL{\log{\cal L}}

\def\I{{\rm Im}}
\def\KS{\ensuremath{K^0_S}}
\def\de{\Delta E}
\def\dt{\deltat}

\def\mprime{m^\prime}
\def\thetaprime{\theta^\prime}
\def\dmd{\Delta m_d}
\def\dmt{\Delta t}
\def\cat{c}
\def\TM{{\rm TM}}
\def\SCF{{\rm SCF}}

\def\fscfave{\kern 0.18em\overline{\kern -0.18em f}_{\rm SCF}}
\def\fscf{f_{\rm SCF}}
\def\fscfi{f_{{\rm SCF},i}}

\def\Qtagi{q_{{\rm tag},i}}
\def\Atagqq{A_{q\bar q}}
\def\Atag{A_{{B^+}}}

\def\Abar{\kern 0.18em\overline{\kern -0.18em A}{}}
\def\Amptp{{\cal A}}

\def\Amptpbar{\kern 0.18em\overline{\kern -0.18em {\cal A}}}

\def\absAmptp{|\Amptp|}
\def\absAmptpbar{|\Amptpbar|}
\def\AmpAll{|{\cal A}^\pm(\dmt)|^2}

\def\detJ{|\det J|}
\def\detJi{|\det J_i|}
\def\Rscf{R_{\rm SCF}}
\def\rar{\rightarrow}

\newcommand{\Kspipi}             {\mbox{$\KS\pip\pim$}}
\newcommand{\BztoKspipi}         {\mbox{$\Bz \to \Kspipi$}}
\newcommand{\BzbtoKspipi}        {\mbox{$\Bzb \to \Kspipi$}}

\def\AmpAllSigp{|{\cal A}_{\rm sig}^+(\dmt)|^2}
\def\AmpAllSigm{|{\cal A}_{\rm sig}^-(\dmt)|^2}
\def\AmpAllSigpm{|{\cal A}_{\rm sig}^\pm(\dmt)|^2}

\def\pipi   {\ensuremath{\pip\pim}\xspace}

\newcommand{\KstarI}             {\mbox{$\Kstar(892)$}}

\newcommand{\KstarpI}            {\mbox{$\Kstarp(892)$}}
\newcommand{\KstarmI}            {\mbox{$\Kstarm(892)$}}
\newcommand{\KstarzI}            {\mbox{$\Kstarz(892)$}}

\newcommand{\KstarII}            {\mbox{$\Kstar(1430)$}}
\newcommand{\KstarpII}           {\mbox{$\Kstarp(1430)$}}

\newcommand{\KstarpmIV}          {\mbox{$\Kstarpm(1680)$}}

\newcommand{\rhoz}               {\mbox{$\rho^0$}}

\newcommand{\rhoI}               {\mbox{$\rhoz(770)$}}

\newcommand{\fI}                 {\mbox{$f_0(980)$}}

\newcommand{\fIKs}               {\mbox{$\fI \KS$}}

\newcommand{\rhoII}              {\mbox{$\rhoz(1450)$}}

\newcommand{\fII}                {\mbox{$f_2(1270)$}}

\newcommand{\fX}                 {\mbox{$f_X(1300)$}}

\newcommand{\D}                  {\mbox{$D$}}

\def\ACP {{\ensuremath{A_{\CP}}\xspace}}
\newcommand{\half}{\mbox{$\frac{1}{2}$}}

\def\eg                 {e.g.}

\def\ie                 {i.e.}

\def\mBz{m_{\Bz}}
\def\spz{s_{+}}
\def\smz{s_{-}}
\def\spm{s_{0}}

\def\mpm{m_{0}}

\def\mpmMax{\mpm^{\rm max}}
\def\mpmMin{\mpm^{\rm min}}

\long\def\inst#1{\par\nobreak\kern 4pt\nobreak
    {\it #1}\par\vskip 10pt plus 3pt minus 3pt}

\begin{document}

\begin{flushleft}
arXiv:\LANLNumber\ [hep-ex] \\
SLAC-PUB-\SLACPubNumber \\
\babar-PUB-\BABARPubYear/\BABARPubNumber
\end{flushleft}

% Title of the paper
\title{
\large \bf
\boldmath
Time-dependent amplitude analysis of $B^0 \to \Kspipi$ 
} % end title

% Input author list file
%% author list as of 09-Jan-2009 (489 authors)
%
\author{B.~Aubert}
\author{Y.~Karyotakis}
\author{J.~P.~Lees}
\author{V.~Poireau}
\author{E.~Prencipe}
\author{X.~Prudent}
\author{V.~Tisserand}
\affiliation{Laboratoire d'Annecy-le-Vieux de Physique des Particules (LAPP), Universit\'e de Savoie, CNRS/IN2P3, F-74941 Annecy-Le-Vieux, France }
\author{J.~Garra~Tico}
\author{E.~Grauges}
\affiliation{Universitat de Barcelona, Facultat de Fisica, Departament ECM, E-08028 Barcelona, Spain }
\author{M.~Martinelli$^{ab}$}
\author{A.~Palano$^{ab}$ }
\author{M.~Pappagallo$^{ab}$ }
\affiliation{INFN Sezione di Bari$^{a}$; Dipartimento di Fisica, Universit\`a di Bari$^{b}$, I-70126 Bari, Italy }
% ...
\collaboration{The \babar\ Collaboration}
\noaffiliation

\date{\today}

% Abstract
\begin{abstract}
   \noindent
   We perform a time-dependent amplitude analysis of $B^0 \rightarrow \Kspipi$ decays to extract the
\CP violation parameters of $\fI \KS$ and $\rhoI \KS$ and the direct \CP asymmetry of $\KstarpI \pi^-$.
The results are obtained from a data sample of $(383\pm3)\times10^{6}$ \BB\ decays, collected with the
\babar\ detector at the \pep2\ asymmetric--energy \B\ factory at SLAC. We find two solutions, with an
equivalent goodness-of-fit. Including systematic and Dalitz plot model uncertainties, the combined
confidence interval for values of the \CP parameter $\beta_{\rm eff}$ in $\Bz$ decays to $\fI\KS$ is
$18^\circ< \beta_{\rm eff}<76^\circ$ at $95\%$ confidence level (C.L).
\CP conservation in $\Bz$ decays to $\fI\KS$ is excluded at  $3.5\sigma$ including systematic uncertainties.
For $\Bz$ decays to $\rhoI\KS$, the combined confidence interval is $-9^\circ< \beta_{\rm eff}<57^\circ$ at
$95\%$  C.L. In decays to $\KstarpI \pim$ we measure the direct \CP asymmetry to be
$\ACP=-0.20 \pm 0.10\pm 0.01\pm 0.02$. The measured phase difference (including $\BzBzb$ mixing) between
decay amplitudes of $\Bz \rightarrow \KstarpI \pim$ and $\Bzb \rightarrow \KstarmI \pip$, excludes the
interval $-137^\circ<\Delta\Phi(\KstarpI \pi^-) < -5^\circ$ at $95\%$ C.L.
\end{abstract}

% PACS

\maketitle

% The body of the paper starts here

\section{INTRODUCTION}
\label{sec:introduction}

The Cabibbo-Kobayashi-Maskawa (CKM)
mechanism~\cite{Cabibbo:1963yz,Kobayashi:1973fv} for quark mixing
describes all transitions between quarks in terms of only four
parameters: three rotation angles and one irreducible phase.
Consequently, the flavor sector of the Standard Model (SM)
is highly predictive.  One particularly interesting prediction is that
mixing-induced \CP asymmetries in decays governed by $b \to
q\bar{q}s$ ($q = u,d,s$) transitions are, to a good
approximation, the same as those found in $b \to c\bar{c}s$
transitions.  Since flavor changing neutral currents are forbidden at
tree-level in the Standard Model, the $b \to s$ transition proceeds
via loop diagrams (penguins), which are affected by new particles in
many extensions of the SM.

Various $b \to s$ dominated charmless hadronic $B$
decays have been studied in order to probe this prediction.  The
values of the mixing-induced \CP asymmetry measured for each
(quasi-)two-body mode can be compared to that measured in $b \to
c\bar{c}s$ transitions (typically using $B^0 \to J/\psi \KS$).  A
recent compilation~\cite{Barberio:2008fa} of results 
shows that they tend to have
central values below that for $b \to c\bar{c}s$. 
Recent theoretical
evaluations~\cite{Grossman:2003qp,Gronau:2003kx,Gronau:2004hp,Cheng:2005bg,Gronau:2005gz,Beneke:2005pu,Engelhard:2005hu,Cheng:2005ug,Williamson:2006hb}
suggest that SM corrections to the $b \to q\bar{q}s$ mixing-induced
\CP violation parameters should be small, in particular for the modes
$\phi K^0$, $\eta^\prime K^0$, and $\KS\KS\KS$, and tend to
{\em increase} the values, \ie~the opposite trend to that seen
in the data. 
However, there is currently no
convincing evidence for new physics effects in these transitions.
Clearly, more precise experimental results are required.

The compilation given in \cite{Barberio:2008fa} includes several
three-body modes, which may be used either by virtue of being \CP
eigenstates ($\KS\KS\KS$, $\KS\piz\piz$)~\cite{Gershon:2004tk} or
because their \CP content can be determined experimentally
($\Kp\Km\Kz$)~\cite{Garmash:2003er,Aubert:2007sd}. 
It also includes quasi-two-body (Q2B) modes, such as $\fI\KS$ and
$\rhoI\KS$, which are reconstructed via their three-body final states
($\KS\pip\pim$ for these modes). The precision
of the Q2B approach is limited as other structures in the
phase space 
may cause interference with the resonances
considered as signal.  Therefore, more precise results can be obtained
using a time-dependent amplitude analysis covering the complete
phase space, or Dalitz plot (DP), of $\Bz \to
\KS\pip\pim$.   Furthermore the
interference terms allow the cosine of the effective weak phase
difference in mixing and decay to be determined, helping to
resolve ambiguities which arise from the Q2B analysis.  This approach
has been successfully used in a time-dependent DP
analysis of $\Bz \to \Kp\Km\Kz$~\cite{Aubert:2007sd}.

The discussion above assumes that the $b \to s$ penguin amplitude
dominates the decay.  However, for each mode contributing to the
$\KS\pip\pim$ final state, there is also the possibility of a $b \to
u$ tree diagram.  These are doubly CKM suppressed compared to the $b
\to s$ penguin diagram (the tree is ${\cal O}(\lambda^4)$ whereas the penguin
is  ${\cal O}(\lambda^2)$, where $\lambda$ is the usual Wolfenstein
parameter~\cite{Wolfenstein:1983yz,Buras:1994ec}).  However, hadronic
factors may enhance the tree amplitudes, resulting in a 
significant ``tree pollution.''
These hadronic factors may be different for each Q2B state, thus the relative
magnitudes of each tree and penguin amplitudes, $\left| T/P \right|$, and the
strong phase difference may be different as well.
Nontheless, the relative weak
phase between these two amplitudes is the same -- and in the Standard Model
is equal to the CKM unitarity triangle angle
$\gamma$.  An amplitude analysis, in contrast
to a Q2B analysis, yields sufficient information to extract relative
phases and magnitudes.  Measurements of decay amplitudes in the DP analysis of  $\Bz \to
\KS\pip\pim$ (and similar modes) can therefore be used to set constraints
on the CKM parameters $(\rhobar,\etabar)$ ~\cite{Deshpande:2002be,Ciuchini:2006kv,Gronau:2006qn,Lipkin:1991st}.

Recently published results on time-dependent DP analysis of  $B^0 \to \Kspipi$ are 
available~\cite{BelleKspipi}. Previous
studies of the $\Bz \to \Kspipi$ decay were  either based
on a Q2B approach~\cite{Aubert:2005wb},  or were amplitude analyses
that did not take into account either time-dependence or flavor-tag dependence~\cite{Garmash:2006fh}.
The available results for  $\Bz \to \Kspipi$ are
consistent with studies 
obtained from other $B\to K\pi\pi$ decay modes:
$\Kp\pim\piz$~\cite{Chang:2004um,Aubert:2007bs} and
$K^+\pi^+\pi^-$~\cite{Aubert:2008bj,Garmash:2005rv}.
The latter results indicate evidence for direct \CP violation in the
$B^+ \to \rho^0(770) K^+$ channel.  If confirmed, this will be the first
observation of \CP violation in the decay of any charged
particle.
The relevance of $B\to K\pi\pi$ is further highlighted by recent theoretical
calculations~\cite{Chang:2008tf} suggesting that large  \CP violation effects are 
expected in several $B\to K^*\pi$ and $B\to K\rho$ resonant modes. 

In this paper we present results from a 
time-dependent amplitude analysis of the $\BztoKspipi$ decay.
In Sec.~\ref{sec:ana_overview} we describe the time-dependent DP
formalism, and introduce the signal parameters that are extracted in the
fit to data.  In Sec.~\ref{sec:babar} we briefly describe the \babar\ detector and
the data set.  In Sec.~\ref{sec:selection}, we explain the selection requirements used
to obtain the signal candidates and suppress backgrounds. In Sec.~\ref{sec:ML} we describe the
fit method and the approach used to control experimental effects such as resolution.
In Sec.~\ref{sec:fitResults} we present the results of the fit,
and extract parameters relevant to the contributing intermediate resonant states. In
Sec.~\ref{sec:Systematics} we discuss systematic uncertainties in the results, and finally we
summarize the results in Sec.~\ref{sec:Summary}.

\section{ANALYSIS OVERVIEW}
\label{sec:ana_overview}

Taking advantage of the interference pattern in the DP, we measure relative
magnitudes and phases for the different resonant decay modes using a
maximum-likelihood fit. Below, we detail the formalism used in the present analysis.

\subsection{Decay amplitudes}
\label{sec:kinematics}

We consider the decay of a spin-zero $\Bz$ with four-momentum
$p_B$ into the three daughters $\pip$, $\pim$, and $\KS$
with $p_+$, $p_-$, and $p_0$ their corresponding four-momenta. Using
as independent (Mandelstam) variables the invariant squared masses
\begin{eqnarray}
\label{eq:dalitzVariables} 
       \spz \;&=&\; m_{\KS\pip}^2 \;=\; (p_+ + p_0)^2~,\\ \nonumber
       \smz \;&=&\; m_{\KS\pim}^2 \;=\; (p_- + p_0)^2~, 
\end{eqnarray}
the invariant squared mass 
$\spm \;=\; m_{\pip\pim}^2  \;=\; (p_+ + p_-)^2$ can be obtained from energy and 
momentum conservation:
\beq
\label{eq:magicSum}
        \spm \;=\; \mBz^2 + 2m_{\pi^+}^2 + m_{\KS}^2
                   - \spz - \smz~.
\eeq
The differential $\Bz$ decay width with respect to the 
variables defined in Eq.~\eqref{eq:dalitzVariables} (\ie~the 
Dalitz plot) reads
\beq
\label{eq:partialWidth}
        d\Gamma(\BztoKspipi) \;=\; 
        \frac{1}{(2\pi)^3}\frac{|\Amptp|^2}{32 \mBz^3}\,d\spz d\smz~,
\eeq
where ${\cal A}$ is the Lorentz-invariant amplitude
of the three-body decay. 
In the following, the amplitudes ${\cal A}$ and $\bar{\cal A}$ correspond to the transitions $\BztoKspipi$ and $\BzbtoKspipi$, respectively.
We describe the distribution of signal events 
in the DP using an isobar approximation,
which models the total amplitude as
resulting from a coherent sum of amplitudes from the $N$ individual decay channels
\begin{eqnarray}
  \label{eq:isobar}
  {\cal A}(\spz,\smz) 
  & = & \sum_{j=1}^{N} c_j F_j(\spz,\smz)~,  \\
  \bar{\cal A}(\spz,\smz) 
  & = & \sum_{j=1}^{N} \bar{c}_j \bar{F}_j(\spz,\smz)~,
\end{eqnarray}
where $F_j$ are DP-dependent dynamical amplitudes described below, 
and $c_j$ complex coefficients describing the relative
magnitude and phase of the different decay channels.
All the weak phase dependence is contained in $c_j$, and $F_j$
contains strong dynamics only; therefore,
\begin{eqnarray}
  F_j(\spz,\smz) & = & \bar{F}_j(\smz,\spz) ~.
\end{eqnarray}
The resonance dynamics are contained within the $F_j$ terms, which are represented
by the product of the invariant mass and angular distribution probabilities, \ie,
\begin{equation}
\label{eq:ResDynEqn}
F_j^L(\spz,\smz) = R_j(m) X_L(|\vec{p}\,^{\star}|\,r') X_L(|\vec{q}\,|\,r) T_j(L,\vec{p},\vec{q}\,)
\end{equation}
where
\begin{itemize}
\item $m$ is the invariant mass of the decay products of the resonance,
\item $R_j(m)$ is the resonance mass term or ``lineshape'' (\eg~Breit--Wigner),
\item $L$ is the orbital angular momentum between the resonance and the
      bachelor particle,
\item $\vec{p}\,^{\star}$ is the momentum of the bachelor particle
      evaluated in the rest frame of the $B$,
\item $\vec{p}$ and $\vec{q}$ are the momenta of the bachelor particle and
      one of the resonance daughters, respectively, both evaluated in the
      rest frame of the resonance
      (for $\KS\pim$, $\KS\pip$, and $\pip\pim$ resonances, $\vec{q}$ is assigned to the momentum
      of the $\KS$, $\pip$, and $\pim$, respectively),
\item $X_L$ are Blatt--Weisskopf barrier factors~\cite{blatt-weisskopf} with parameters $r'$
      (taken to be $2\,(\gevc)^{-1}$) and $r$ (given in Table~\ref{tab:model}), and
\item $T_j(L,\vec{p},\vec{q})$ is the angular distribution:
\begin{eqnarray}
L=0 &:& T_j = 1~,\\
L=1 &:& T_j = -4\vec{p}\cdot\vec{q}~,\\
L=2 &:& T_j = \frac{8}{3} \left[3(\vec{p}\cdot\vec{q}\,)^2 - (|\vec{p}\,||\vec{q}\,|)^2\right]~.
\end{eqnarray}
\end{itemize}

The helicity angle of a resonance is defined as the angle between $\vec{p}$ and $\vec{q}$.
Explicitly, for $\KS\pim$, $\KS\pip$, and $\pip\pim$ resonances the helicity angle is defined
between the momenta of the bachelor particle and of the $\KS$, $\pip$, and $\pim$, respectively,
in the resonance rest frame.

For most resonances in this analysis the $R_j$ are taken to be relativistic
Breit--Wigner (RBW)~\cite{Amsler:2008zz} lineshapes:
\begin{equation}
R_j(m) = \frac{1}{(m^2_0 - m^2) - i m_0 \Gamma(m)},
\label{eqn:BreitWigner}
\end{equation}
where $m_0$ is the nominal mass of the resonance and $\Gamma(m)$ is the
mass-dependent width.
In the general case of a spin-$J$ resonance, the latter can be expressed as
\begin{equation}
\Gamma(m) = \Gamma_0 \left( \frac{q}{q_0}\right)^{2J+1} 
\left(\frac{m_0}{m}\right) \frac{X^2_J(q)}{X^2_J(q_0)}.
\label{eqn:resWidth}
\end{equation}
The symbol $\Gamma_0$ denotes the nominal width of the resonance.
The values of $m_0$ and $\Gamma_0$ are listed in Table~\ref{tab:model}.
The symbol $q_0$ denotes the value of $q$ when $m = m_0$.

For the \fI\ lineshape the Flatt\'e form~\cite{Flatte} is used.
In this case the mass-dependent width is given by the sum 
of the widths in the $\pi\pi$ and $KK$ systems:
\begin{equation}
\Gamma(m) = \Gamma_{\pi\pi}(m) + \Gamma_{KK}(m),
\label{eqn:FlatteW1}
\end{equation}
where
\begin{eqnarray}
\Gamma_{\pi\pi}(m) &=&
g_{\pi} \Bigg(\frac{1}{3}\sqrt{1 - 4m_{\piz}^2/m^2} + \\ \nonumber
&& \phantom{g_{\pi} \Bigg(\frac{1}{3}} \frac{2}{3}\sqrt{1 - 4m_{\pipm}^2/m^2}\Bigg)\,,\\
\Gamma_{KK}(m) &=&
g_{K} \Bigg(\frac{1}{2}\sqrt{1 - 4m_{\Kpm}^2/m^2} + \\ \nonumber
&& \phantom{g_{K} \Bigg(\frac{1}{2}} \frac{1}{2}\sqrt{1 - 4m_{\Kz}^2/m^2}\Bigg)\,.
\label{eqn:FlatteW2}
\end{eqnarray}
The fractional coefficients arise from isospin conservation and $g_{\pi}$
and $g_{K}$ are coupling constants for which the values are given in Table~\ref{tab:model}.

For the \rhoI\ we use the  
Gounaris--Sakurai (GS) parameterization~\cite{Gounaris:1968mw}, that describes the $P$-wave
scattering amplitude for a broad resonance, decaying to two pions:
\begin{equation}
\label{eq:rhoGS}
R_{j}(m) = \frac{1+d\cdot\Gamma_0/m_0}
               	{(m_0^2 - m^2) + f(m) - i\, m_0 \Gamma(m)}~,
\end{equation}
where
\begin{eqnarray}
f(m) =
\Gamma_0 \frac{m_0^2}{q_0^3}
       \bigg[&&\!\!\!\!\!
             q^2 \left(\,h(m)-h(m_0)\,\right) + \\ \nonumber
&&           \left(\,m_0^2-m^2\,\right)\,q^2_0\,
             \frac{dh\ }{dm^2}\bigg|_{m=m_0}\,
       \bigg]~,
\end{eqnarray}
and the function $h(m)$ is defined as
\begin{equation}
h(m) = \frac{2}{\pi}\,\frac{q}{m}\,
       \ln\left(\frac{m+2q}{2m_\pi}\right)~,
\end{equation}
with 
\begin{equation}
\frac{dh\ }{dm^2}\bigg|_{m=m_0} =
h(m_0)\left(\frac{1}{8q_0^2}-\frac{1}{2m_0^2}\right) \,+\, \frac{1}{2\pi m_0^2}~. 
\end{equation}
The normalization condition at $R_j(0)$ fixes the parameter
$d=f(0)/(\Gamma_0 m_0)$. It is found to be:
\begin{equation}
d = \frac{3}{\pi}\frac{m_\pi^2}{q_0^2}\,
    \ln\left(\frac{m_0+2q_0}{2m_\pi}\right) 
    + \frac{m_0}{2\pi\,q_0} 
    - \frac{m_\pi^2 m_0}{\pi\,q_0^3}~.
\end{equation}

The $0^+$ component of the $K\pi$ spectrum is not well
understood~\cite{LASS,Bugg:2003kj}; we dub this component $(K\pi)^{*\pm}_0$ and
use the LASS parameterization~\cite{LASS} which consists of the
\KstarII\ resonance together with an effective range nonresonant (NR) component:
\begin{eqnarray}
\label{eq:LASSEqn}
R_j(m)  &=& \frac{m_{K\pi}}{q \cot{\delta_B} - iq} \\ &+& e^{2i \delta_B} 
\frac{m_0 \Gamma_0 \frac{m_0}{q_0}}
     {(m_0^2 - m_{K\pi}^2) - i m_0 \Gamma_0 \frac{q}{m_{K\pi}} \frac{m_0}{q_0}},
\nonumber
\end{eqnarray}
where $\cot{\delta_B} = \frac{1}{aq} + \frac{1}{2} r q$.
The values we have used 
for the scattering length ($a$) and effective range ($r$) parameters  of this
distribution are given in Table~\ref{tab:model}. The effective range part of the amplitude is cut off at $m_{K\pi}^{\rm cutoff} = 1800\mevcc$.
Integrating separately the resonant part, the effective range part, and the coherent sum we find that the $\KstarII$ resonance accounts for $81.7\%$, the effective range term $44.1\%$, and destructive interference between the two terms is responsible for the excess $25.8\%$.

A flat phase space term has been included in the signal model to account 
for NR $\Bz \to \Kspipi$ decays.

We determine a nominal signal Dalitz-plot model using
information from previous studies~\cite{Aubert:2005wb,Garmash:2006fh} and the
change in the fit likelihood value observed when omitting or adding resonances.
The components of the nominal signal model are summarized in Table \ref{tab:model}.
Other components, taken into account only to estimate the DP model uncertainty,
are discussed in Sec.~\ref{sec:Systematics}.

\begin{table}[htbp]
\begin{center}
\caption{Parameters of the DP model used in the fit. Values are given in  $\mev{\rm (}/c^2{\rm)}$, unless mentioned otherwise. The mass and width for the $f_X(1300)$ are averaged from results in $B^+\to K^+\pi^-\pi^+$ Dalitz analyses~\cite{Aubert:2008bj,Garmash:2005rv}.
\label{tab:model}}
\resizebox{\columnwidth}{!}{
\begin{tabular}{cccc}
\hline\hline
Resonance      & Parameters                      & Lineshape   & Ref. for         \\
               &                                 &             & Parameters \\ \hline\\[-9pt]
$\fI$          & $m_0=965 \pm 10$                & Flatt\'e    & \cite{valFlatte} \\
               & $g_{\pi}=165 \pm 18$            &             &                  \\
               & $g_{K}=695 \pm 93$              &             &                  \\ \hline\\[-9pt]
$\rhoI$        & $m_0=775.5 \pm 0.4$             & GS          & \cite{Amsler:2008zz}       \\
               & $\Gamma_0=146.4 \pm 1.1$        &             &                  \\ 
	        & $r=5.3^{+0.9}_{-0.7}\,(\gevc)^{-1}$  &             &                  \\ \hline\\[-9pt]
$K^{*+}(892)$  & $m_0=891.66 \pm 0.26$           & RBW         & \cite{Amsler:2008zz}       \\
$K^{*-}(892)$  & $\Gamma_0=50.8 \pm 0.9$         &             &                  \\ 
               & $r=3.6\pm 0.6\,(\gevc)^{-1}$    &             &                  \\  \hline\\[-9pt]
$(K\pi)^{*+}_0$& $m_0=1415 \pm 3$                & LASS        & \cite{Aubert:2008bj}   \\
$(K\pi)^{*-}_0$& $\Gamma_0=300 \pm 6$            &             &                  \\
               & $m_{K\pi}^{\rm cutoff}=1800$    &             &                  \\ 
               & $a=2.07\pm0.10\,(\gevc)^{-1}$   &             &                  \\
               & $r=3.32\pm0.34\,(\gevc)^{-1}$   &             &                  \\ \hline\\[-9pt]
$f_2(1270)$    & $m_0= 1275.4\pm1.1$             & RBW         & \cite{Amsler:2008zz}   \\
               & $\Gamma_0=185.2^{+3.1}_{-2.5}$  &             &                  \\ 
               & $r=3.0\,(\gevc)^{-1}$           &             &                  \\ \hline\\[-9pt]
$f_X(1300)$    & $m_0=1471 \pm 7$                & RBW         & \cite{Aubert:2008bj,Garmash:2005rv}\\
               & $\Gamma_0= 97\pm 15$            &             &                  \\ \hline\\[-9pt]
NR decays      &                                 & flat phase space &             \\ \hline\\[-9pt]
$\chiczero$    & $m_0=3414.75\pm0.35$            & RBW         & \cite{Amsler:2008zz}   \\
               & $\Gamma_0=10.4 \pm 0.7$         &             &                  \\ 
\hline \hline
\end{tabular}
}
\end{center}
\end{table}

\subsection{Time dependence}

With $\deltat \equiv t_{\rm sig} - t_{\rm tag}$ defined as the proper 
time interval between the decay of the fully reconstructed $\BztoKspipi$
($B^0_{{\rm sig}}$)
and that of the  other meson ($\Bz_{\rm tag}$) from the \FourS,  the time-dependent decay
rate $\AmpAllSigp$ ($\AmpAllSigm$) when the $\Bz_{\rm tag}$ is a $\Bz$ ($\Bzb$) 
is given by 
\beqn
\label{eq:dt}
    \AmpAllSigpm&
        =&
                \frac{e^{-|\dmt|/\tau_{B^0}}}{4\tau_{B^0}}
        \bigg[\absAmptp^2 + \absAmptpbar^2\nonumber\\
&&            \mp \left(\absAmptp^2 - \absAmptpbar^2\right)\cos(\dmd\dmt)\nonumber\\
&&            \pm\,2\I\left[\Amptpbar\Amptp^*\right]\sin(\dmd\dmt)   
        \bigg]~,
\eeqn
where $\tau_{B^0}$ is the  neutral $B$ meson lifetime and $\deltamd$ is the $\BzBzb$ 
mass difference. In the last formula and in the following, the DP dependence 
of the amplitudes is implicit.
Here, we have assumed that 
there is no \CP violation in mixing, and have used a
convention whereby the phase from $\BzBzb$ mixing is absorbed into
the $\Bzb$ decay amplitude (\ie~into the $\bar{c}_j$ terms). In other words, we assume that the $\BzBzb$ mixing parameters satisfy $|q/p|=1$ and absorb $q/p$ into $\bar{c}_j$.
Lifetime differences in the neutral $B$ meson system are assumed to be negligible.

\subsection{The square Dalitz plot}
\label{sec:SquareDP}

Both the signal events and the combinatorial $\epem\to q\bar q$ ($q=u,d,s,c$) 
continuum background events populate the kinematic boundaries of the 
DP due to the low final state masses compared with the $\Bz$ mass. 
The representation in Eq.~\eqref{eq:partialWidth} is inconvenient when
empirical reference shapes are to be used.
Large variations occurring in small areas of the DP are very difficult to describe in detail.
We therefore apply the transformation
\beq
\label{eq:SqDalitzTrans}
        d\spz \,d\smz \;\longrightarrow \detJ\, d\mprime\, d\thetaprime~,
\eeq
which defines the square Dalitz plot (SDP).
The new coordinates 
are
\beq
\label{eq:SqDalitzVars}
        \mprime \equiv \frac{1}{\pi}
                \arccos\left(2\frac{\mpm - \mpmMin}{\mpmMax - \mpmMin}
                        - 1
                      \right),~
        \thetaprime \equiv \frac{1}{\pi}\theta_{0}~,
\eeq
where $\mpm=\sqrt{s_0}$ is the $\pip\pim$ invariant mass,
$\mpmMax=\mBz - m_{\KS}$ and $\mpmMin=2m_{\pi^+}$ are the kinematic
limits of $\mpm$, $\theta_{0}$ is the $\pi^+ \pi^-$ resonance helicity angle
and $J$ is the Jacobian of the transformation.
Both variables  range between 0 and 1.
The determinant of the Jacobian is given by
\beq
\label{eq:detJ}
        \detJ \;=\;     4 \,|{\bf p}^*_+||{\bf p}^*_0| \,\mpm
                        \cdot   
                        \frac{\partial \mpm}{\partial \mprime}
                        \cdot   
                        \frac{\partial \cos\theta_{0}}{\partial \thetaprime}~,
\eeq
where 
$|{\bf p}^*_+|=\sqrt{E^{*\,2}_+ - m_{\pi^+}^2}$ and
$|{\bf p}^*_0|=\sqrt{E^{*\,2}_0 - m_{\KS}^2}$, and where the $\pi^+$ ($\KS$) energy 
$E^*_+$ ($E^*_0$), is defined in the $\pi^+\pi^-$ rest frame.
This transformation was introduced in ~Ref.~\cite{Aubert:2005sk},
and has been used in several $B$ decay DP analyses.

\section{THE \babar\ DETECTOR AND DATASET}
\label{sec:babar}

The data used in this analysis were collected with the \babar\ 
detector at the \pep2\ asymmetric-energy $e^+e^-$ storage ring at 
SLAC between October 1999 and August 2006. The sample consists of
an integrated luminosity of
$347.3\;\mathrm{fb}^{-1}$, corresponding to $(383\pm3)\times10^{6}$ 
$B\Bbar$ pairs collected at the \FourS resonance (``on-resonance''), 
and $36.6$~\invfb collected about $40$~\mev 
below the~\FourS (``off-resonance'').

A detailed description of the \babar\ detector is presented in 
Ref.~\cite{babar}. The tracking system used for track and vertex 
reconstruction has two  components: a silicon vertex tracker 
(SVT) and a drift chamber (DCH), both operating within a 1.5~T 
magnetic field generated by a superconducting solenoidal magnet. 
Photons are identified in an electromagnetic calorimeter (EMC).
It surrounds a detector of internally reflected Cherenkov light 
(DIRC), which associates Cherenkov photons with tracks for particle 
identification. Muon candidates are identified with the
use of the instrumented flux return (IFR) of the solenoid.

\section{EVENT SELECTION AND BACKGROUNDS}
\label{sec:selection}

We reconstruct $\BztoKspipi$ candidates 
from pairs of 
oppositely-charged tracks and a $\KS\to\pip\pim$ candidate, which are required to form a good quality vertex.
In order to ensure that all events are within 
the DP boundaries, we constrain the  invariant mass of the final state to the $B$ mass.  
For the $\pi^+\pi^-$ pair from the $B$, 
we use information  from the tracking system, EMC, and DIRC to 
remove tracks consistent with electron, kaon, and proton hypotheses.  
In addition we require at least one track to be inconsistent with
the muon hypothesis based on information from the IFR.
The $\KS$ candidate is required to have a mass within $15\mevcc$ of 
the nominal $K^0$ mass~\cite{Amsler:2008zz},
%and a decay vertex well separated from the $B^0$ decay vertex.
and a lifetime significance of at least five standard deviations. The last requirement ensures that the decay vertices of the \Bz and the \KS\ are well separated.
In addition, combinatorial background is suppressed
by requiring the cosine 
of the angle between the $\KS$ flight direction and the vector connecting 
the $B$-daughter pions and the $\KS$ vertices to be greater than $0.999$.

A $B$-meson candidate is characterized kinematically by the energy-substituted 
mass $\mes\equiv\sqrt{(s/2+{\mathbf {p}}_i\cdot{\mathbf{p}}_B)^2/E_i^2-p_B^2}$
and energy difference $\de \equiv E_B^*-\half\sqrt{s}$, 
where $(E_B,\pvec_B)$ and $(E_i,\pvec_i)$ are the four-vectors
of the $B$-candidate and the initial electron-positron system,
respectively. The asterisk denotes the \FourS\  frame,
and $s$ is the square of the invariant mass of the electron-positron system.  
We require $5.272 < \mes <5.286\gevcc$ and $|\de|<0.065\gev$.
Following the calculation of these kinematic variables,
each of the $B$ candidates is refitted with its mass
constrained to the world average value of the $B$-meson
mass~\cite{Amsler:2008zz} in order to improve the DP position resolution,
and ensure that Eq.~\eqref{eq:magicSum} holds.

Backgrounds arise primarily from random combinations in continuum events.
To enhance discrimination between signal and continuum, we 
use a neural network (NN)~\cite{Gay:1995sm} to combine four discriminating variables: 
the angles with respect to the beam axis of the $B$ momentum and $B$ thrust 
axis in the \FourS\ frame, and the zeroth and second order monomials
$L_{0,2}$ of the energy flow about the $B$ thrust axis.  The monomials
are defined by $ L_n = \sum_i p_i\times\left|\cos\theta_i\right|^n$,
where $\theta_i$ is the angle with respect to the $B$ thrust axis of track
or neutral cluster $i$ and $p_i$ is the magnitude of its momentum. The sum excludes the
$B$ candidate and all quantities are calculated in the \FourS\ frame.
The NN is trained using off-resonance data as well as
simulated signal events, all of which passed the selection criteria.
The final sample of signal candidates 
is selected with a requirement on the NN output that retains $90\%$ of the signal
and rejects $71\%$ of the continuum.

The time difference $\deltat$ is obtained from the measured distance between 
the  positions of the $\Bz_{\rm sig}$ and 
$\Bz_{\rm tag}$ decay vertices, using the boost $\beta\gamma=0.56$ of 
the \epem\ system.
$\Bz$ candidates with $|\deltat|>20$~ps are rejected, as are candidates
for which the error on $\deltat$ is higher than $2.5$~ps.
To determine the flavor of $\Bz_{\rm tag}$ 
we use the $B$ flavor tagging algorithm of Ref.~\cite{BabarS2b}.
This algorithm combines several different signatures, such as charges, momenta, and decay angles of charged particles in the event to achieve optimal separation
between the two $B$ flavors.
This produces six mutually exclusive tagging categories:
lepton, two different kaon categories, slow pion, kaon-slow pion, and a category that uses a combination of other signatures.
We also retain untagged events in a seventh category since although these
events do not contribute to the measurement of the time-dependent
\CP\ asymmetries they do provide additional statistics for the measurements
of direct \CP\ violation and \CP-conserving quantities such as the
branching fractions~\cite{Gardner:2003su}.
Multiple \B 
candidates passing the full selection occur 
between $\sim1\%$ of the time for NR signal events and $\sim8\%$
of the time for $\Bz \to \fIKs$ signal events.
If an event has more than one candidate, 
we select one using a reproducible pseudo-random procedure based
on the event timestamp.

With the above selection criteria, we  obtain a signal efficiency determined 
from Monte Carlo (MC) simulation of $21-25\%$, depending on the position in the 
DP.  

Of the selected signal events, $8\%$ of $B^0 \to \rho^0\KS$, 
$6\%$ of $B^0 \to \KstarI^+\pim$ and $4\% $ of $B^0 \to \fI\KS$ events are
misreconstructed.  Misreconstructed events occur when a track
from the tagging $B$ is assigned to the reconstructed signal candidate. 
This occurs most often for  low-momentum tracks  and hence the misreconstructed events 
are concentrated in the corners of the DP.  Since these are also where the low-mass resonances 
overlap strongly with other resonances, it is important to model the misreconstructed events correctly. 
The model used to account for misreconstructed events is detailed in Sec.~\ref{sec:deltaT}.

We use MC events to study the background from other $B$ 
decays ($B$ background). More than fifty channels were considered in 
preliminary studies, of which twenty are  included
in the final likelihood model -- those with at least two events expected after
selection.
These exclusive \B background modes are grouped into ten 
different classes that gather decays with similar kinematic and topological
properties: nine for neutral \B decays, one of which accounts for inclusive decays,
and one for inclusive charged \B decays.

Table \ref{tab:bbackground} summarizes the ten $B$ background classes that are
used in the fit.
The yields of those classes that have a clear signature in the DP are
allowed to float in the maximum-likelihood fit, the remainder are fixed.
When the yield of a class is varied in the maximum-likelihood fit
the quoted number of events corresponds to the fit results.
For the other modes, the expected numbers of selected events are
computed by multiplying the selection efficiencies (estimated using MC
simulated decays) by the world average branching fractions~\cite{Barberio:2008fa,Amsler:2008zz},
scaled to the data set luminosity ($347\;\mathrm{fb}^{-1}$).

\begin{table*}[t]
\begin{center}
\caption{ \label{tab:bbackground}
	Summary of \B background modes included in the fit model.
	When the yield is varied in the fit, the quoted number of events
	corresponds to the fit results. Otherwise, the expected number, taking into account the
	branching ratios and efficiency, is given.}
\setlength{\tabcolsep}{0.0pc}
\begin{tabular*}{\textwidth}{@{\extracolsep{\fill}}lccc}
\hline\hline
Mode                                  & Varied & BR                                    & Number of events \\
\hline\\[-9pt]
 $B^0 \rar \D^-(\to\KS\pim)\pip$         & yes & $\cdots$                              & $ 3377 \pm 60  $ \\
 $B^0 \rar \jpsi(\to l^+ l^-)\KS$        & yes & $\cdots$                              & $ 1803 \pm 43  $ \\
 $B^0 \rar \psitwos \KS$  		 & yes & $\cdots$                              & $ 142  \pm 13  $ \\
 $B^0 \rar \eta^\prime\KS$               & yes & $\cdots$                              & $ 37   \pm 16  $ \\
 $B^0 \rar a_1^{\pm} \pi^{\mp}$          & no  & $(39.7 \pm 3.7) \times 10^{-6}$ & $ 7.3  \pm 0.7 $ \\
 $B^0 \rar \D^{*-}(\to D\pi) \pip $      & no  & $(2.57 \pm 0.10)\times 10^{-3}$ & $ 43.8 \pm 2.5 $ \\
 $B^0 \rar \D^- h^+ \ \ \text{;}\ \ B^0 \rar \D^-\mu^+\nu_{\mu}$
                                         & no  & $(2.94 \pm 0.19)\times 10^{-3}$ & $ 281  \pm 20  $ \\
 $B^0 \rar \D^{*-} \rho^+$               & no  & $(14.2 \pm 1.4) \times 10^{-3}$ & $ 34.5 \pm 4.6 $ \\
\hline\\[-9pt]
$B^0 \rar \{\text{neutral generic decays}\}$ 	& no  & not applicable             & $114\pm 7$  \\
$B^+ \rar \{\text{charged generic decays}\}$ 	& no  & not applicable             & $282\pm 11$ \\
\hline\hline
\end{tabular*}

\end{center}
\end{table*}
\section{THE MAXIMUM-LIKELIHOOD FIT}
\label{sec:ML}

We perform an unbinned extended maximum-likelihood fit to extract
the inclusive $\BztoKspipi$ event yield and the resonant amplitudes.  
The fit uses the variables $\mes$, $\de$, the NN output, and the 
SDP to discriminate signal from background. The 
$\dt$ measurement allows the determination of mixing-induced \CP violation
and provides additional continuum background rejection. 

The selected on-resonance data sample is assumed to consist of signal,
continuum background, and \B background components.
The signal likelihood consists of the sum of a correctly 
reconstructed (``truth-matched,'' TM) term and a misreconstructed 
(``self-cross-feed,'' SCF) term.
Generally, the components in the fit are separated by the flavor and tagging category of the tag side \B decay.

The probability density function (PDF) ${\cal P}_i^\cat$ for an
event $i$ in tagging category $\cat$ is the sum of the probability densities 
of all components, namely
\begin{eqnarray}
\label{eq:theLikelihood}
        {\cal P}_i^\cat
        &\equiv& 
                N_{\rm sig} f^\cat_{\rm sig}
                \left[  (1-\fscfave^\cat){\cal P}_{{\rm sig}-\TM,i}^\cat +
                        \fscfave^\cat{\cal P}_{{\rm sig}-\SCF,i}^\cat 
                \right] 
                \nonumber\\[0.3cm]
        &&
                +\; N^\cat_{q\bar q}\frac{1}{2}
                \left(1 + \Qtagi\Atagqq\right){\cal P}_{q\bar q,i}^\cat
                \nonumber \\[0.3cm]
        &&
                +\; N_{B^+} f^\cat_{B^+}
                \frac{1}{2}\left(1 + \Qtagi \Atag\right){\cal P}_{B^+,i}^\cat
                \nonumber \\[0.3cm]
        &&
                +\; \sum_{j=1}^{N^{B^0}_{\rm class}}
                N_{B^0j} f^\cat_{B^0j}
                {\cal P}_{B^0,ij}^\cat~.
\end{eqnarray}
The variables are defined in Table~\ref{tab:DefVarLik}.
\renewcommand{\arraystretch}{1.15}
\begin{table*}[htbp]
\begin{center}
\caption{
Definitions of the different variables in the likelihood function given in Eq.~\eqref{eq:theLikelihood}.
\label{tab:DefVarLik}}
\begin{tabular}{ll}
\hline\hline
Variable           & Definition \\
\hline
$N_{\rm sig}$      & total number of $\Kspipi$ signal events in the data sample              \\
$f^\cat_{\rm sig}$ & fraction of signal events that are tagged in category $\cat$            \\
$\fscfave^\cat$    & fraction of SCF events in tagging category $\cat$, averaged over the DP \\
${\cal P}_{{\rm sig}-\TM,i}^\cat$  & product of PDFs of the discriminating variables used in tagging category $\cat$ for TM events  \\
${\cal P}_{{\rm sig}-\SCF,i}^\cat$ & product of PDFs of the discriminating variables used in tagging category $\cat$ for SCF events \\
$N^\cat_{q\bar q}$ & number of continuum events that are tagged in category $\cat$ \\
$\Qtagi$           & tag flavor of the event, defined to be $+1$ for a $\Bz_{\rm tag}$ and $-1$ for a $\Bzb_{\rm tag}$ \\
$\Atagqq$          & parameterizes possible asymmetry in continuum events \\
${\cal P}_{q\bar q,i}^\cat$        & continuum PDF for tagging category $\cat$ \\
$N^{B^0}_{\rm class}$              & number of neutral $B$-related background classes considered in the fit, namely nine \\
$N_{B^+}$          & number of expected charged $B$ background events \\
$N_{B^0j}$         & number of expected events in the neutral $B$ background class $j$ \\
$f^\cat_{B^+}$     & fraction of charged $B$ background events that are tagged in category $\cat$ \\
$f^\cat_{B^0j}$    & fraction of neutral $B$ background events of class $j$ that are tagged in category $\cat$ \\
$\Atag$            & describes a possible asymmetry in the charged $B$ background \\
${\cal P}_{B^+,i}^\cat$            & $B^+$ background PDF for tagging category $\cat$ \\
${\cal P}_{B^0,ij}^\cat$           & neutral $B$ background PDF for tagging category $\cat$ and class $j$\\[2pt]
\hline\hline
\end{tabular}
\end{center}
\end{table*}
\renewcommand{\arraystretch}{1.}
The PDFs ${\cal P}_{X}^{\cat}$ ($X=\{{\rm sig}\!-\!\TM,\  {\rm sig}\!-\!\SCF,\  q\bar q,\  B^+,\  B^0$\})
are the product of the four PDFs of the discriminating variables,
$x_1 = \mes$, $x_2 = \de$, $x_3 = {\rm NN~output}$, and the triplet
$x_4 = \{\mprime, \thetaprime, \deltat\}$:
\begin{equation}
\label{eq:likVars}
        {\cal P}_{X,i(j)}^{\cat} \;\equiv\; 
        \prod_{k=1}^4 P_{X,i(j)}^\cat(x_k)~,
\end{equation}
where $i$ is the event index and $j$ is a $B$ background class.
Not all the PDFs depend on the tagging category;
the general notations $P_{X,i(j)}^\cat$ and ${\cal P}_{X,i(j)}^{\cat}$ are used for simplicity.
Correlations between the tag and the position in the DP are absorbed in tag-flavor-dependent
SDP PDFs that are used for continuum and charged \B
backgrounds. The parameters $\Atag$ and $\Atagqq$ parametrize any potential asymmetry between these PDFs.
The extended likelihood over all tagging categories is given by
\begin{equation}
        {\cal L} \;\equiv\;  
        \prod_{\cat=1}^{7} e^{-\bar N^\cat}\,
        \prod_{i}^{N^\cat} {\cal P}_{i}^\cat~,
\end{equation}
where $\bar N^\cat$ is the total number of events expected in category 
$\cat$. 

A total of $75$ parameters are varied in the fit. They include the $12$ inclusive yields (signal, four \B background
classes, and seven continuum yields, one per tagging category), $30$ parameters for the complex amplitudes from
Eq.~\eqref{eq:dt}, and $33$ parameters of the different PDFs. The latter include most of the
parameters describing the continuum distributions.

\subsection{The $\dt$ and Dalitz plot PDFs}
\label{sec:deltaT}

        The SDP PDFs require as input the DP-dependent 
        selection efficiency, $\varepsilon=\varepsilon(\mprime,\thetaprime)$, 
        and SCF fraction, $\fscf=\fscf(\mprime,\thetaprime)$.
        Both quantities are taken from MC simulation. 
        Away from the DP corners the efficiency is uniform. It 
        decreases when approaching the corners, where one of the 
        three particles in the final state is nearly at rest so that the 
        acceptance requirements on the particle reconstruction become 
        restrictive.
        Combinatorial backgrounds and hence SCF fractions are large in
        the corners
        of the DP due to the presence of soft tracks.

        For an event~$i$ we define the time-dependent SDP PDFs
        \begin{eqnarray}
        \label{eq:DPpdfSigTM}
                \lefteqn{ P_{{\rm sig}-\TM,i}(\mprime, \thetaprime, \deltat) = } \\ \nonumber
                && \varepsilon_i\,(1 - \fscfi)\,\detJi\,\AmpAll~,
        \end{eqnarray}   
        \begin{eqnarray}
        \label{eq:DPpdfSigSCF}
                \lefteqn{ P_{{\rm sig}-\SCF,\,i}(\mprime, \thetaprime, \deltat) =}\\ \nonumber
                && \varepsilon_i\,\fscfi\,\detJi\,\AmpAll~,
        \end{eqnarray}   
        where $P_{{\rm sig}-\TM,i}(\mprime, \thetaprime, \deltat)$
        and $P_{{\rm sig}-\SCF,\,i}(\mprime, \thetaprime, \deltat)$ are normalized to unity. The 
        phase space integration involves the expectation values        
        $\langle \varepsilon\,(1-\fscf)\,\detJ \,F_k F^*_{k^\prime}\rangle$
        and 
        $\langle \varepsilon\,\fscf\,\detJ\, F_k F^*_{k^\prime}\rangle$
        for TM and SCF events, where the indices $k$, $k^\prime$ 
        run over all resonances belonging to the signal model.
        The expectation values are model-dependent and are 
        computed by MC integration over the SDP:
        \begin{eqnarray}
        \label{eq:normAverage}
               \lefteqn{\langle \varepsilon\,(1-\fscf)\,\detJ\, F_k F^*_{k^\prime}\rangle\;=}\\ \nonumber
               && \frac{\int_0^1\int_0^1 	
                            \varepsilon\,(1-\fscf)\,\detJ\, F_k F^*_{k^\prime}
                        \,d\mprime d\thetaprime}
                       {\int_0^1\int_0^1 \varepsilon\,\detJ\, F_k F^*_{k^\prime}
                        \,d\mprime d\thetaprime}~,
        \end{eqnarray}
        and similarly for 
        $\langle \varepsilon\,\fscf\,\,\detJ\, F_k F^*_{k^\prime}\rangle$,
        where all quantities in the integrands are DP-dependent.

        Equation~\eqref{eq:theLikelihood} invokes the phase 
        space-averaged SCF fraction 
        $\fscfave\equiv\langle\fscf\,\detJ\, F_k F^*_{k^\prime}\rangle$. 
        The PDF normalization  is decay-dynamics-dependent
        and is computed iteratively. We 
        determine the average SCF fractions separately for each tagging category 
        from MC simulation. 
        
        The width of the dominant 
        resonances are large compared 
        to the mass resolution for TM events (about $8\mevcc$ core Gaussian
        resolution). We  therefore neglect resolution effects in the TM 
        model.  
        Misreconstructed events have a poor mass resolution that strongly 
        varies across the DP. It is described in the fit by a 
        $2\times 2$-dimensional resolution function
        \begin{equation}
        \label{eq:rscf}
                \Rscf(\mprime_r,\thetaprime_r,\mprime_t,\thetaprime_t)~,
        \end{equation}
        which represents the probability to reconstruct at the coordinates
        $(\mprime_r,\thetaprime_r)$ an event that has the true coordinates 
        $(\mprime_t,\thetaprime_t)$. It obeys the unitarity condition
        \begin{equation}
                \intl_0^1\intl_0^1 
                \Rscf(\mprime_r,\thetaprime_r,\mprime_t,\thetaprime_t)
                \,d\mprime_r d\thetaprime_r = 1~,
        \end{equation}
        and is convolved with the signal model. 
        The $\Rscf$ function is obtained from MC simulation.

\begin{figure*}[htbp]
  \centerline{  \epsfxsize7cm\epsffile{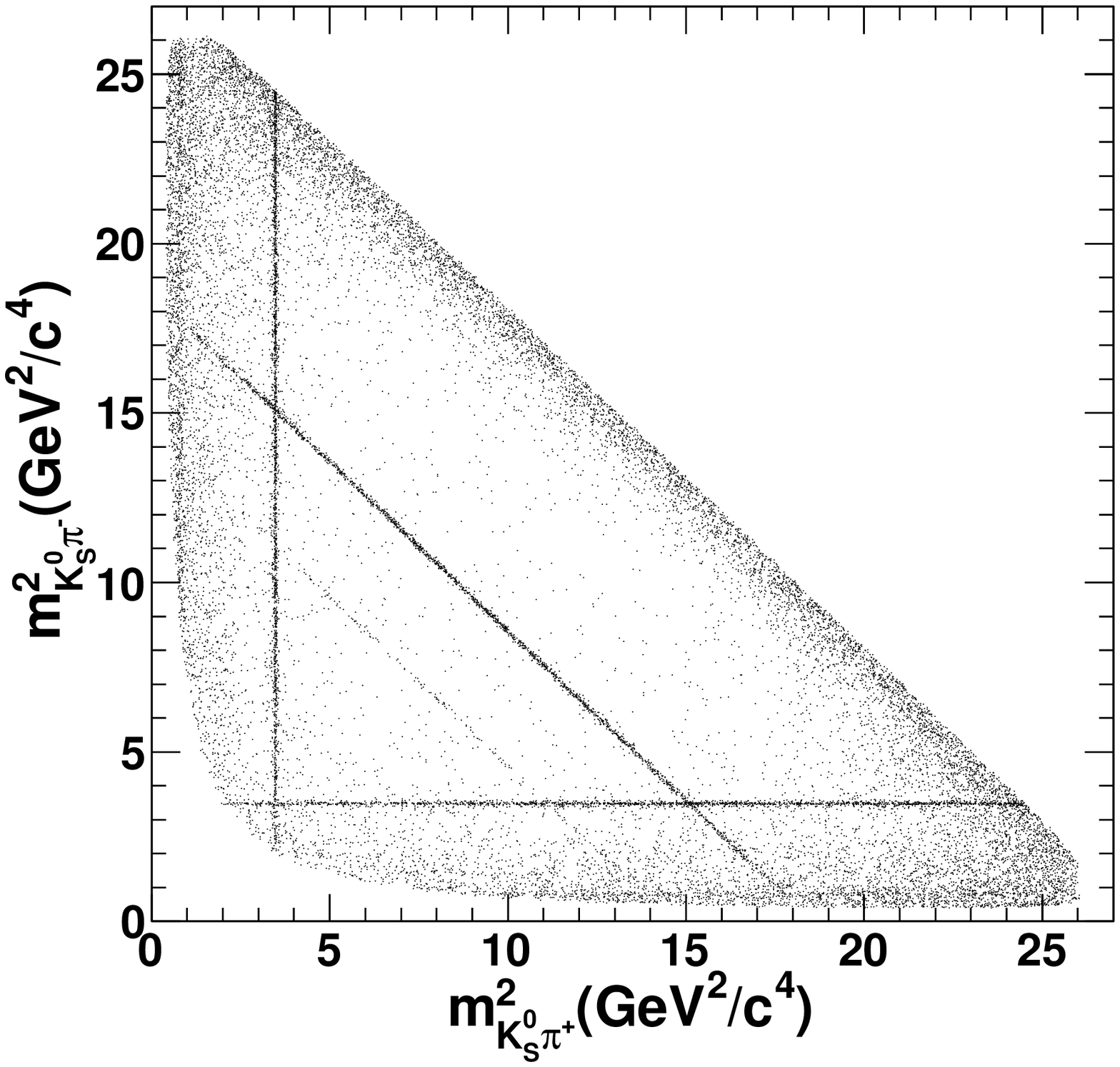}
 \epsfxsize7cm\epsffile{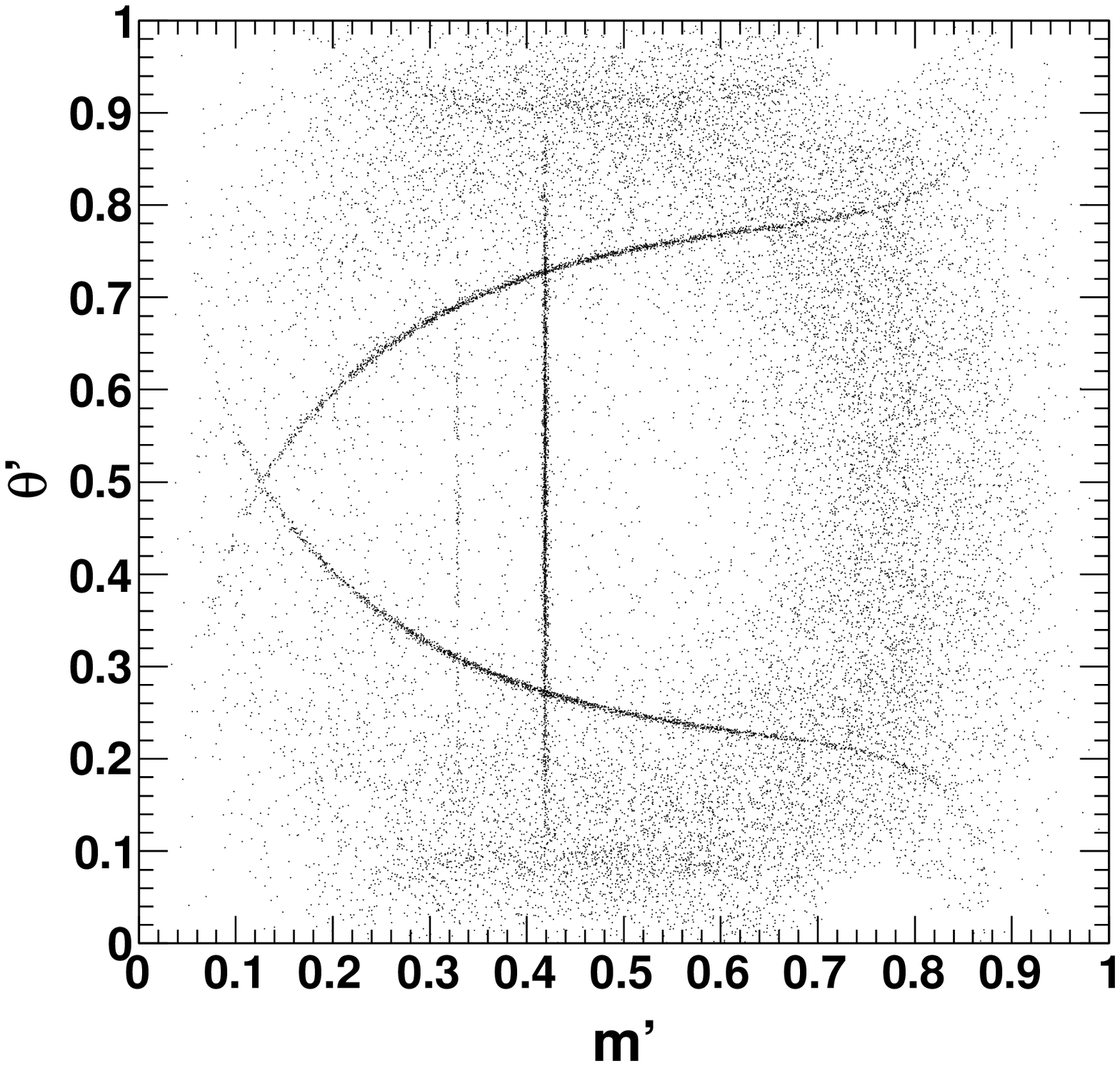}}
  \caption{\label{fig:dataSample} Standard (left) and square (right) Dalitz plots of the selected data sample of $22525$ $\B\to\Kspipi$
candidates. The narrow bands correspond to  $\D^\mp \pipm$,  $J/\psi K^0_S$ and $\psi(2S)K^0_S$
background events.}
\end{figure*}

        We use the signal model described in Sec.~\ref{sec:kinematics}. 
        It contains the dynamical information and is connected with $\dt$ via 
        the matrix element in Eq.~\eqref{eq:dt}, which intervenes in the signal PDFs defined in
        Eq.~\eqref{eq:DPpdfSigTM} and~\eqref{eq:DPpdfSigSCF}.
        The PDFs are diluted 
        by the effects of mistagging and the limited vertex 
        resolution~\cite{Aubert:2007jn}. 
        The $\deltat$ resolution function for signal (both TM and SCF) and \B background 
        events is a sum of three Gaussian distributions. The parameters of the signal resolution function are 
        determined by a fit to fully reconstructed $\Bz$ 
        decays~\cite{BabarS2b}. 

        The charged \B background
                contribution to the likelihood, given in Eq.~\eqref{eq:theLikelihood},
                uses distinct SDP PDFs for each 
                reconstructed $B$ flavor tag, and a flavor-tag-averaged PDF for 
                untagged events. The PDFs are obtained from MC simulation and are 
                described by histograms.
                The $\dt$ resolution parameters are determined by a fit to fully 
                reconstructed $\Bp$ decays. 
                For the $\Bp$ background class we adjust the 
                effective lifetime to account for the misreconstruction of the 
                event that modifies the nominal $\dt$ resolution function.

        The neutral $B$ background is parameterized with PDFs that
                depend on the flavor tag of the event. In the case of \CP
                eigenstates, correlations between the flavor tag and the Dalitz 
                coordinates are expected to be small. However, non-\CP  eigenstates,
                such as $a_1^\pm\pi^\mp$, may exhibit such correlations. Both types 
                of decays can have direct
                and mixing-induced \CP  violation. A third type of decay
                involves charged $\D$ mesons and does not exhibit mixing-induced
                \CP  violation, but usually has a strong correlation between the
                flavor tag and the DP coordinates because 
                it consists of $B$-flavor eigenstates. Direct \CP violation is also
                possible in these decays, though it is set to zero in the nominal model.
                The DP PDFs are obtained from MC simulation and are 
                described by histograms.
                For neutral $B$ background, the signal $\dt$ resolution model 
		            is assumed. Note that the SDP- and $\dt$-dependent PDFs factorize for the 
                charged $B$ background modes, but not necessarily
                for the neutral $B$ background due to $\BzBzb$ mixing.

        The DP
                treatment of the continuum events is similar to that used
                for charged $B$ background. 
                The SDP PDF for continuum background is 
                obtained from on-resonance events selected in the
                $\mes$ sidebands and corrected for feed-through
                from \B decays. A large number of cross checks have been 
                performed to validate the empirical shape 
                used.
                The continuum $\deltat$ distribution is parameterized as the sum of 
                three Gaussian distributions with common mean and 
                three distinct widths that scale with the $\dt$ per-event error. 
                This introduces six shape parameters that are determined by 
                the fit.
                The model is motivated by the observation that 
                the mean of the $\dt$ distribution is independent of the per-event error, and that the 
                width depends linearly on this error.

\subsection{Description of the other variables}
\label{sec:likmESanddE}

   The \mes\ distribution of TM signal events is
     parameterized by a bifurcated Crystal Ball 
     function~\cite{Skwarnicki:1986xj,Oreglia:1980cs,Gaiser:1982yw},
     which is a combination of a one-sided Gaussian and 
     a Crystal Ball function. The mean and the two widths of this function
     are determined by the fit.
   The \DeltaE\ distribution of TM signal events is
     parameterized by a double Gaussian function.
     The five parameters of this function are determined by the fit.
   Both \mes\ and \DeltaE\ PDFs are described by histograms, taken from the
     distributions found in appropriate MC samples, for SCF signal events and
     all \B\ background classes. Exceptions to this are the \mes\ PDFs
     for the $\Bz\to\Dm\pip$ and $\Bz\to\jpsi\KS$ components,
     and the \DeltaE\ PDF for $\Bz\to\Dm\pip$, which are the same as the
     corresponding distributions of TM signal events.
   The \mes\ and \DeltaE\ PDFs for continuum events are
     parameterized by an ARGUS shape function~\cite{argus} and
     a first-order polynomial, respectively, with parameters
     determined by the fit.

   We use histograms to empirically describe the distributions of the NN output
     found in the MC simulation for TM and SCF signal events
     and for all \B\ background classes. We distinguish tagging categories
     for TM signal events to account for differences observed in the shapes.
   The continuum NN distribution is parameterized by a third-order polynomial that
     is constrained to take positive values in the range populated by the data.
     The coefficients of the polynomial are determined by the fit. 
     Continuum events exhibit a correlation between the DP coordinates and the
     shape of the event that is exploited in the NN. 
     To correct for this effect, we introduce a linear dependence of the polynomial
     coefficients on the variable $\Delta_{\rm DP}$, defined as the smallest of the
     three invariant masses, and is thus a measure of the distance of the DP coordinates
     from the kinematic boundaries of the DP. 
     The parameters describing this dependence are determined by the fit.
\section{FIT RESULTS}
\label{sec:fitResults}

\begin{figure*}[htbp]
\includegraphics[width=8.7cm,keepaspectratio]{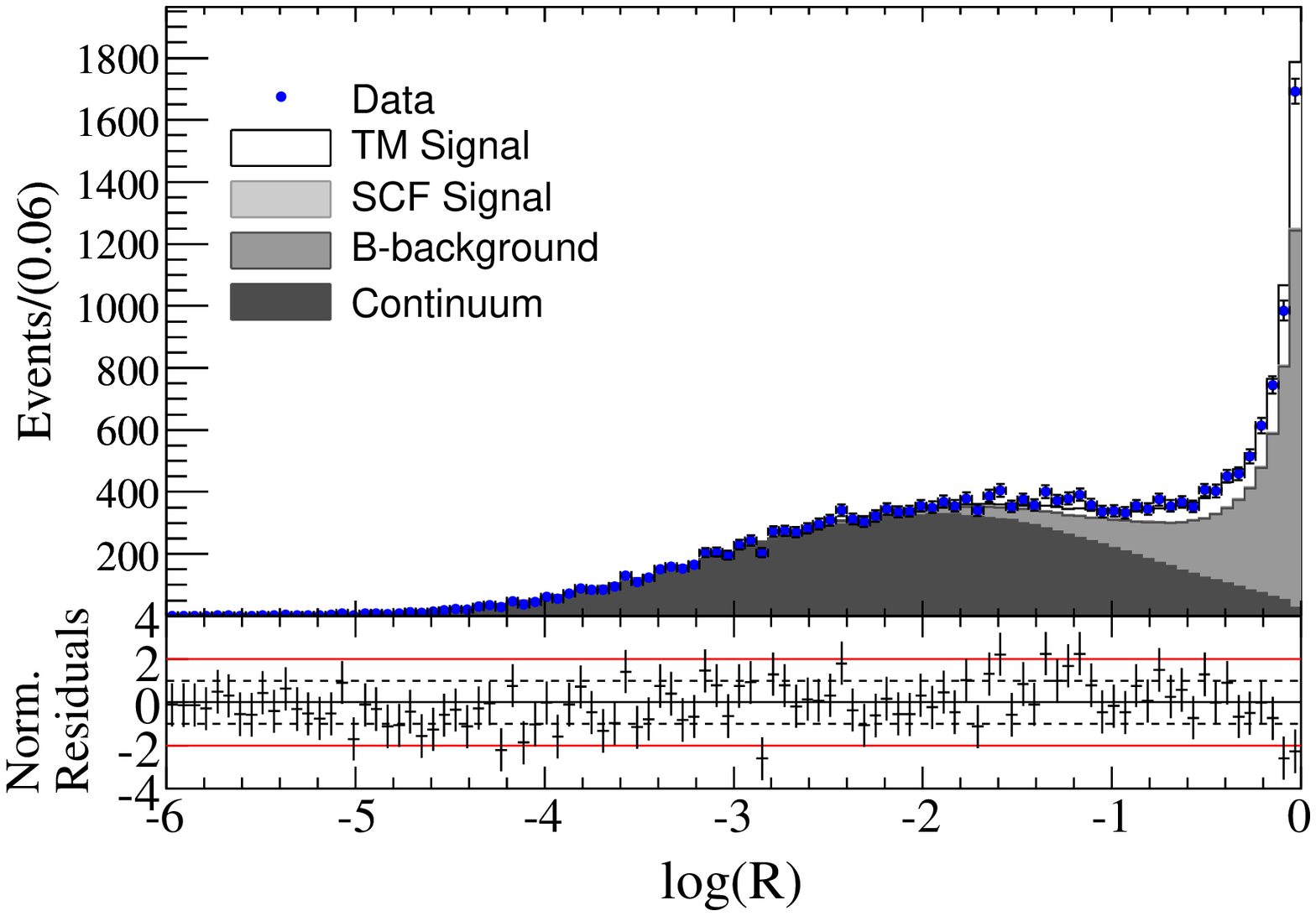}
\includegraphics[width=8.7cm,keepaspectratio]{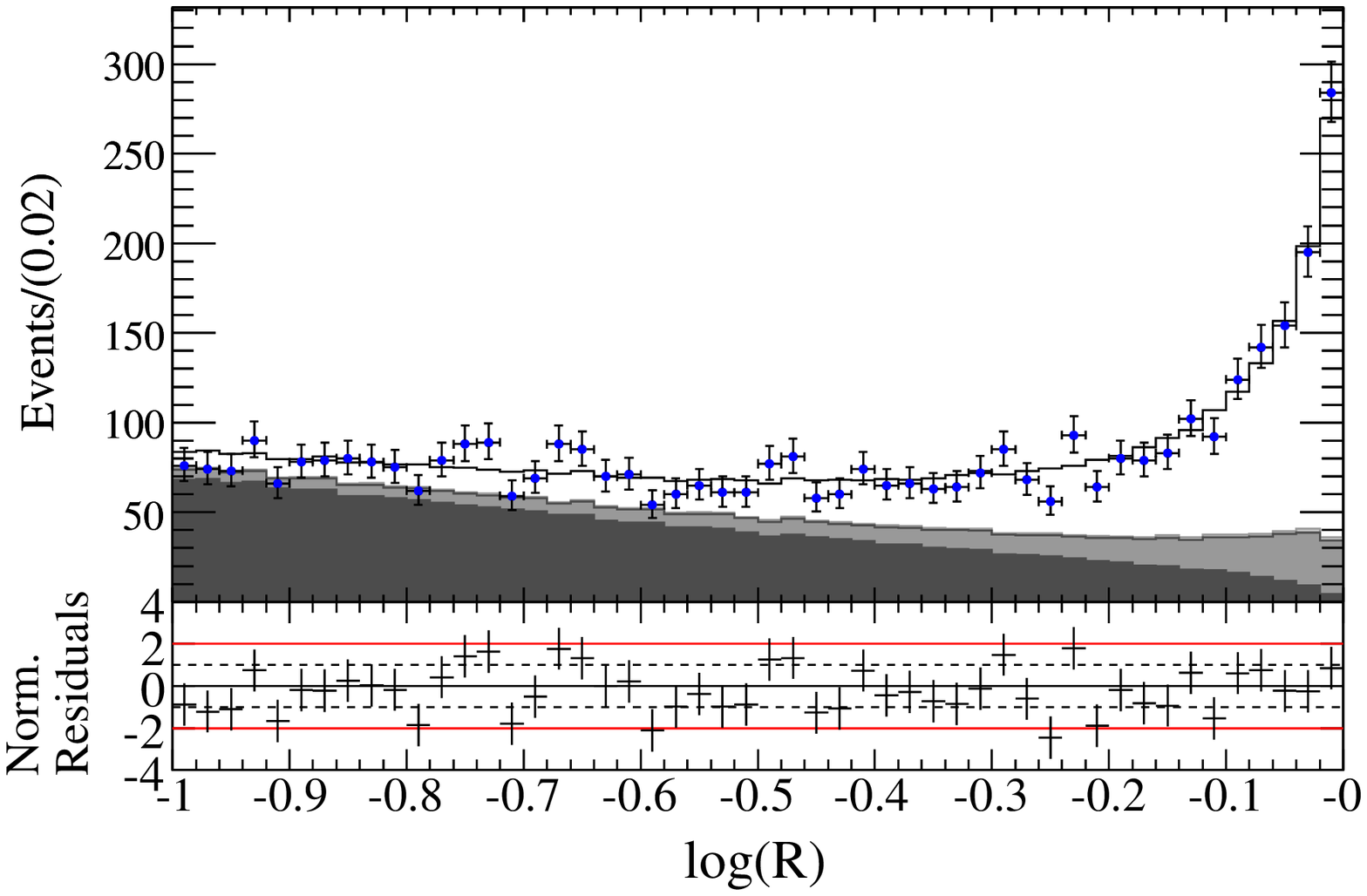}
  \caption{\label{fig:lkl} Distributions of the logarithm of likelihood ratio ($\log R$)
        for all events entering the fit (left)
        and in the signal-like region (right).
        In the right hand side plot, a veto in the $\Dm\pip$, $\jpsi \KS$,
        and $\psi(2S)\KS$ bands has been applied.
	      Points with error bars give 
        the on-resonance data. The solid histogram shows the
        projection of the fit result. The dark,
        medium, and light shaded areas represent respectively the contribution
        from continuum events, the sum of continuum events 
        and the $B$ background expectation, and the sum of these and 
        the misreconstructed signal events. The last contribution is hardly visible due to its
        small fraction. Below each bin are shown the residuals, normalized in error units. 
        The parallel dotted and full lines are the $1\sigma$ and $2\sigma$ deviations.
        Points, histograms, shaded areas, and residual plots have similar definitions
        in Fig.~\ref{fig:projections} to~\ref{fig:dtAssym}}
\end{figure*}

\begin{table*}[htbp]
\begin{center}
\caption{
Results of fit to data for the isobar amplitudes with statistical uncertainties.  Both solutions are shown.
\label{tab:resultsAmps}}
\begin{tabular}{ccccc}
\hline\hline
                                     & \multicolumn{2}{c}{Solution I}           & \multicolumn{2}{c}{Solution II}        \\ 
Isobar Amplitude                     & Magnitude          & Phase    ($^\circ$)&  Magnitude          & Phase ($^\circ$)  \\
\hline
$c_{f_0(980)K^0_S}$            & $4.0$              &  $0.0$             & $4.0$              & $0.0$              \\
$\bar{c}_{f_0(980)K^0_S}$      & $3.7 \pm 0.4$      & $-73.9 \pm 19.6$   & $3.2 \pm 0.6$      &  $-112.3 \pm 20.9$ \\[3pt]
\hline
$c_{\rho(770)K^0_S}$           & $0.10 \pm 0.02$    & $35.6 \pm 14.9$    & $0.09 \pm 0.02$    & $66.7 \pm 18.3$\\
$\bar{c}_{\rho(770)K^0_S}$     & $0.11 \pm 0.02$    & $15.3 \pm 20.0$    & $0.10 \pm 0.03$    & $-0.1 \pm 18.2$\\[3pt]
\hline
$c_{K^{*+}(892)\pi^-}$         & $0.154 \pm 0.016$  & $-138.7 \pm 25.7$  & $0.145 \pm 0.017$  & $-107.0 \pm 24.1$\\
$\bar{c}_{K^{*-}(892)\pi^+}$   & $0.125 \pm 0.015$  & $163.1  \pm 23.0$  & $0.119 \pm 0.015$  & $76.4   \pm 23.0$\\[3pt]
\hline
$c_{(K\pi)_0^{*+}\pi^-}$       & $6.9 \pm 0.6$      & $-151.7 \pm 19.7$  & $6.5 \pm 0.6$      & $-122.5 \pm 20.3$\\
$\bar{c}_{(K\pi)_0^{*-}\pi^+}$ & $7.6 \pm 0.6$      & $136.2  \pm 19.8$  & $7.3 \pm 0.7$      & $52.6   \pm 20.3$\\[3pt]
\hline
$c_{f_2(1270)K^0_S}$           & $0.014 \pm 0.002$  & $5.8   \pm 19.2$   & $0.012 \pm 0.003$  & $23.9  \pm 22.7$\\
$\bar{c}_{f_2(1270)K^0_S}$     & $0.011 \pm 0.003$  & $-24.0 \pm 28.0$   & $0.011 \pm 0.003$  & $-83.3 \pm 24.3$\\[3pt]
\hline
$c_{f_X(1300) K^0_S}$          & $1.41 \pm 0.23$    & $43.2 \pm 22.0$    & $1.40 \pm 0.28$    & $85.9  \pm 24.8$\\
$\bar{c}_{f_X(1300) K^0_S}$    & $1.24 \pm 0.27$    & $31.6 \pm 23.0$    & $1.02 \pm 0.33$    & $-67.9 \pm 22.1$\\[3pt]
\hline
$c_{NR}$                       & $2.6 \pm 0.5$      & $35.3 \pm 16.4$    & $1.9 \pm 0.7$      & $56.7  \pm 23.6$\\
$\bar{c}_{NR}$                 & $2.7 \pm 0.6$      & $36.1 \pm 18.3$    & $3.1 \pm 0.6$      & $-45.2 \pm 17.8$\\[3pt]
\hline
$c_{\chi_{c0}K^0_S}$           & $0.33 \pm 0.15$    & $61.4 \pm 44.5$    & $0.28 \pm 0.16$    & $51.9  \pm 38.4$\\
$\bar{c}_{\chi_{c0}K^0_S}$     & $0.44 \pm 0.09$    & $15.1 \pm 30.0$    & $0.43 \pm 0.08$    & $-58.5 \pm 27.9$\\[3pt]
\hline
\hline
\end{tabular}
\end{center}
\end{table*}

The standard and square Dalitz plots of the selected data sample are shown in Fig.~\ref{fig:dataSample}.
The maximum-likelihood fit of $22525$ candidates results in a $\BztoKspipi$ event yield of
$2182\pm 64$ and a continuum yield of $14240\pm 126$, where the uncertainties are statistical only.
The remaining number of events is covered by the yields of backgrounds from charged and neutral $B$ decays,
where the dominant contributions are $3361 \pm 60$ $B^0 \to \D^-\pip$ and $1804 \pm 44$ $B^0 \to \jpsi \KS$ events.

When the fit is repeated starting from input parameter values randomly chosen within
wide ranges above and below the nominal values for the magnitudes and within the
$[0-360^{\circ}]$ interval for the phases, we observe convergence toward two solutions with 
minimum values of the negative log likelihood function $-2\logL$
that are equal within $0.32$ units. 
In the following, we refer to them as solution I (the global minimum) and solution II (a local minimum).
Between the two solutions, the fit values for most free parameters are very similar. Exceptions occur among isobar
parameters, and most particularly isobar phases, some of which can differ significantly.

\begin{figure*}[htbp]
\includegraphics[width=5.9cm,keepaspectratio]{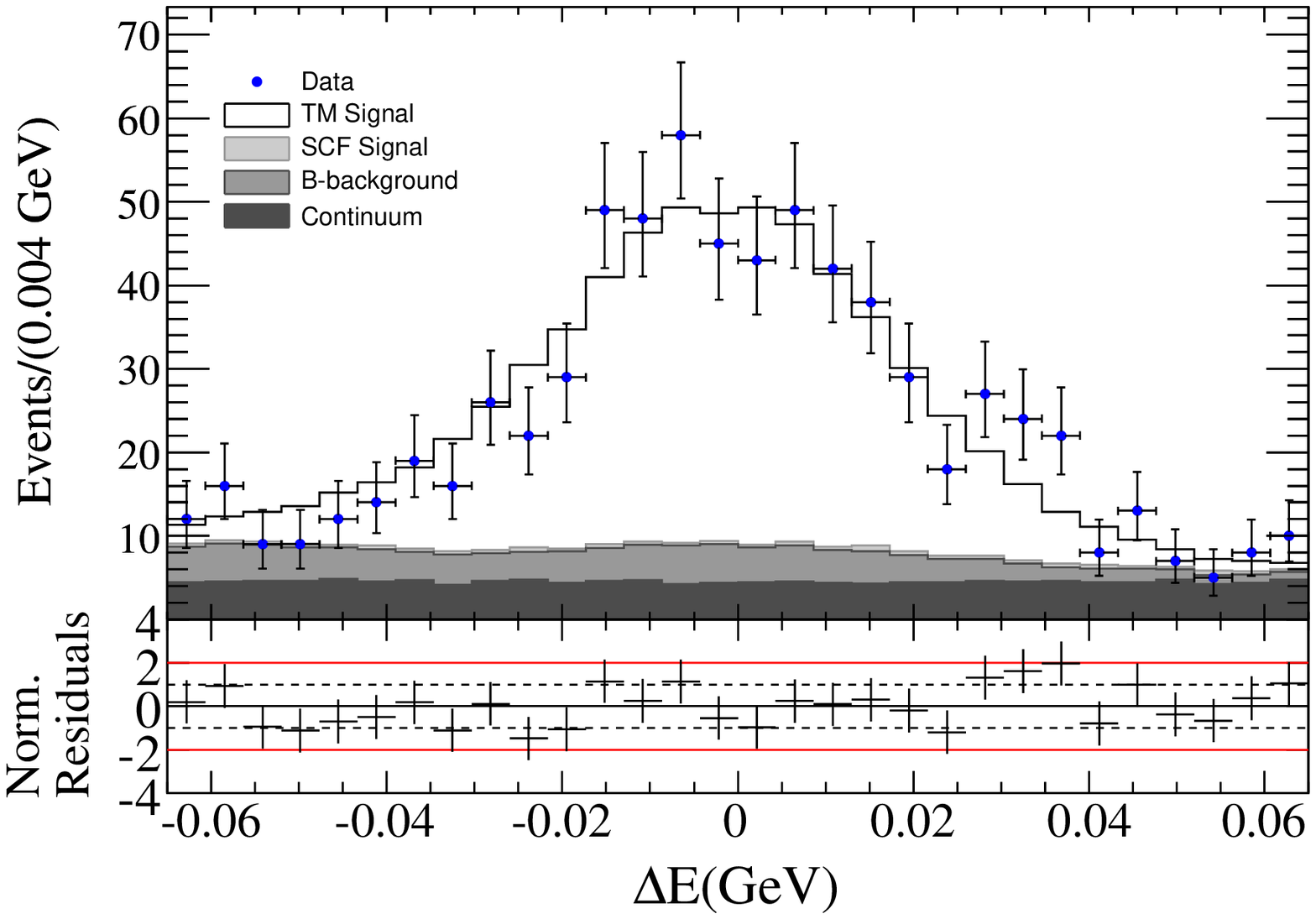}
\includegraphics[width=5.9cm,keepaspectratio]{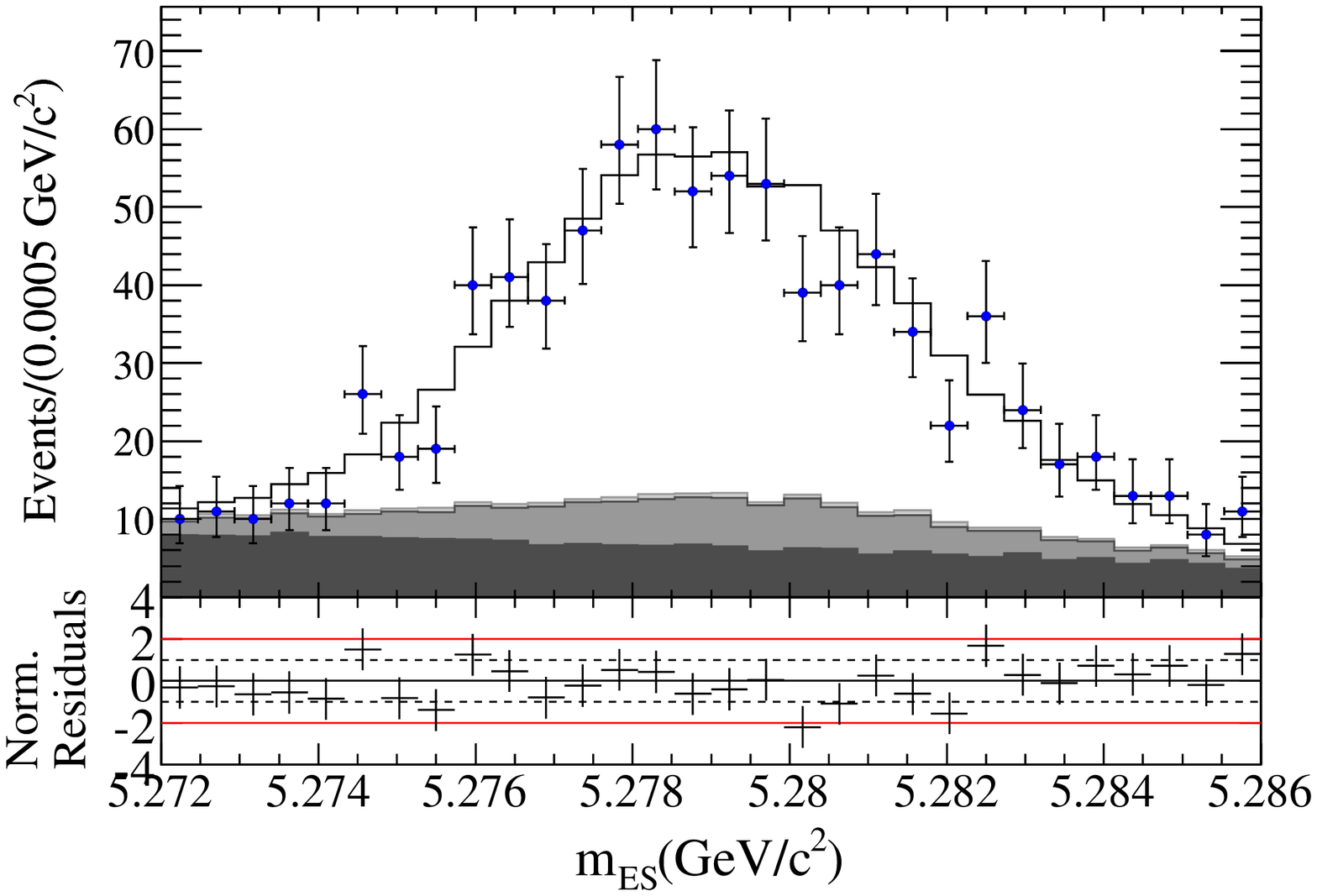}
\includegraphics[width=5.9cm,keepaspectratio]{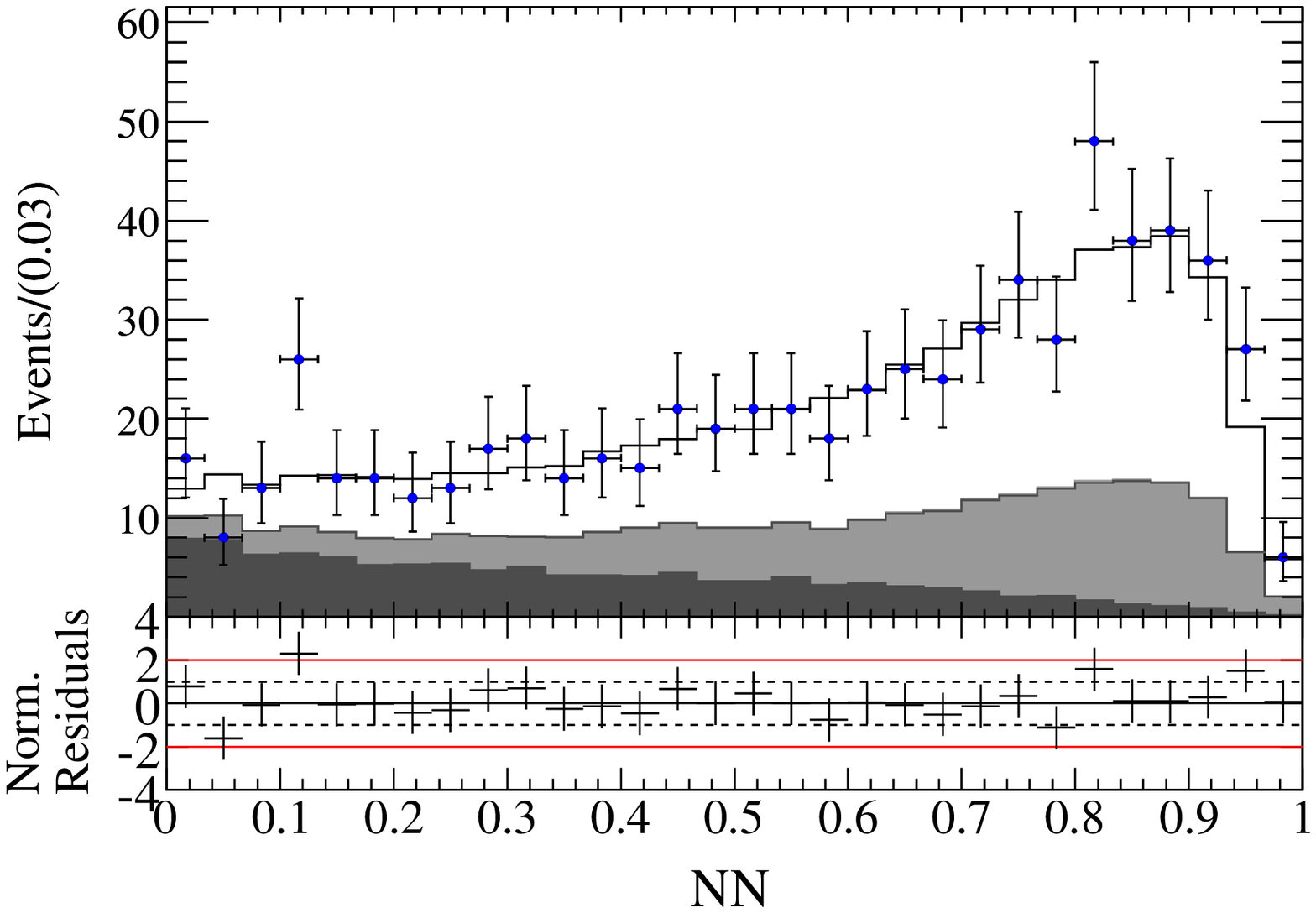}
  \caption{\label{fig:projections} Distributions of $\de$ (left), 
	$\mes$ (center), and $NN$ output (right) for  a sample 
        enhanced in $\BztoKspipi$ signal with a requirement on the likelihood ratio $R$
	computed without the variable being plotted. In each case the applied cut rejects $99\%$ of continuum background,
	while retaining $28\%$ of signal for $\de$ and $\mes$, and $16\%$ for $NN$. A veto in the $\Dm\pip$ and $\jpsi \KS$ bands has been applied.}
\end{figure*}

\begin{figure*}[htbp]
\begin{center}
\includegraphics[width=8.7cm,keepaspectratio]{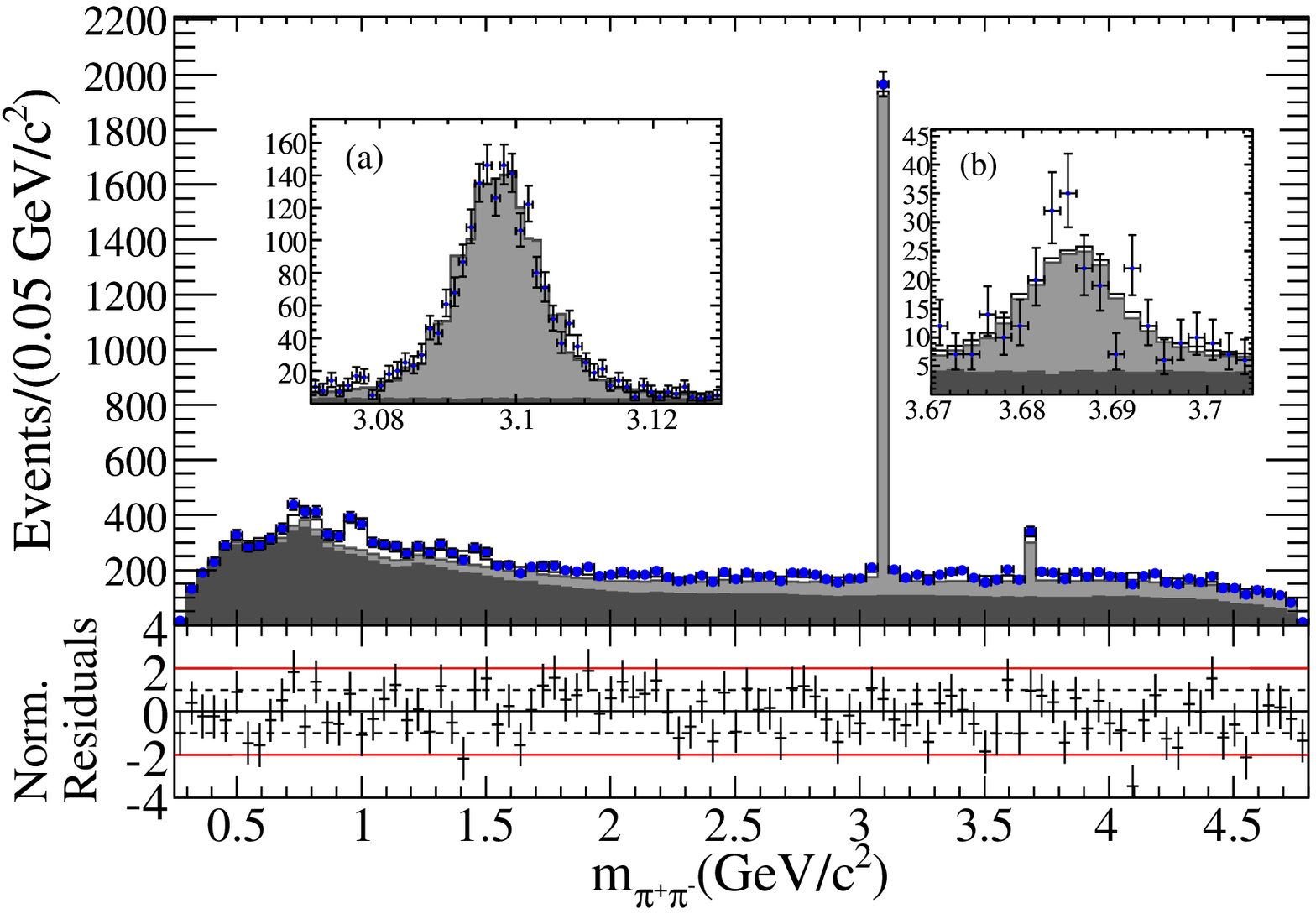}
\includegraphics[width=8.7cm,keepaspectratio]{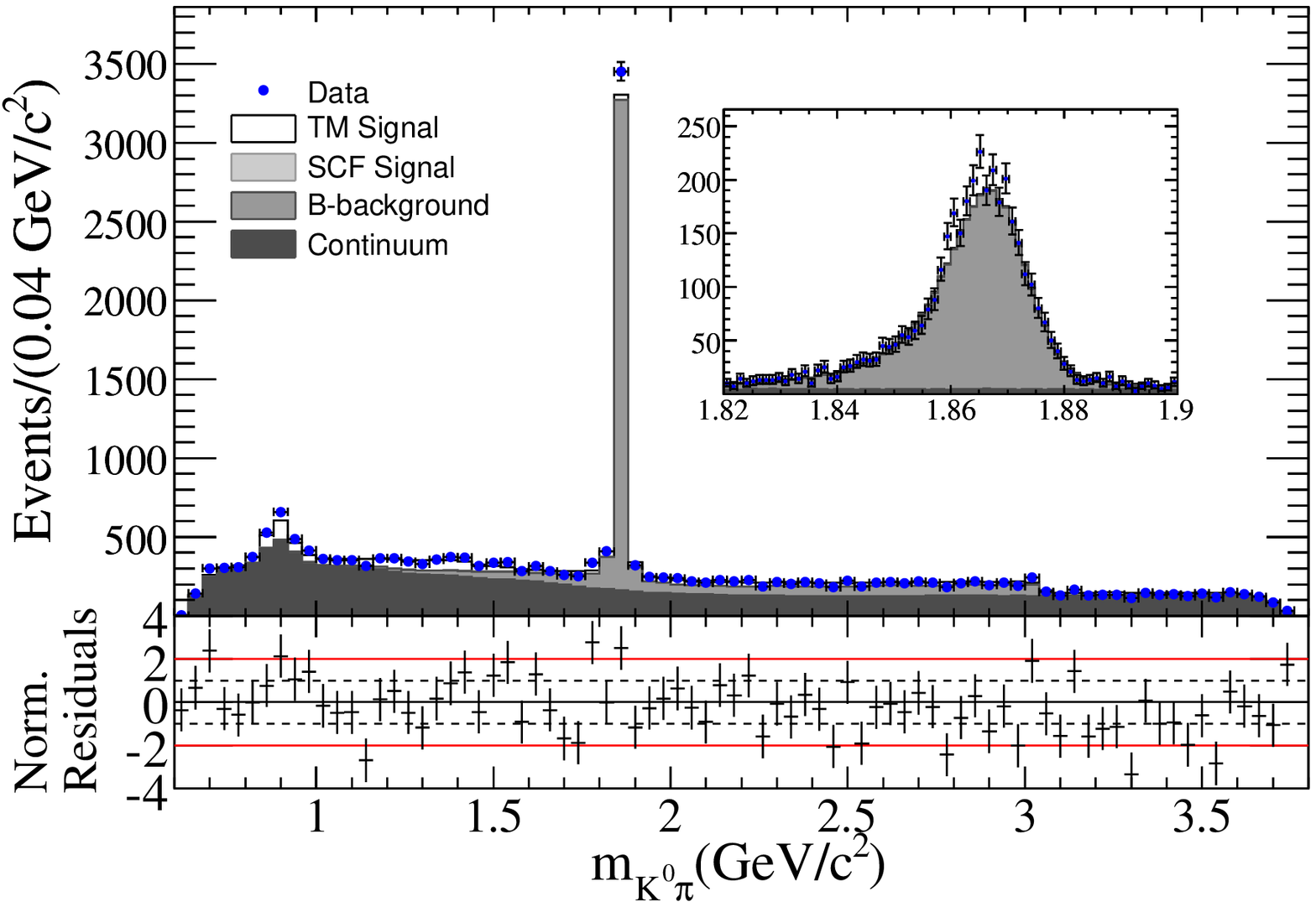}
\caption{\label{fig:mpipi_Signal}{
	Spectra of $m_{\pip\pim}$ (left) and symmetrized $m_{\KS\pi}$ (right)
	for the whole data sample.
	For $m_{\pip\pim}$, the insets show the $\jpsi$ region (a) 
	and in the $\psitwos$ region (b).
	The symmetrized $m_{\KS\pi}$ is obtained by folding
	the SDP with respect to the $\thetaprime$ variable at $0.5$.
	The inset shows the $D$ region.}}
\end{center}
\end{figure*}

\begin{figure*}[htbp]
\begin{center}
\includegraphics[width=8.7cm,keepaspectratio]{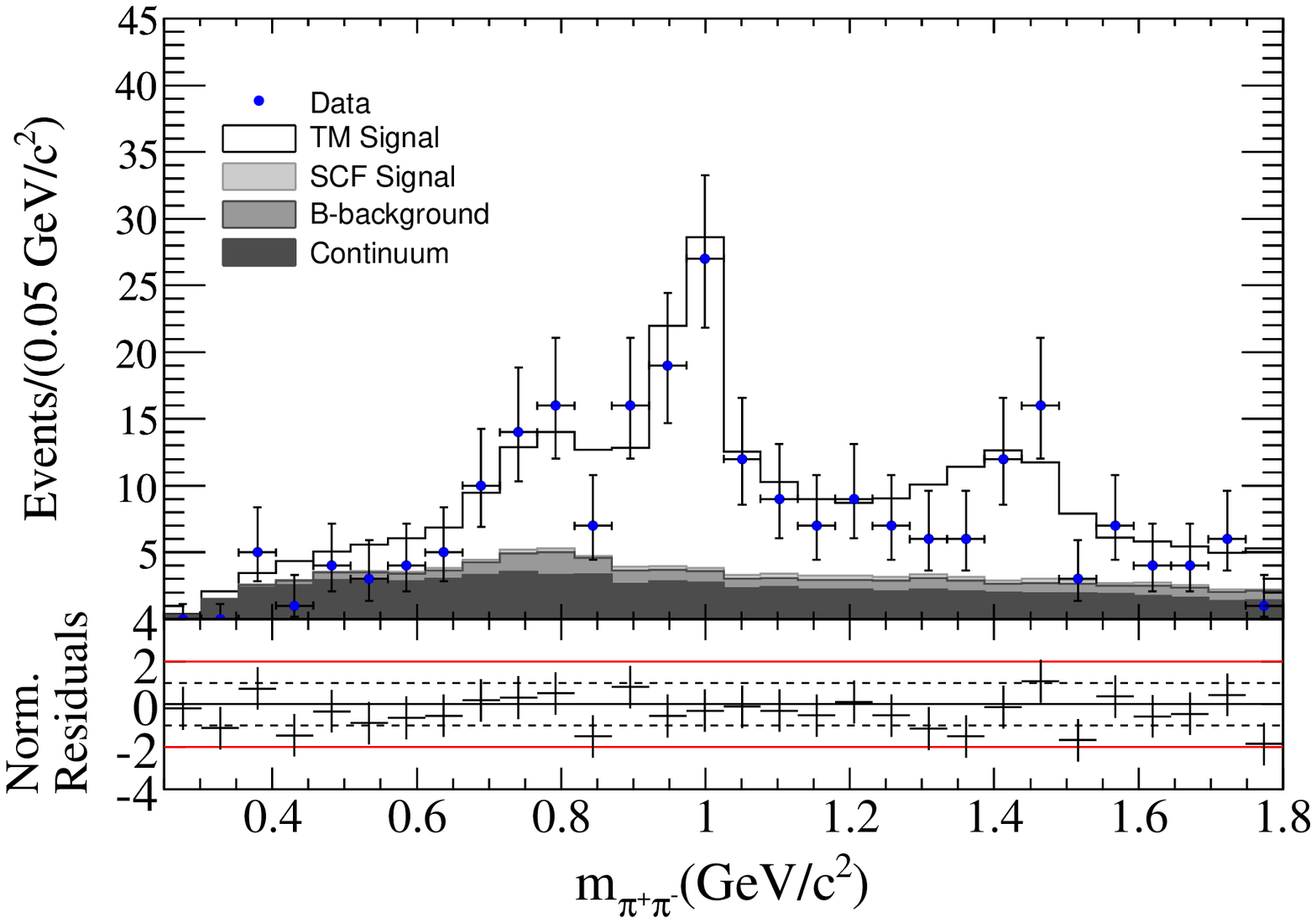}
\includegraphics[width=8.7cm,keepaspectratio]{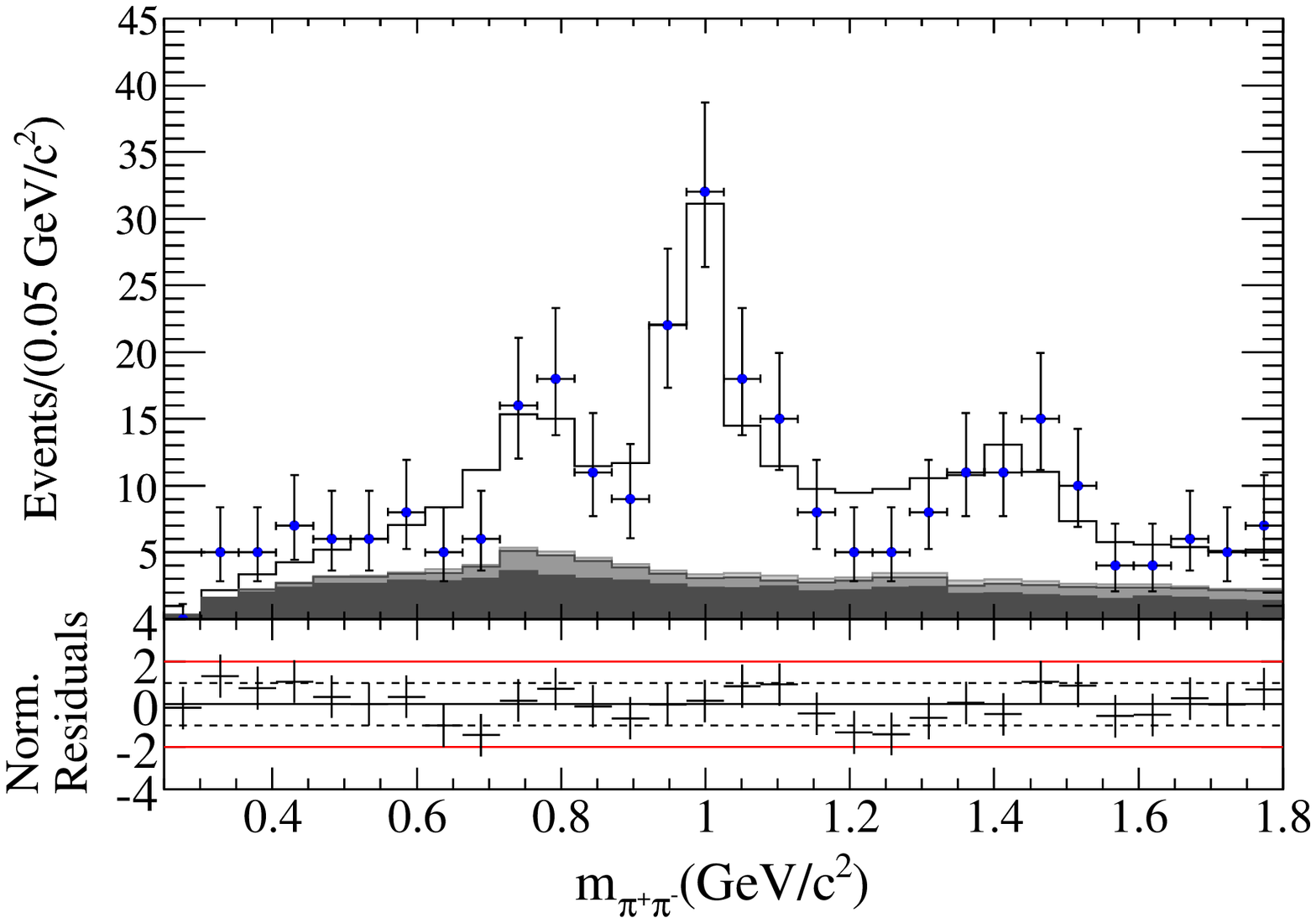}
\caption{\label{fig:mpipi_Zooming_Signal}{Distribution of $m_{\pip\pim}$ for  a sample 
        enhanced in $\BztoKspipi$ signal, showing the $f_0(980)K^0_S$ and 
$\rho^0(770)K^0_S$ signal region for positive (left) and negative (right) $\pi^+\pi^-$ helicity. 
The contribution from $f_X(1300)\KS$ and  $f_2(1270)\KS$ are also visible.
A veto in the $D\pi$ band has been applied. 
The $\Delta t$ and DP PDFs have been 
excluded from the likelihood ratio $R$ used to enhance the sample in signal events.
The cut on $R$ retains $21\%$ of signal, while rejecting $99\%$ of continuum.
}} 
\end{center}
\end{figure*}

\begin{figure*}[htbp]
\begin{center}
\includegraphics[width=8.7cm,keepaspectratio]{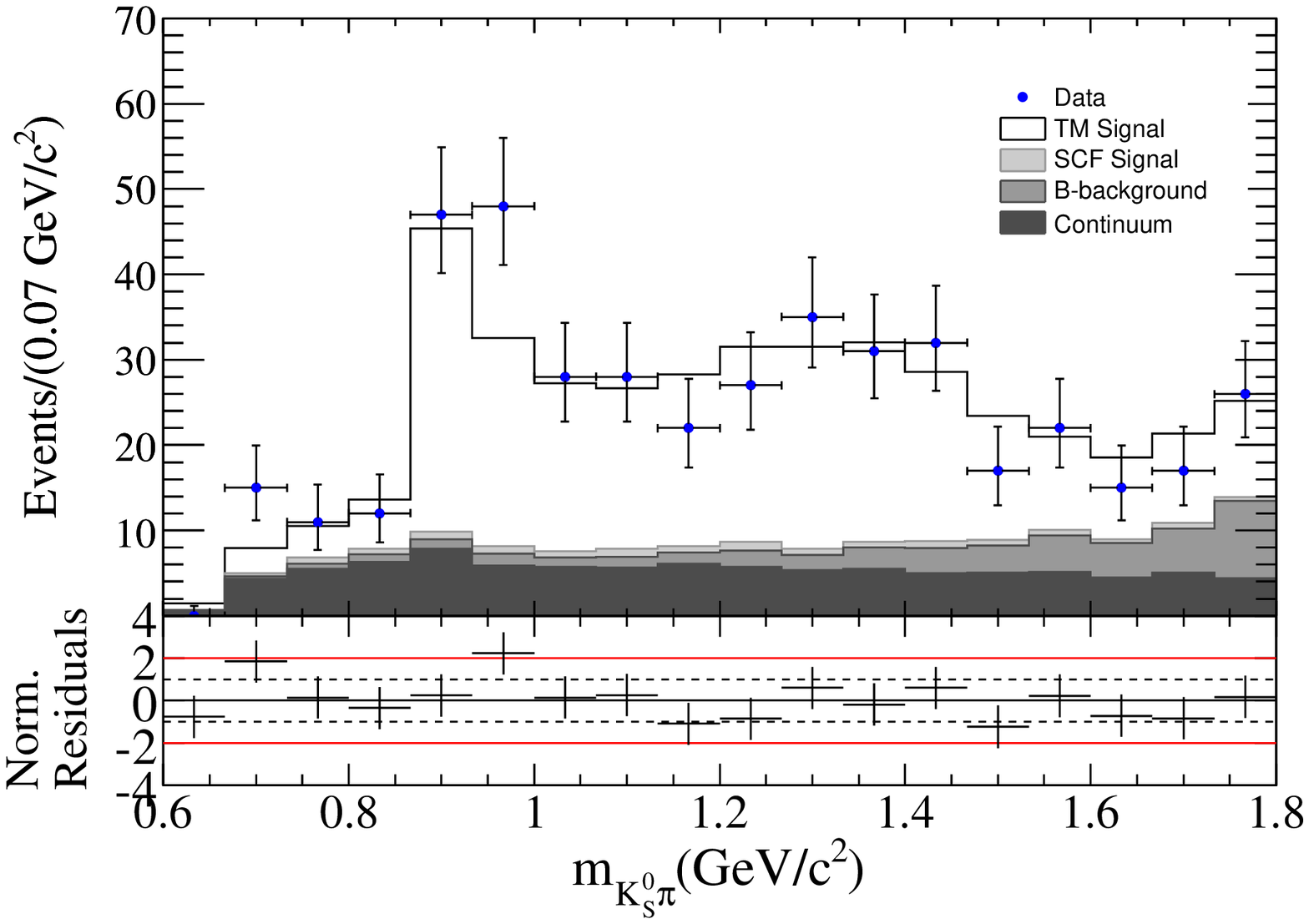}
\includegraphics[width=8.7cm,keepaspectratio]{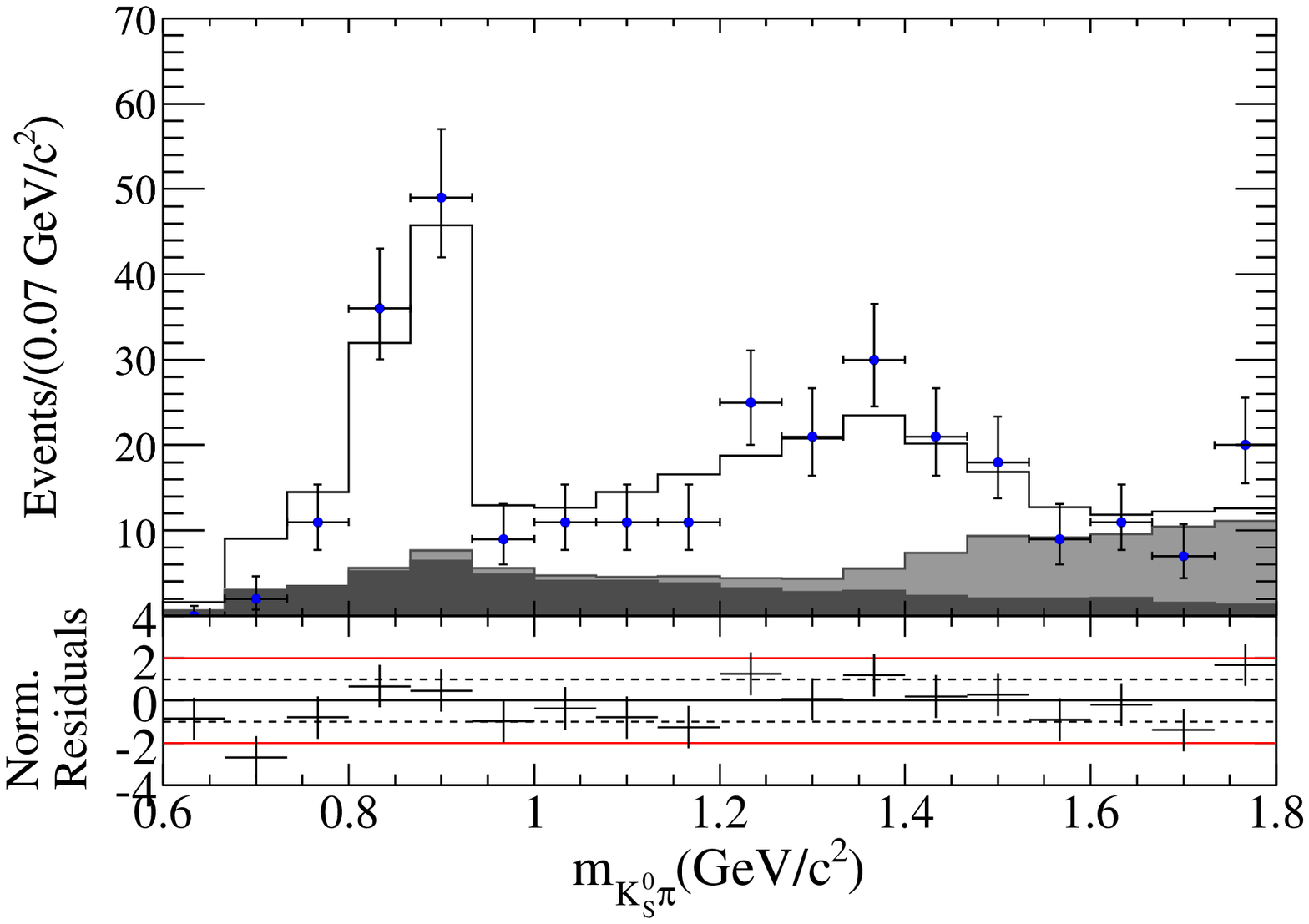}
\caption{\label{fig:mKspiSim_Zooming_Signal}{
Distributions of $m_{\KS\pi}$ for a sample 
enhanced in $\BztoKspipi$ signal, showing the 
$K^{*}(892)\pi$ and $K^{*}(1430)\pi$ signal region for positive (left) and negative (right) 
$K^0_S\pi$ helicity.
A veto in the $J/\psi K^0_S$ and $\psi(2S)K^0_S$ bands has been applied. 
The $\Delta t$ and DP PDFs have been excluded
from the definition of the likelihood ratio used to enhance
the sample in signal events.
The cut on $R$ retains $18\%$ of signal while rejecting $94\%$ of continuum.
An interference between the vector and scalar $K^{*+}$ is apparent through a positive
(negative) forward-backward asymmetry
below (above) the $K^{*}(892)$.}
}
\end{center}
\end{figure*}

\begin{figure*}[htbp]
\begin{center}
\includegraphics[width=9.0cm,keepaspectratio]{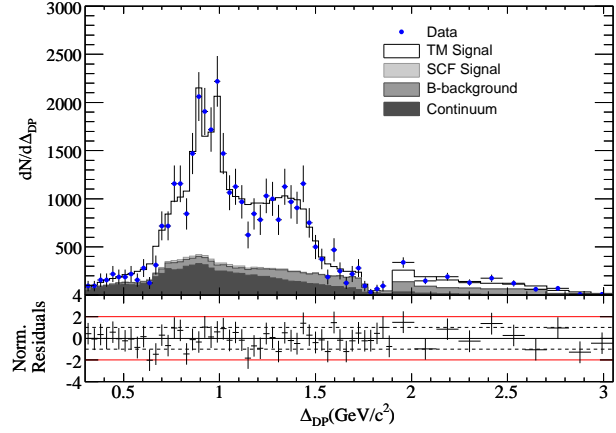}
\caption{\label{fig:deltaDalitz}{
Distributions of the $\Delta_{\rm DP}$ variable, for a sample enhanced in $\BztoKspipi$ signal. 
The  $\Delta_{\rm DP}$ variable is defined as
$\min(m_{\KS\pip}, m_{\KS\pim}, m_{\pip\pim})$.
Small (large) values of $\Delta_{\rm DP}$ correspond to the edges (center) of the DP.
On the left (right) side of the figure, for $\Delta_{\rm DP}< 1.9\gevcc$ ($>1.9\gevcc$),
the dominant contribution to the signal is from the light resonances (the NR) component of the signal model. 
A veto in the  $D\pi$, $J/\psi K^0_S$, and $\psi(2S)K^0_S$ bands has been applied.
The $\Delta t$ and DP PDFs have been
excluded from the likelihood ratio $R$ used to enhance the sample in signal events.
The cut on $R$ retains $37\%$ of signal while rejecting $88\%$ of continuum.
}}
\end{center}
\end{figure*}

\begin{figure*}[htbp]
  \centerline{  \epsfxsize8.5cm\epsffile{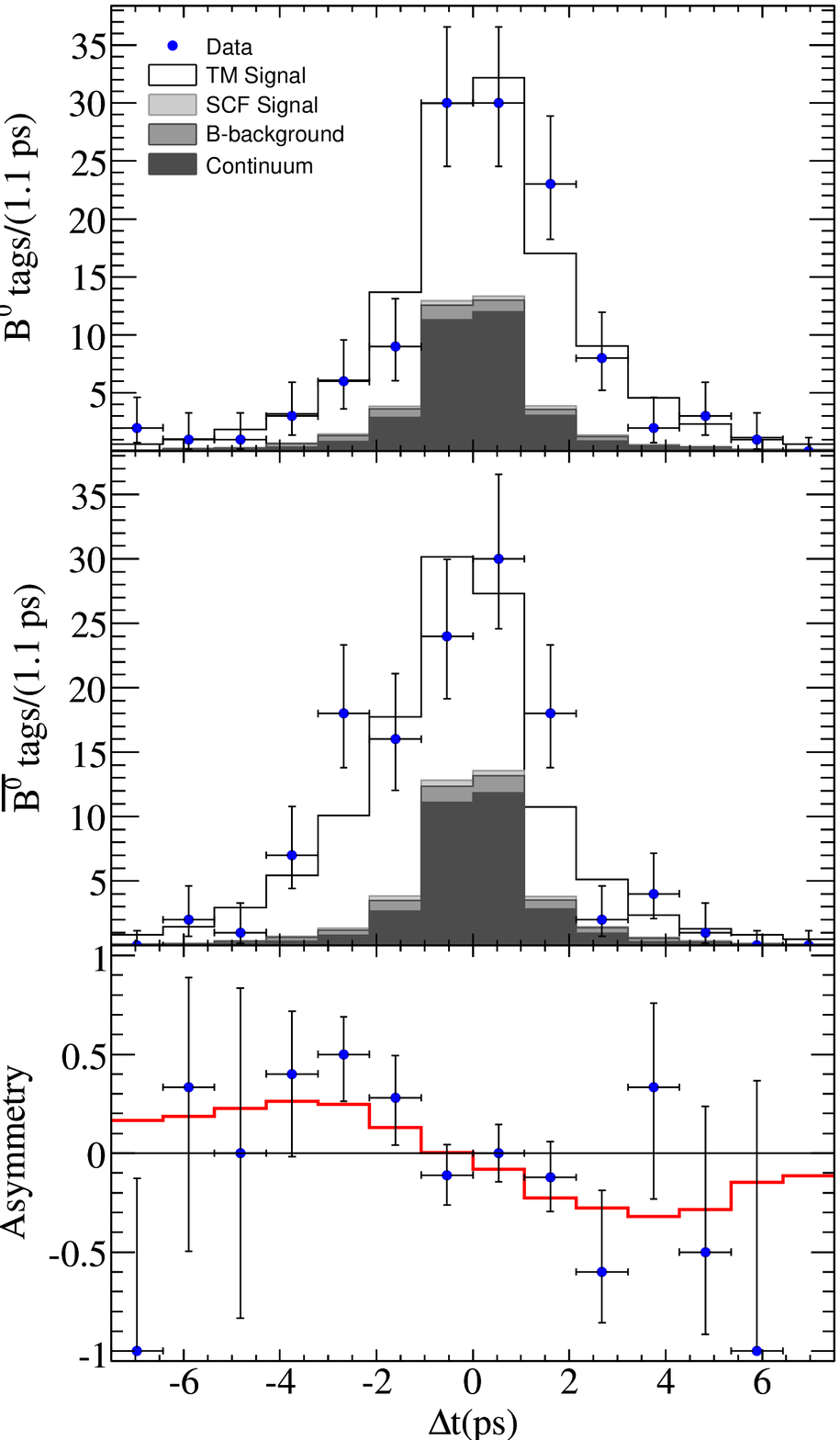}
	\epsfxsize8.5cm\epsffile{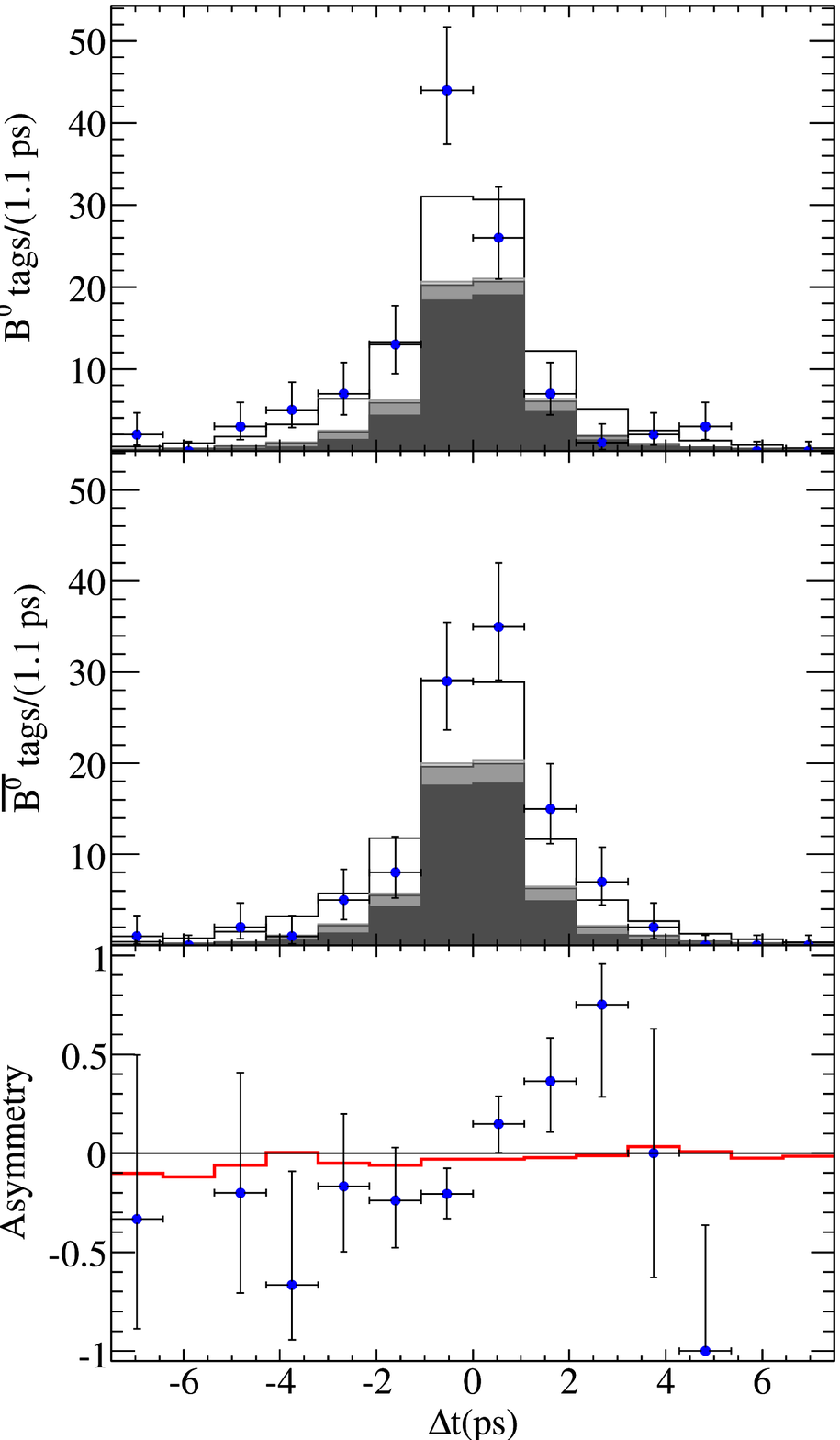}}
  \caption{\label{fig:dtAssym} Distributions of $\dt$ when the $\Bz_{\rm tag}$ is a $\Bz$ (top), $\Bzb$ (middle), and the derived $\dt$ asymmetry (bottom). 
Plots on the left (right) hand side, correspond to events in the $\fI \KS$ ($\rhoI \KS$) region. 
These distributions correspond to samples where the $\D^- \pip$ and $\jpsi \KS$ 
bands are removed from the DP, and the  $\Delta t$ and DP PDFs have been
excluded from the likelihood ratio $R$ used to enhance the sample in signal events.
The cut on $R$ retains $24\%$ of signal while rejecting $98\%$ of continuum.}
\end{figure*}

For a given event $i$, we define the likelihood ratio as $R \equiv {\cal P}_{{\rm sig-TM},i}/{\cal P}_i$
(see Eq.~\eqref{eq:theLikelihood} and explanations below). 
Figure~\ref{fig:lkl} shows distributions of $\log R$ for all
the events entering the fit, and for the signal-like region. 
We obtain signal enriched samples that are used in some of the figures below,
by removing events with small values of
$R$; in  each case $R$ is computed excluding the variable being plotted.
Figure~\ref{fig:projections} shows
distributions of $\de$, $\mes$, and the NN output which are enhanced in signal content by requirements on
$R$.
Figures~\ref{fig:mpipi_Signal} to~\ref{fig:deltaDalitz}
show similar distributions for $m(\pi^+\pi^-)$, $m(K_s^0 \pi)$, and $\Delta_{\rm DP}$.
These distributions illustrate the good quality of the fit in the signal-enhanced regions.
Signal enriched distributions of $\dt$ and $\dt$ asymmetry for events in the regions of $\fI \KS$ and $\rhoI \KS$
are shown in Fig.~\ref{fig:dtAssym}.

In the fit, we measure directly the relative magnitudes and phases 
of the different components of the signal model. The magnitude and phase of the $\Bz \to \fI \KS$ amplitude are fixed to $4$ and $0$, respectively, as a reference.
The results corresponding to the two solutions
are given together with their statistical uncertainties in Table~\ref{tab:resultsAmps}.
The full (statistical, systematic and model dependent) correlation matrices between the magnitudes and the phases for the two solutions are given in the Appendix.
The measured relative amplitudes $c_k$, where the index represents an intermediate 
resonance, are used to extract the Q2B parameters defined below.

For a resonant decay mode $k$
which is a \CP eigenstate,
the following Q2B parameters are extracted: 
the angle
$\beta_{\rm eff}$ defined as
     \begin{equation}
       \beta_{\rm eff}(k) = \frac{1}{2}\arg(c_k\bar c_k^*)~
	\label{eq:2betaeff}
     \end{equation}
and the direct and mixing-induced \CP asymmetries, defined as:
    \begin{equation}
    C(k) = \frac{|c_k|^2 - |\bar c_k|^2}{|c_k|^2 + 
|\bar c_k|^2}~, 
    \end{equation} \label{eq:C}
    \begin{equation}
    S(k) = \frac{  2\,{\cal I}m(\bar c_kc_k^*)}{|c_k|^2 + 
|\bar c_k|^2}~.
    \label{eq:S}
    \end{equation} 
For a flavor-specific resonant decay mode $k$ such as $\Bz \to \KstarpI \pim$, it is customary to define
the direct \CP asymmetry parameter $A_{\CP}$ as:
    \begin{equation}
       A_{\CP}(k) =  \frac{|\bar c_{\bar{k}}|^2 - |c_{k}|^2}
       {|\bar c_{\bar{k}}|^2 + |c_{k}|^2}.
    \end{equation} \label{eq:Acp}
For a pair of resonances $k$ and $k^\prime$,
the  phase $\phi(k, k^\prime)$
relating their amplitudes $c_k$ and $c_{k^\prime}$,  defined as
    \begin{equation}
      \phi(k, k^\prime) = \arg(c_{k}c_{k^\prime}^*),
    \end{equation}
can be accessed by exploiting the interference pattern in the DP areas where $k$ and $k^\prime$ overlap;
correspondingly, the phase $\bar{\phi}(k, k^\prime)$  
for the \CP-conjugated amplitudes $\bar{c}_k$ and $\bar{c}_{k^\prime}$  is
    \begin{equation}
      \bar{\phi}(k, k^\prime) = \arg(\bar{c}_k \bar{c}_{k^\prime}^*).
    \end{equation}
From these two phases, the difference $\Delta \phi (k, k^\prime) = \bar{\phi}(k, k^\prime) -
\phi(k, k^\prime)$, can be extracted. This parameter
is a direct \CP violation observable, and can only be accessed
in an amplitude analysis.

For a resonant decay mode $k$, the phase relating its amplitude $c_k$ to
its charge conjugate $\bar{c}_k$ is defined as
   \begin{equation}
     \Delta\Phi(k) = \arg(c_k\bar{c}_k^*);
	\label{eq:DeltaPhi}
   \end{equation}
here it is worth recalling that we use a
convention in which the $\Bzb$ decay amplitudes have absorbed the phase from $\BzBzb$ mixing,
and so the phase of $q/p$ is implicit in the $ \Delta\Phi(k)$ parameter.
Although the definition of this parameter is technically similar to the
$\beta_{\rm eff}$ phase defined in Eq.~\eqref{eq:2betaeff}, they differ in their physical
interpretation. The parameter $\beta_{\rm eff}$ quantifies the time-dependent mixing-induced \CP 
asymmetry, and therefore is most relevant for the \CP\ eigenstate modes, such as $\rhoI\KS$ and $\fI\KS$.
On the other hand the $\Delta\Phi(k)$ parameter concerns mostly  flavor-specific
modes, such as $\Bz\to\KstarpI\pim$, for which there is no interference between decays with and without
mixing. For such modes, sensitivity to $\Delta\Phi(k)$  is provided indirectly by the interference
pattern of the resonance $k$ with other
modes that are accessible both to \Bz and \Bzb decays.

We also extract the relative fit fraction $FF$ of a Q2B channel $k$, which is calculated as:
      \begin{equation}
      FF(k) = \frac{(|c_{k}|^2 + |\bar c_{k}|^2)\langle F_{k}F^*_{k}\rangle}
{\sum_{\mu\nu}{(c_{\mu}c_{\nu}^* + \bar c_{\mu}\bar c_{\nu}^*)\langle F_{\mu}F^*_{\nu}\rangle}}~,
      \end{equation}
where the terms
      \begin{equation}
         \langle F_{\mu}F^*_{\nu}\rangle = \int\!\!\!\int{F_{\mu}F^*_{\nu}ds_+ds_-}
      \end{equation}
are obtained by integration over the complete Dalitz plot.
The total fit fraction is defined as the algebraic sum of all fit fractions. This quantity is not
necessarily unity due to the potential presence of net constructive or destructive interference.
Using the relative fit fractions, we calculate the branching fraction $\mathcal B$ for the intermediate mode $k$ as
	\begin{equation}
	\label{eq:BF_definition}
	FF(k)\times \mathcal{B}(\BztoKspipi )~, 
	\end{equation}
where $\mathcal{B}(\BztoKspipi )$ is the total inclusive branching fraction
	\begin{equation}
	\mathcal{B}(\BztoKspipi)=\frac{N_{sig}}{\bar{\varepsilon} N_{B\bar{B}}}~. 
	\end{equation}
We compute the average efficiency, $\bar{\varepsilon}$, by weighting MC events with the measured
intensity  distribution of signal events, $(|{\cal A}({\rm DP})|^2+|\bar{\cal A}({\rm DP})|^2)/2$.  The term $N_{B\bar{B}}$ is the total number of $B\bar{B}$ pairs in the sample.
Finally, we use the following integrals of amplitudes over the complete Dalitz plot to measure the inclusive direct \CP-asymmetry:
      \begin{equation}
         A_{\CP}^{incl} = \frac{\int\!\!\int{\left(\left|\bar{\cal A}\right|^2-\left|{\cal A}\right|^2\right)ds_+ds_-}}
                            {\int\!\!\int{\left(\left|\bar{\cal A}\right|^2+\left|{\cal A}\right|^2\right)ds_+ds_-}}~.
      \end{equation}
The Q2B parameters and fit fractions are given in Table~\ref{tab:resultQ2BPar},
together with their statistical and systematic errors.
The branching fractions are shown in Table \ref{tab:Final_Res_BFs}.

\begin{table*}[hbtp]
\begin{center}
\caption{
Summary of measurements of the Q2B parameters for solutions I and II.
The first uncertainty is statistical, the second is systematic, and the third represents the DP signal model dependence.
We also show the total (statistical and systematic) linear correlations between the parameters $\beta_{\rm eff}$ ($S$) and $C$. 
Phases are given in degrees and $FF$s in percent.
\label{tab:resultQ2BPar}}
\begin{tabular}{lcc}
\hline\hline
Parameter                                       &  Solution I                                & Solution II                       \\
\hline\\[-9pt]
$C(\fI\KS)$                                     & $0.08  \pm 0.19  \pm 0.03 \pm 0.04$        & $0.23  \pm 0.19 \pm 0.03 \pm 0.04$\\
$\beta_{\rm eff}(\fI\KS)$                           & $36.0  \pm 9.8   \pm 2.1 \pm 2.1$          & $56.2  \pm 10.4 \pm 2.1 \pm 2.1$\\
$S(\fI\KS)$                                     & $-0.96^{+0.21}_{-0.04} \pm 0.03 \pm 0.02$  & $-0.90^{+0.26}_{-0.08} \pm 0.03 \pm 0.02$\\
${\rm Corr}[\beta_{\rm eff}(\fI\KS),C(\fI\KS)]$ & $-3.1\%$                                   & $-17.0\%$\\
${\rm Corr}[S(\fI\KS),C(\fI\KS)]$               & $ 19.7\%$                                  & $ 12.5\%$\\
$FF(\fI\KS)$                                    & $13.8^{+1.5}_{-1.4} \pm 0.8 \pm 0.6$       & $13.5^{+1.4}_{-1.3} \pm 0.8 \pm 0.6$\\
\hline\\[-9pt]
$C(\rhoI\KS)$                                   & $-0.05 \pm 0.26 \pm 0.10 \pm 0.03$         & $-0.14 \pm 0.26 \pm 0.10 \pm 0.03$\\
$\beta_{\rm eff}(\rhoI\KS)$                     & $10.2 \pm 8.9   \pm 3.0 \pm 1.9$           & $33.4  \pm 10.4 \pm 3.0 \pm 1.9$\\
$S(\rhoI\KS)$                                   & $0.35^{+0.26}_{-0.31}  \pm 0.06 \pm 0.03$  & $0.91^{+0.07}_{-0.19} \pm 0.06 \pm 0.03$\\
${\rm Corr}[\beta_{\rm eff}(\rhoI\KS),C(\rhoI\KS)]$ & $-23.0\%$                              & $-34.0\%$\\
${\rm Corr}[S(\rhoI\KS),C(\rhoI\KS)]$               & $-21.3\%$                              & $-10.4\%$\\
$FF(\rhoI\KS)$                                  & $8.6^{+1.4}_{-1.3} \pm 0.5 \pm 0.2$        & $8.5^{+1.3}_{-1.2} \pm 0.5 \pm 0.2$\\
\hline\\[-9pt]
$\ACP(\KstarI\pi)$                              & $-0.21 \pm 0.10 \pm 0.01 \pm 0.02$         & $-0.19^{+0.10}_{-0.11} \pm 0.01 \pm 0.02$\\
$\Delta\Phi(\KstarI\pi)$                        & $58.3 \pm 32.7 \pm 4.6 \pm 8.1$            & $176.6 \pm 28.8 \pm 4.6 \pm 8.1$\\
$FF(\KstarI\pi)$                                & $11.0^{+1.2}_{-1.0} \pm 0.6 \pm 0.8$       & $10.9^{+1.2}_{-1.0} \pm 0.6 \pm 0.8$\\
\hline\\[-9pt]
$\ACP((K\pi)^*_0\pi)$                           & $ 0.09 \pm 0.07 \pm 0.02 \pm 0.02$         & $ 0.12^{+0.07}_{-0.06} \pm 0.02 \pm 0.02$\\
$\Delta\Phi((K\pi)^*_0\pi)$                     & $72.2 \pm 24.6 \pm 4.1 \pm 4.4$            & $-175.1 \pm 22.6 \pm 4.1 \pm 4.4$\\
$FF((K\pi)^*_0\pi)$                             & $45.2 \pm 2.3 \pm 1.9 \pm 0.9$             & $46.1 \pm 2.4 \pm 1.9 \pm 0.9$\\
\hline\\[-9pt]
$C(\fII\KS)$                                    & $ 0.28^{+0.35}_{-0.40} \pm 0.08 \pm 0.07$  & $ 0.09 \pm 0.46 \pm 0.08 \pm 0.07$ \\
$\beta_{\rm eff}(\fII\KS)$                          & $ 14.9  \pm 17.9 \pm 3.1  \pm 5.2 $        & $53.6  \pm 16.7 \pm 3.1  \pm 5.2$  \\
$S(\fII\KS)$                                    & $ -0.48 \pm 0.52 \pm 0.06 \pm 0.10$        & $-0.95 \pm 0.17 \pm 0.06 \pm 0.10$ \\
${\rm Corr}[\beta_{\rm eff}(\fII\KS),C(\fII\KS)]$ & $11.5\%$                                 & $-2.8\%$                           \\
${\rm Corr}[S(\fII\KS),C(\fII\KS)]$               & $0.9\%$                                  & $21.2\%$                           \\
$FF(\fII\KS)$                                   & $2.3^{+0.8}_{-0.7} \pm 0.2 \pm 0.7$        & $2.3^{+0.9}_{-0.7} \pm 0.2 \pm 0.7$\\
\hline\\[-9pt]
$C(\fX\KS)$                                     & $  0.13^{+0.33}_{-0.35} \pm 0.04 \pm 0.09$ & $  0.30^{+0.34}_{-0.41} \pm 0.04 \pm 0.09$\\
$\beta_{\rm eff}(\fX\KS)$                           & $  5.8  \pm 15.2 \pm 2.2  \pm 2.3        $ & $76.9  \pm 13.8 \pm 2.2  \pm 2.3$         \\
$S(\fX\KS)$                                     & $ -0.20 \pm 0.52 \pm 0.07 \pm 0.07       $ & $-0.42 \pm 0.41 \pm 0.07 \pm 0.07$        \\
${\rm Corr}[\beta_{\rm eff}(\fX\KS),C(\fX\KS)]$ & $ -27.0\%$                                 & $-9.3\%$                                  \\
${\rm Corr}[S(\fX\KS),C(\fX\KS)]$               & $  28.5\%$                                 & $ 6.1\%$                                  \\
$FF(\fX\KS)$                                    & $3.6^{+1.0}_{-0.9} \pm 0.3 \pm 0.9$        & $3.5^{+1.0}_{-0.8} \pm 0.3 \pm 0.9$       \\
\hline\\[-9pt]
$C(NR)$                                         & $  0.01 \pm 0.25 \pm 0.06 \pm 0.05$         &   $-0.45^{+0.28}_{-0.24} \pm 0.06 \pm 0.05$\\
$\beta_{\rm eff}(NR)$                               & $  0.4  \pm 8.8  \pm 1.9  \pm 3.8 $         &   $51.0  \pm 13.3 \pm 1.9  \pm 3.8$        \\
$S(NR)$                                         & $ -0.01 \pm 0.31 \pm 0.05 \pm 0.09$         &   $-0.87 \pm 0.18 \pm 0.05 \pm 0.09$       \\
${\rm Corr}[\beta_{\rm eff}(NR),C(NR)]$         & $-10.6\%$                                   &   $-37.9\%$                                \\
${\rm Corr}[S(NR),C(NR)]$                       & $ 10.6\%$                                   &   $-91.5\%$                                \\
$FF(NR)$                                        & $11.5 \pm 2.0 \pm 1.0 \pm 0.6$              & $12.6 \pm 2.0 \pm 1.0 \pm 0.6$\\
\hline\\[-9pt]
$C(\chiczero\KS)$                                           & $ -0.29^{+0.53}_{-0.44} \pm 0.03 \pm 0.05$  &   $ -0.41^{+0.54}_{-0.42} \pm 0.03 \pm 0.05$ \\
$\beta_{\rm eff}(\chiczero\KS)$                                 & $ 23.2  \pm 22.4 \pm 2.3  \pm 4.2$          &   $55.2  \pm 23.3 \pm 2.3  \pm 4.2$          \\
$S(\chiczero\KS)$                                           & $ -0.69 \pm 0.52 \pm 0.04 \pm 0.07$         &   $-0.85 \pm 0.34 \pm 0.04 \pm 0.07$         \\
${\rm Corr}[\beta_{\rm eff}(\chiczero\KS),C(\chiczero\KS)]$ & $ -5.8\%$                                   &   $-5.8\%$                                   \\
${\rm Corr}[S(\chiczero\KS),C(\chiczero\KS)]$               & $-19.1\%$                                   &   $-74.2\%$                                  \\
$FF(\chiczero\KS)$                                          & $1.04^{+0.41}_{-0.33} \pm 0.04 \pm 0.11$   & $0.99^{+0.37}_{-0.30} \pm 0.04 \pm 0.11$\\
\hline\\[-9pt]
total $FF$                                      & $97.2^{+1.7}_{-1.3} \pm 2.1 \pm 1.15$       & $98.3^{+1.5}_{-1.3} \pm 2.1 \pm 1.15$\\
$A_{\CP}^{incl}$                                & $-0.01 \pm 0.05 \pm 0.01  \pm 0.01$         & $0.01 \pm 0.05 \pm 0.01 \pm 0.01$\\
\hline\\[-9pt]
$\phi(f^0(980)K^0_S,\rho(770)K^0_S)$            & $-35.6  \pm 14.9 \pm 6.1 \pm 4.4$           & $-66.7  \pm 18.3 \pm 6.1 \pm 4.4$\\
$\phi(K^{*}(892)\pi,(K\pi)^*_0\pi)$             & $13.0   \pm 10.9 \pm 4.6 \pm 4.7$           & $15.5   \pm 10.2 \pm 4.6 \pm 4.7$\\
$\phi(\rho(770)K^0_S,K^{*}(892)\pi)$            & $174.3  \pm 28.0 \pm 8.7 \pm 12.7$          & $-173.7 \pm 29.8 \pm 8.7 \pm 12.7$\\
$\phi(\rho(770)K^0_S,(K\pi)^*_0\pi)$            & $-172.8 \pm 22.6 \pm 10.1 \pm 8.7$          & $-170.8 \pm 26.8 \pm 10.1 \pm 8.7$\\
\hline\\[-9pt]
$\bar{\phi}(f^0(980)K^0_S,\rho(770)K^0_S)$       & $ -89.2  \pm 17.1 \pm  8.5 \pm  7.2$       &   $-112.2  \pm 17.8 \pm  8.5  \pm  7.2$\\
$\bar{\phi}(K^{*}(892)\pi,(K\pi)^*_0\pi)$        & $  26.9  \pm  9.2 \pm  4.9 \pm  6.1$       &   $  23.8  \pm  9.1 \pm  4.9  \pm  6.1$\\
$\bar{\phi}(\rho(770)K^0_S,K^{*}(892)\pi)$       & $-147.8  \pm 24.7 \pm 11.3 \pm 11.9$       &   $ -76.5  \pm 24.0 \pm 11.3  \pm 11.9$\\
$\bar{\phi}(\rho(770)K^0_S,(K\pi)^*_0\pi)$       & $-120.9  \pm 21.6 \pm  8.7 \pm  7.3$       &   $ -52.7  \pm 21.4 \pm  8.7  \pm 7.3 $\\
\hline\hline
\end{tabular}
\end{center}
\end{table*}
								 	      					  
\renewcommand\arraystretch{1.15}
\begin{table*}[htbp]
\begin{center}
\caption{
Summary of measurements of branching fractions averaged over charge conjugate states. The quoted numbers were obtained by multiplying the corresponding fit fractions by the measured inclusive $\Bz \to \Kz\pim\pim$ branching fraction. $R$ denotes an intermediate resonant state and $h$ stands for a final state hadron: a charged pion or a $\Kz$. To correct for the secondary branching fractions we used the values from Ref.~\cite{Amsler:2008zz} and ${\mathcal B}(\KstarpI\to\Kz\pip)=\frac{2}{3}$. The first uncertainty is statistical, the second is systematic, and the third represents the DP signal model dependence. The fourth errors, when applicable, are due to the uncertainties on the
secondary branching fractions. The quoted central values correspond to the global minimum, and errors account for the presence of the second solution.
\label{tab:Final_Res_BFs}}
\begin{tabular}{lcc}
\hline\hline
Mode          & $\ \ \ {\mathcal B}(\Bz \to {\rm Mode})\times {\mathcal B}(R \to hh)\times 10^{-6}\ \ \ $
                                                               & $\ \ \ {\mathcal B}(\Bz \to {\rm Mode})\times 10^{-6}$   \\
\hline
Inclusive $\Bz \to \Kz\pip\pim$ & $\cdots$                       & $ 50.15 \pm 1.47 \pm 1.60 \pm 0.73 $                       \\
\hline
$\fI\Kz$            & $ 6.92 \pm 0.77 \pm 0.46 \pm 0.32 $        & $\cdots$                                                   \\
$\rhoI\Kz$          & $ 4.31^{+0.70}_{-0.61} \pm 0.29 \pm 0.12 $ & $ 4.36^{+0.71}_{-0.62} \pm 0.29 \pm 0.12 \pm 0.01 $        \\
$\KstarpI\pim$      & $ 5.52^{+0.61}_{-0.54} \pm 0.35 \pm 0.41 $ & $ 8.29^{+0.92}_{-0.81} \pm 0.53 \pm 0.62 $                 \\
$(K\pi)_0^{*+}\pim$ & $ 22.7^{+1.7}_{-1.3} \pm 1.2 \pm 0.6 $     & $\cdots$                                                   \\
$\fII\Kz$           & $ 1.15^{+0.42}_{-0.35} \pm 0.11 \pm 0.35 $ & $ 2.71^{+0.99}_{-0.83} \pm 0.26 \pm 0.83 ^{+0.08}_{-0.04}$ \\
$\fX\Kz$            & $ 1.81^{+0.55}_{-0.45} \pm 0.16 \pm 0.45 $ & $\cdots$                                                   \\
flat NR             & $\cdots$                                   & $ 5.77^{+1.61}_{-1.00} \pm 0.53 \pm 0.31 $                 \\
$\chiczero\Kz$      & $ 0.52^{+0.20}_{-0.16} \pm 0.03 \pm 0.06 $ & $ 142^{+55}_{-44} \pm 8 \pm 16 \pm 12 $                    \\
\hline \hline
\end{tabular}
\end{center}
\end{table*}
\renewcommand\arraystretch{1.}

\begin{figure*}[htbp]
\begin{center}
\includegraphics[width=7.5cm,keepaspectratio]{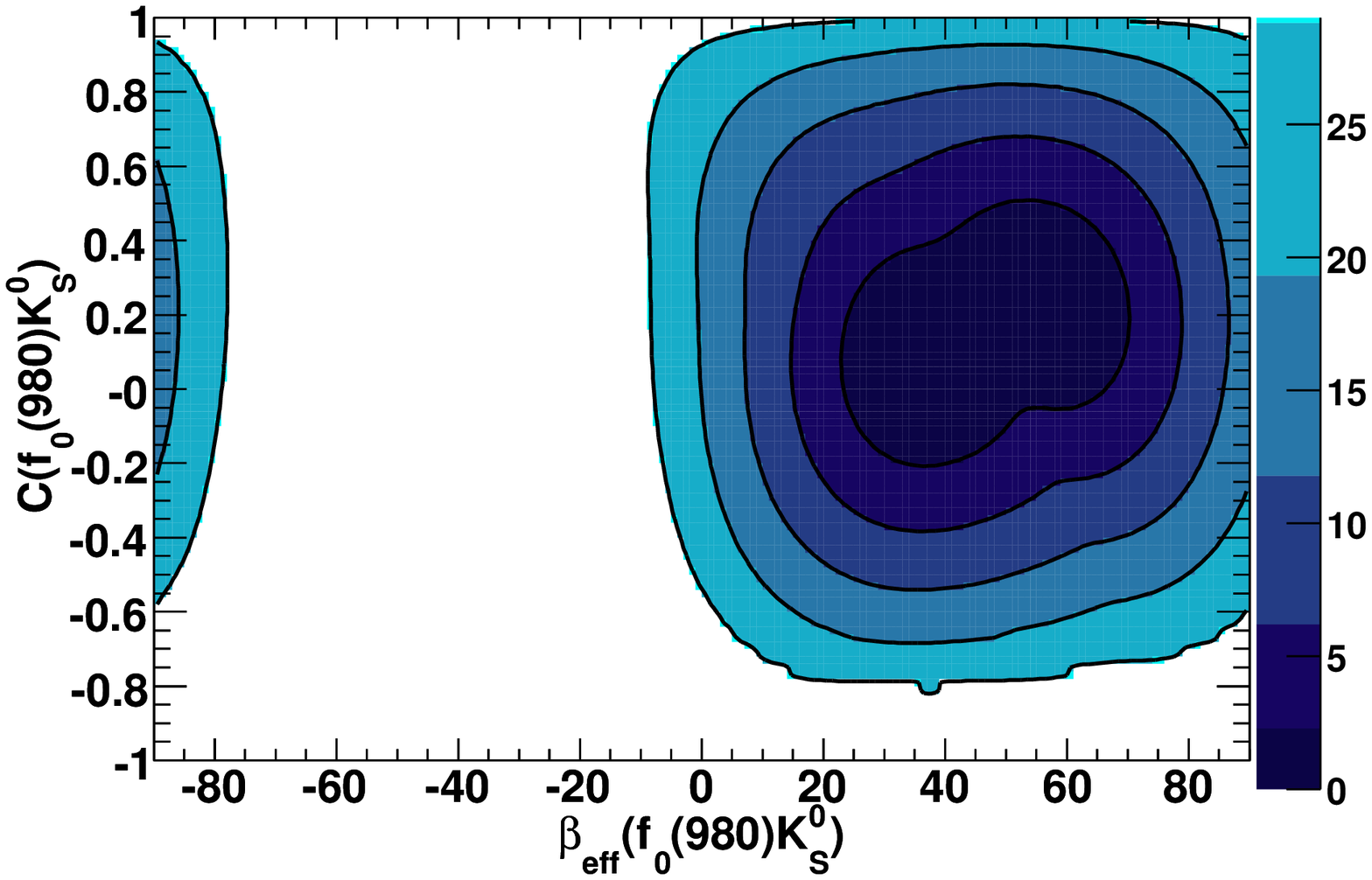}
\includegraphics[width=7.5cm,keepaspectratio]{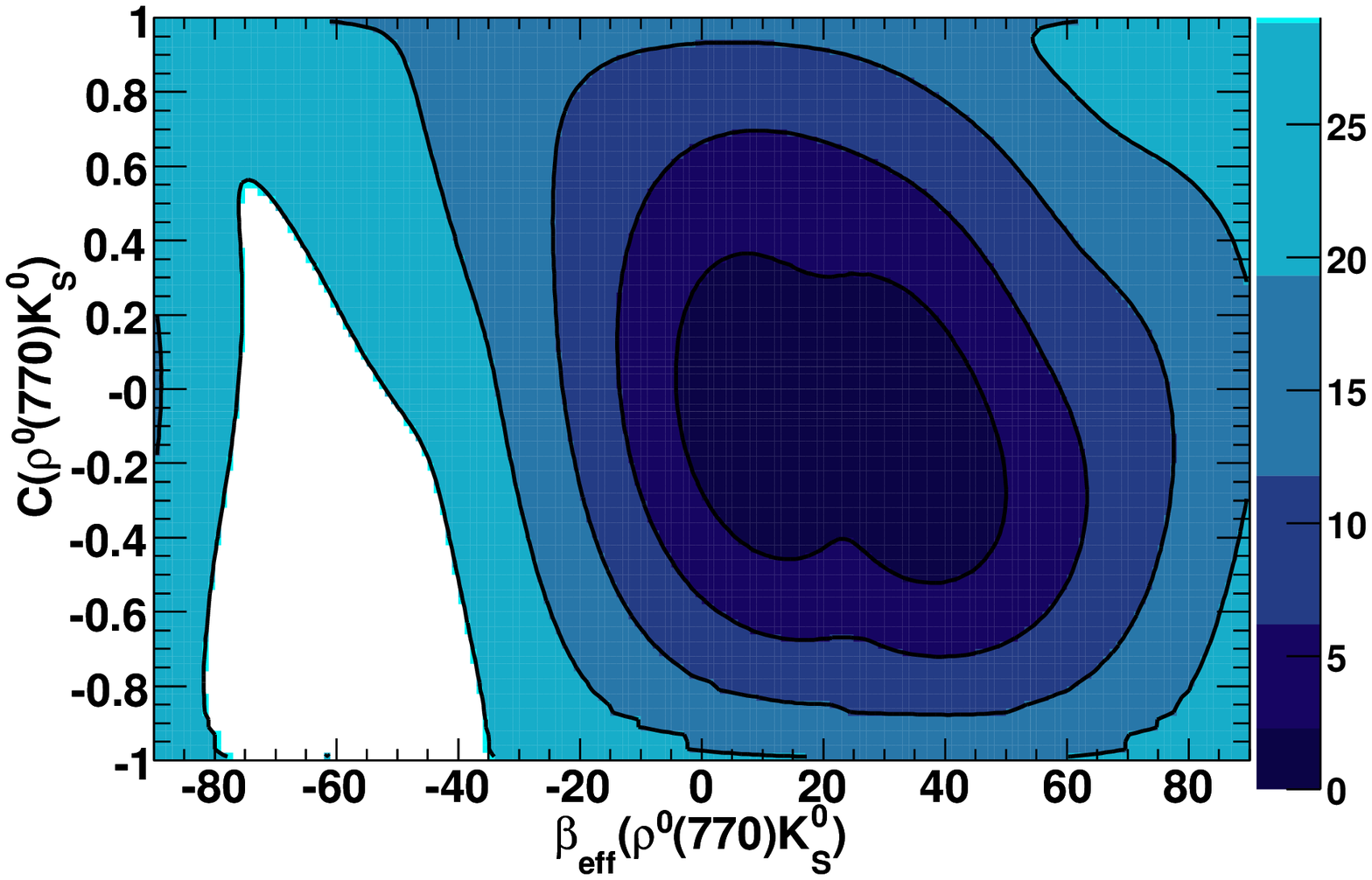}
\includegraphics[width=7.5cm,keepaspectratio]{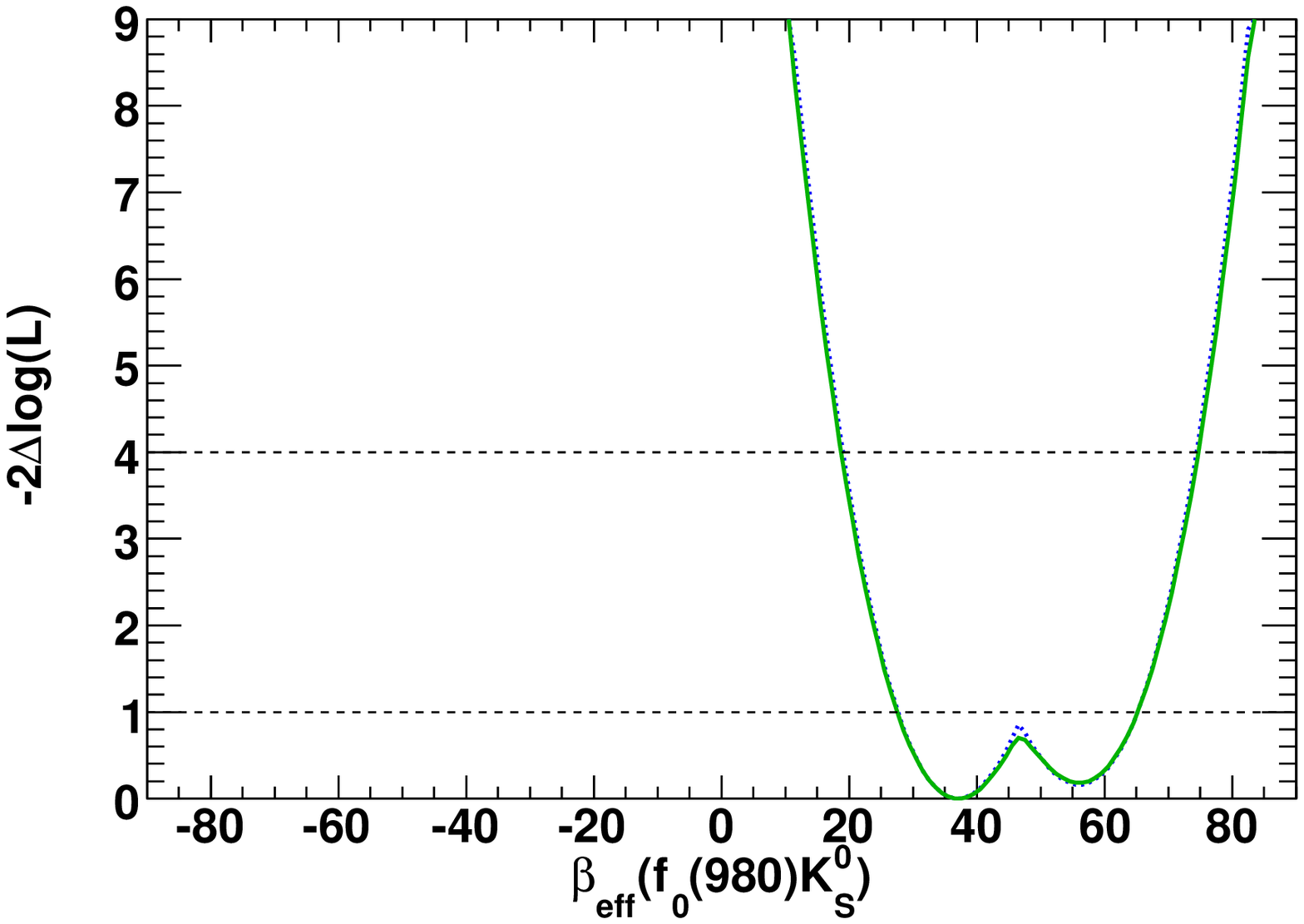}
\includegraphics[width=7.5cm,keepaspectratio]{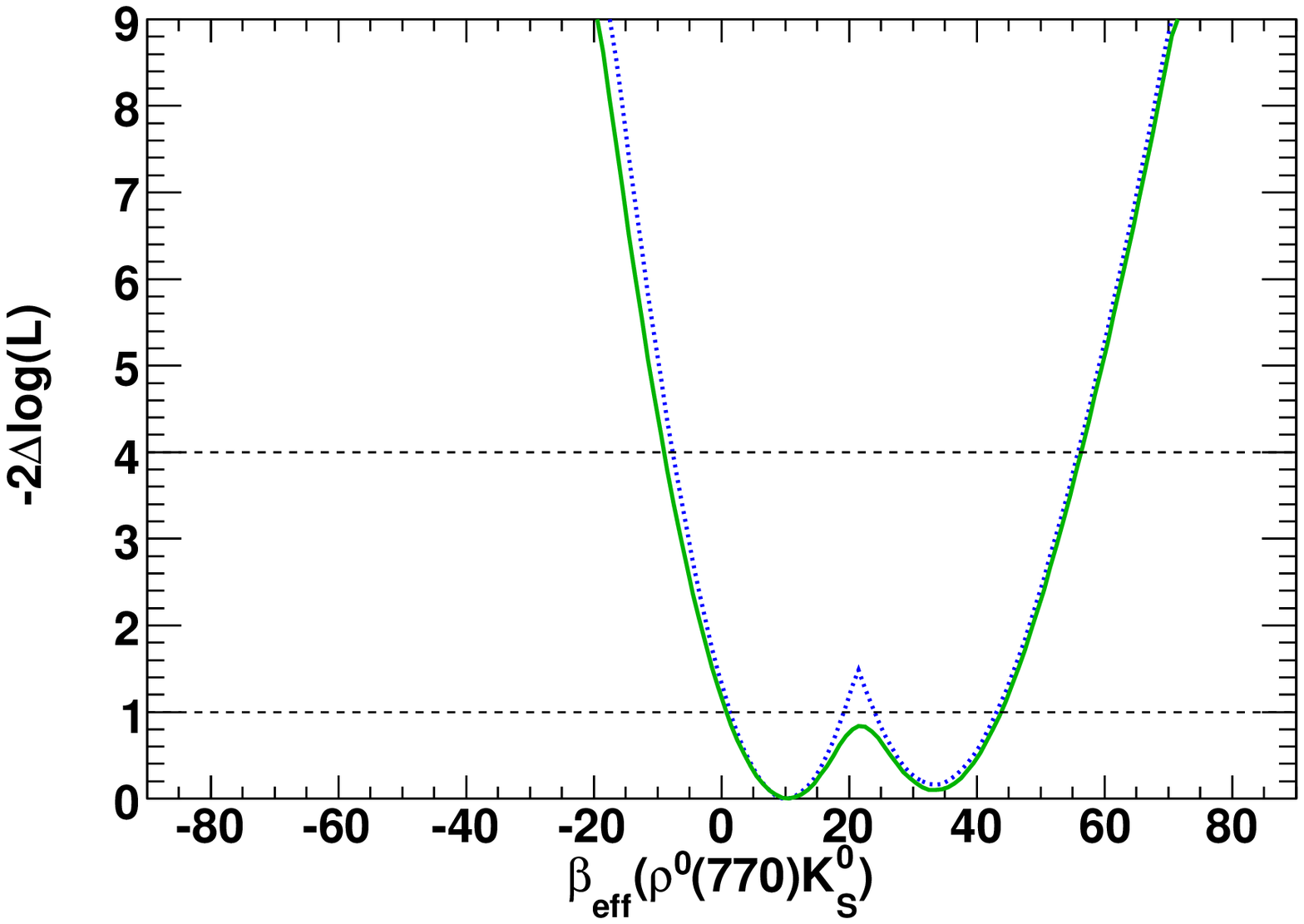}
\caption{\label{fig:2DLikeScans_f0rho_CBeta}{
Two-dimensional scans of $-2\Delta\logL$ as a function of
$\beta_{\rm eff}$ and $C$ (top) and the one-dimensional scans as a function of $\beta_{\rm eff}$ (bottom)
for the $f_0(980)K^0_S$ (left) and  $\rho^0(770)K^0_S$ (right) isobar components.
The value $-2\Delta\logL$ is computed including systematic uncertainties.
On the two-dimensional scans, shaded areas, from the darkest to the lightest, represent
the one to five standard deviations contours.
The statistical (dashed line), and total (solid line) $-2\Delta\logL$ are shown on the one-dimensional scans,
where horizontal dotted lines mark the one and two standard deviation levels.
}}
\end{center}
\end{figure*}

\begin{figure*}[htbp]
\begin{center}
\includegraphics[width=7.5cm,keepaspectratio]{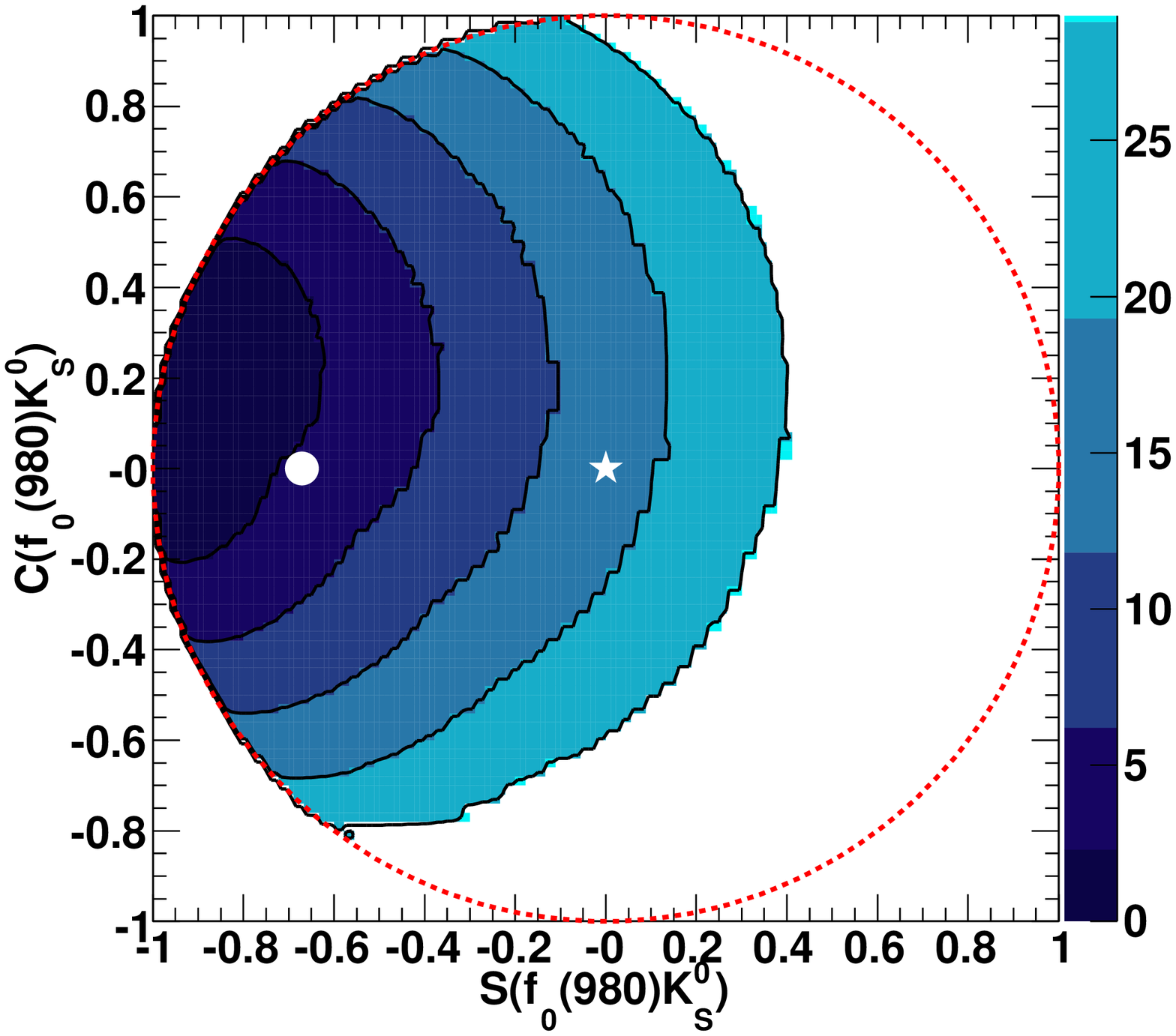}
\includegraphics[width=8.0cm,keepaspectratio]{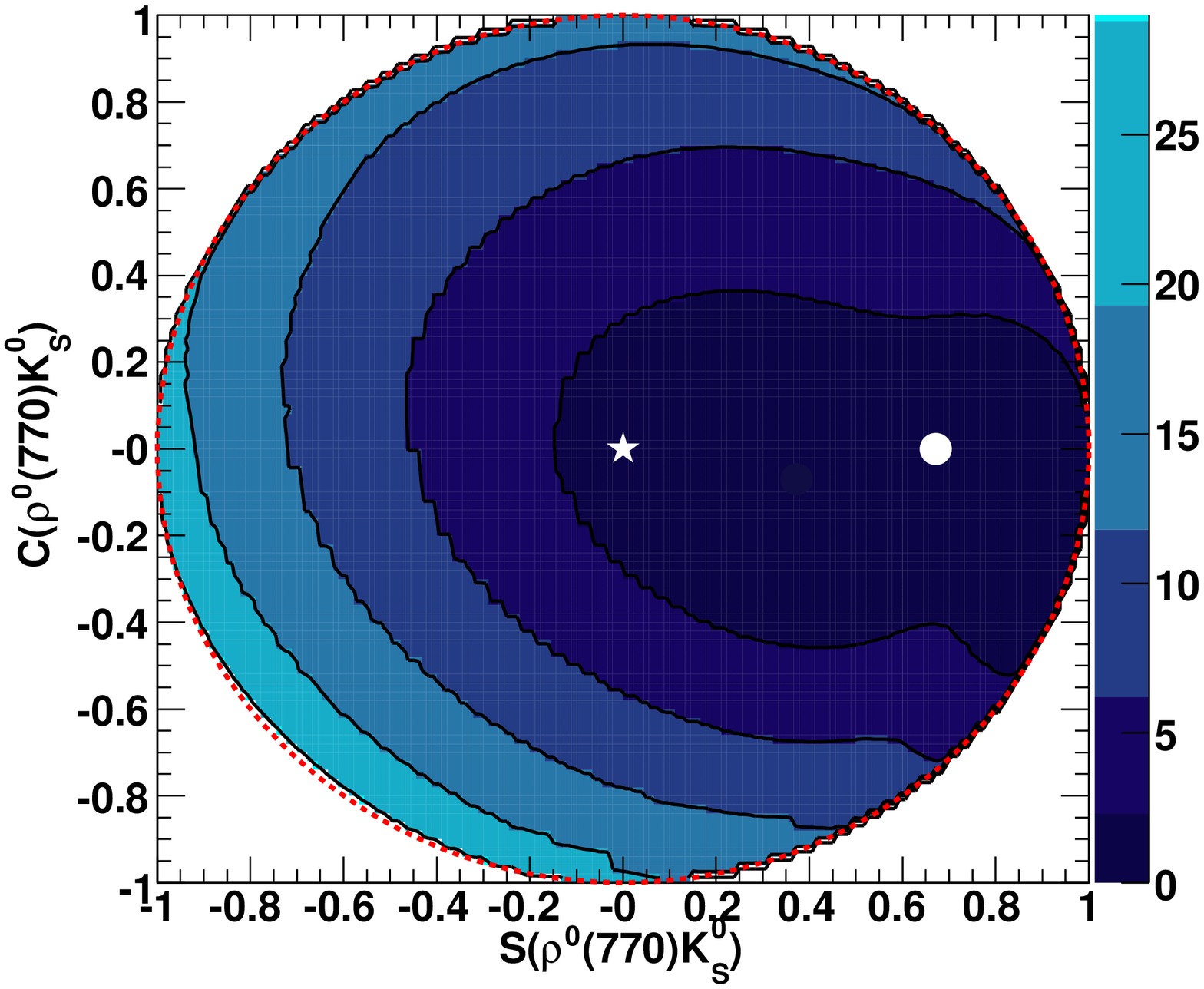}
\caption{\label{fig:2DLikeScans_f0rho_CS}{
Two-dimensional scans of $-2\Delta\logL$ as a function of
$(S,C)$, for the $f_0(980)K^0_S$ (left) and $\rho^0(770)K^0_S$ (right) isobar components.
The value $-2\Delta\logL$ is computed including systematic uncertainties.
Shaded areas, from the darkest to the lightest, represent the one to five standard deviations contours.
The $\bullet$ ($\star$) marks the expectation based on
the current world average from $b \to c \overline{c} s$
modes~\cite{Barberio:2008fa} (zero point). The dashed circle represents the physical border $S^2+C^2=1$.
}}
\end{center}
\end{figure*}

\begin{figure*}[htbp]
\begin{center}
\includegraphics[width=7.5cm,keepaspectratio]{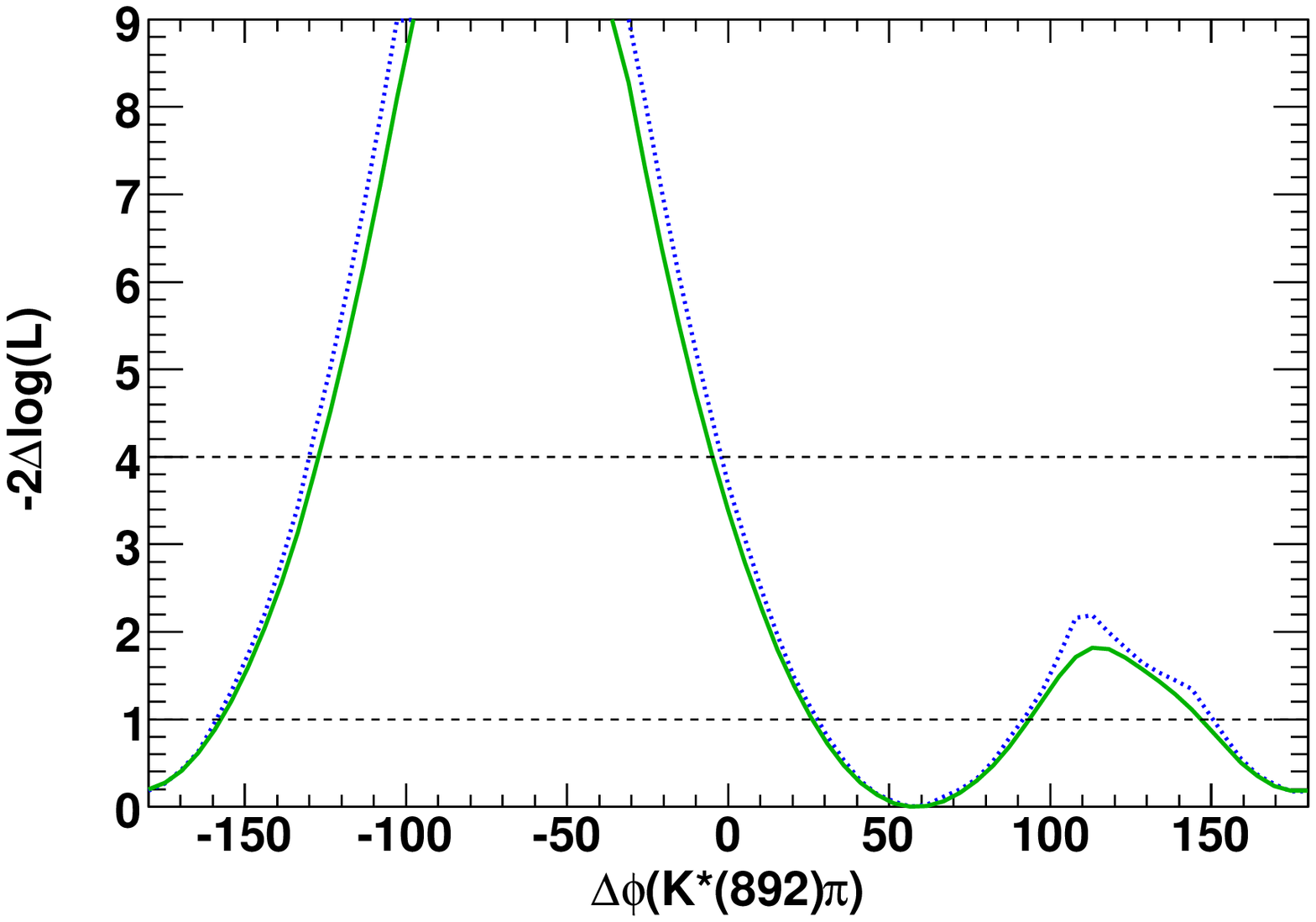}
\includegraphics[width=7.5cm,keepaspectratio]{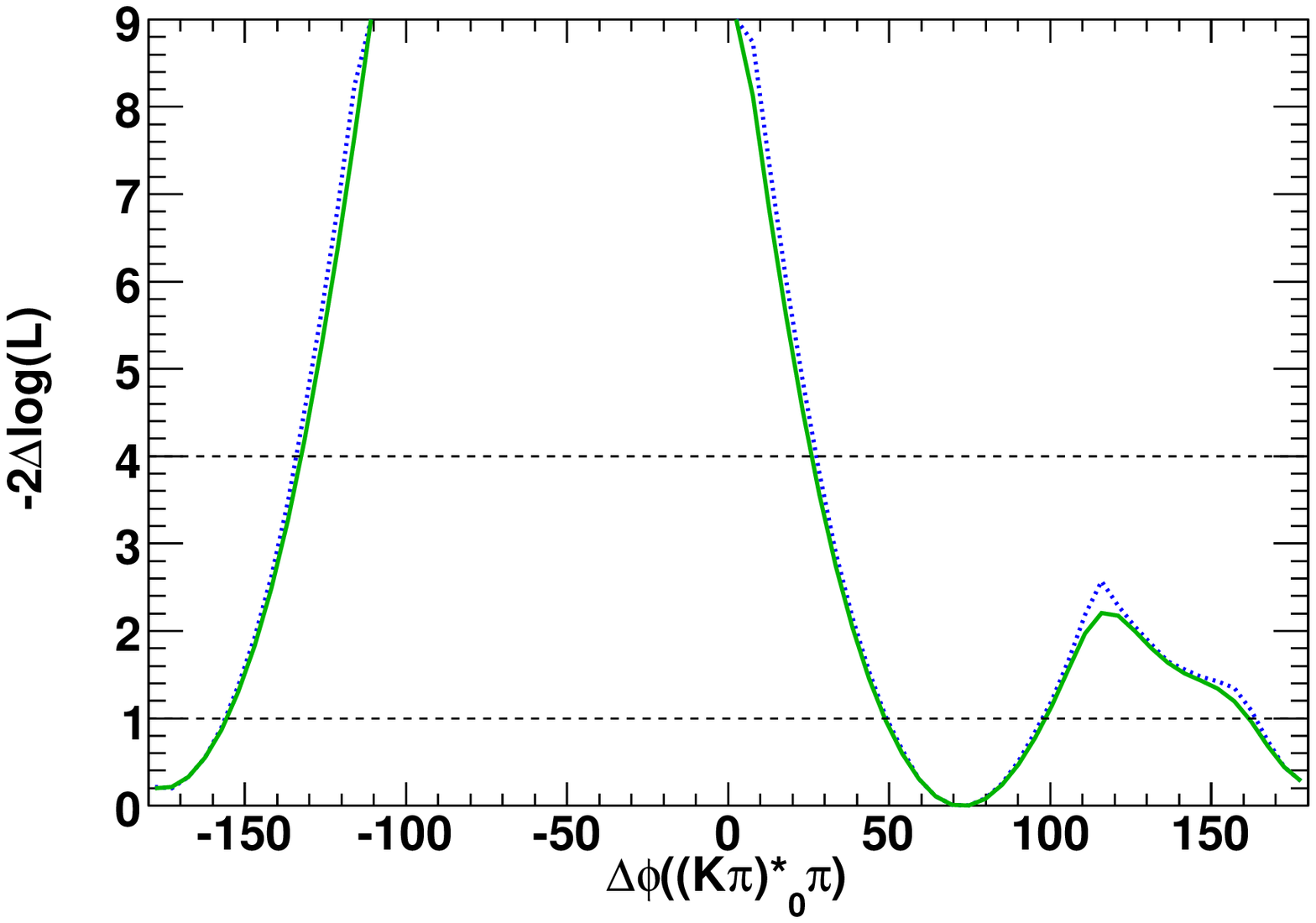}
\caption{\label{fig:LikeScans_InterPhases_1}{Statistical (dashed line) and total (solid line) scans of $-2\Delta\logL$
as a function of the relative phases 
$\Delta\Phi(K^{*}(892)\pi)$ (left) and $\Delta\Phi( (K\pi)^{*}_0 )$ (right). Horizontal dotted lines mark the 
one and two standard deviation levels.
}}
\end{center}
\end{figure*}

\begin{figure*}[htbp]
\begin{center}
\includegraphics[width=5.9cm,keepaspectratio]{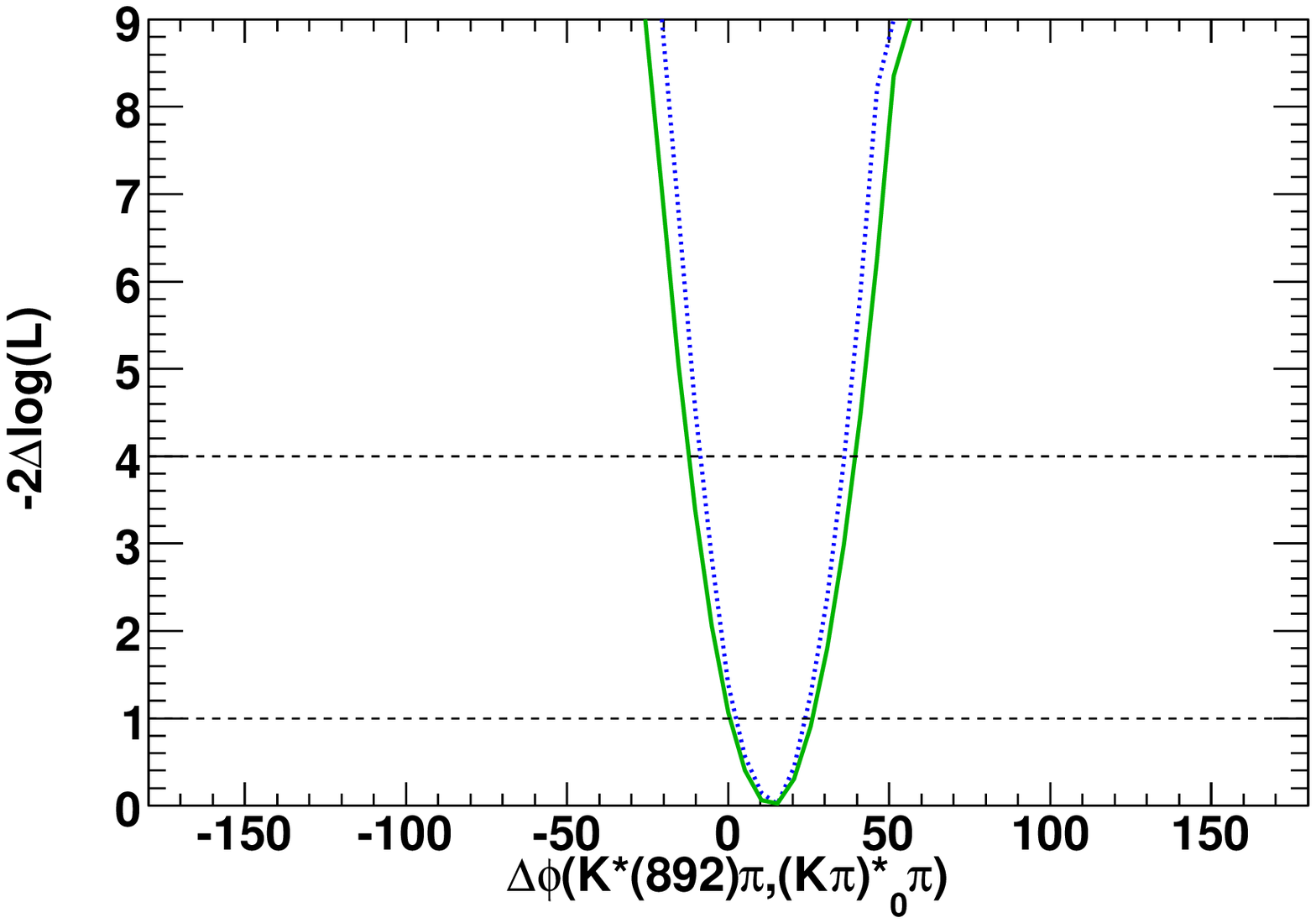}
\includegraphics[width=5.9cm,keepaspectratio]{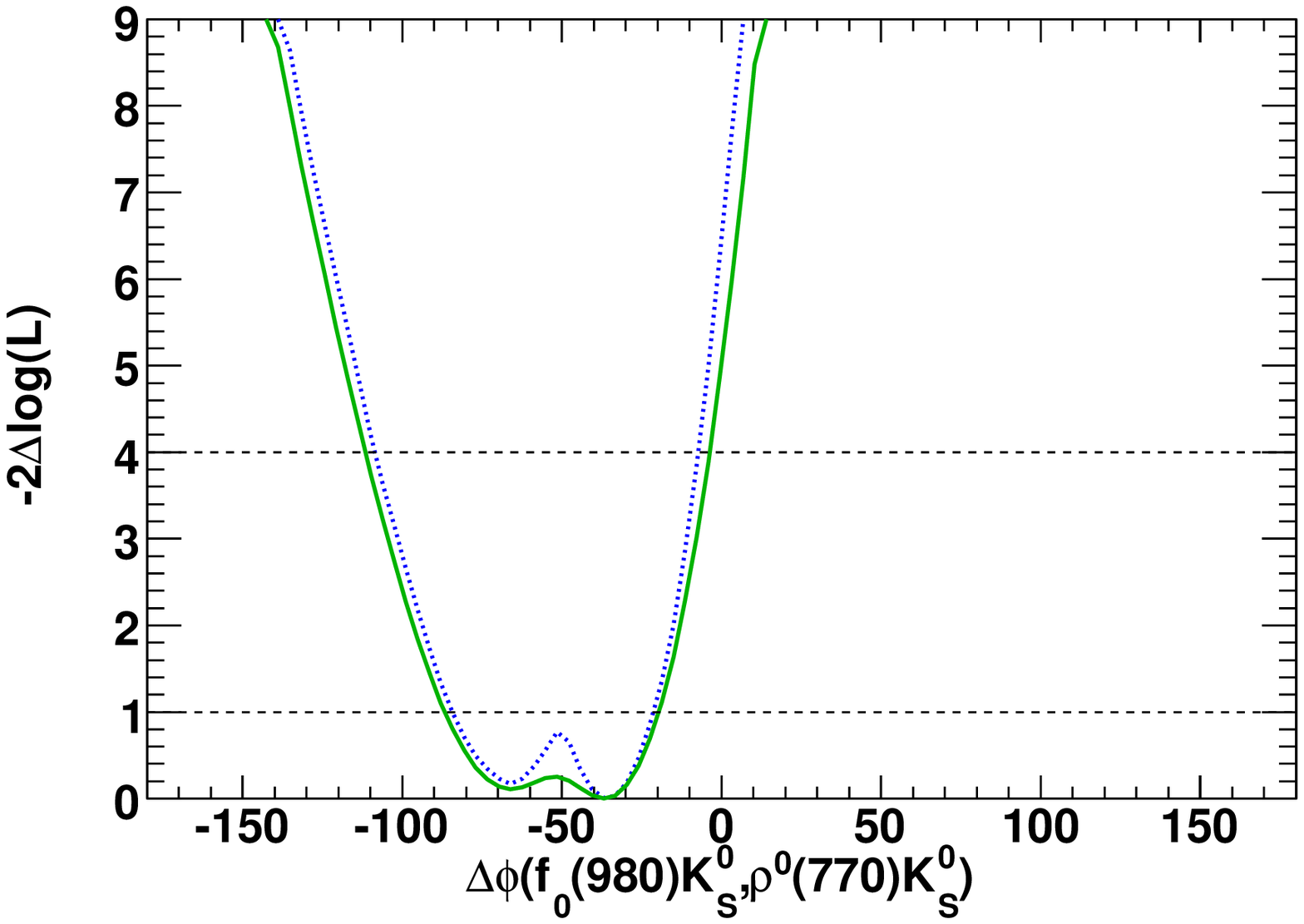}
\includegraphics[width=5.9cm,keepaspectratio]{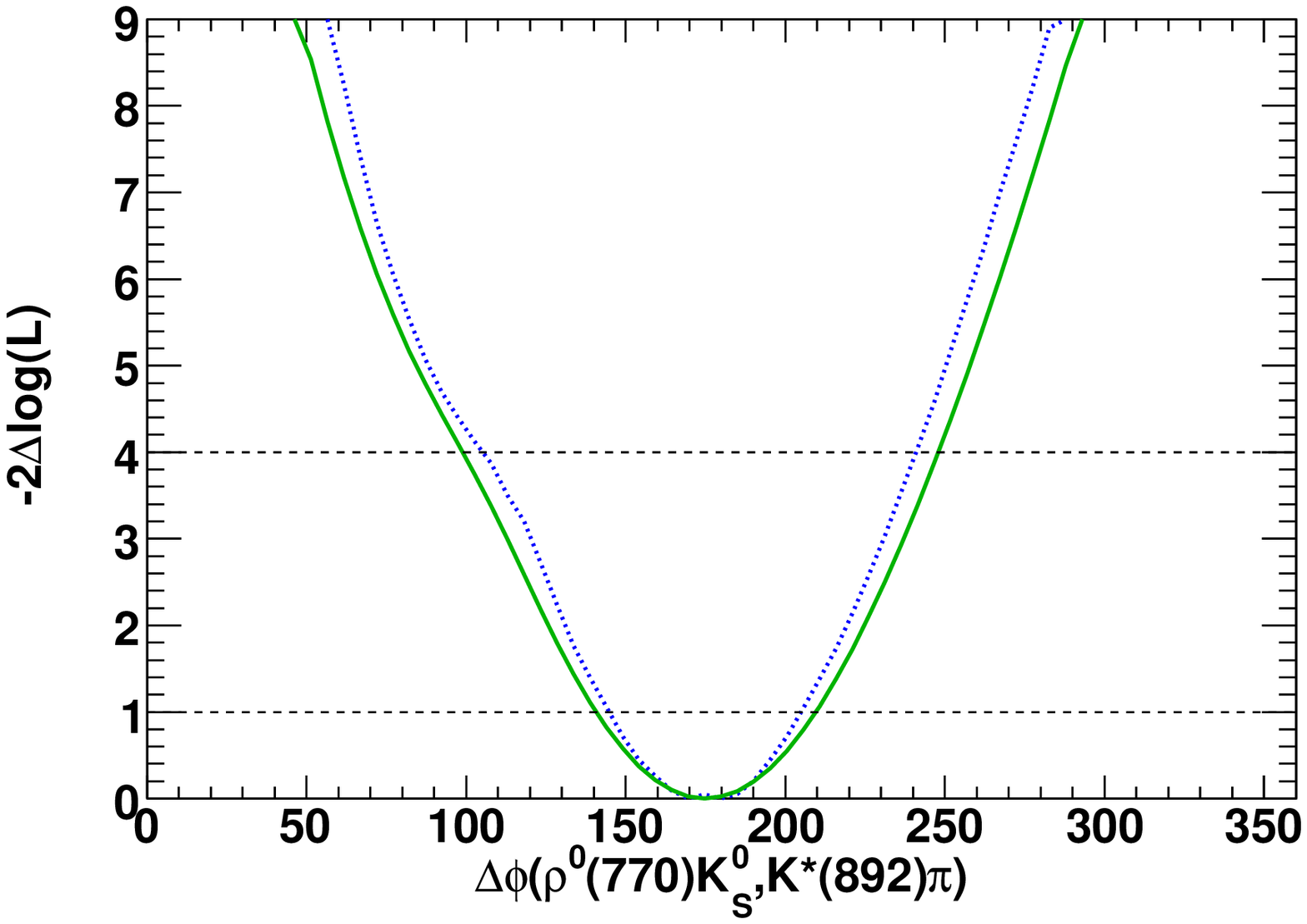}
\includegraphics[width=5.9cm,keepaspectratio]{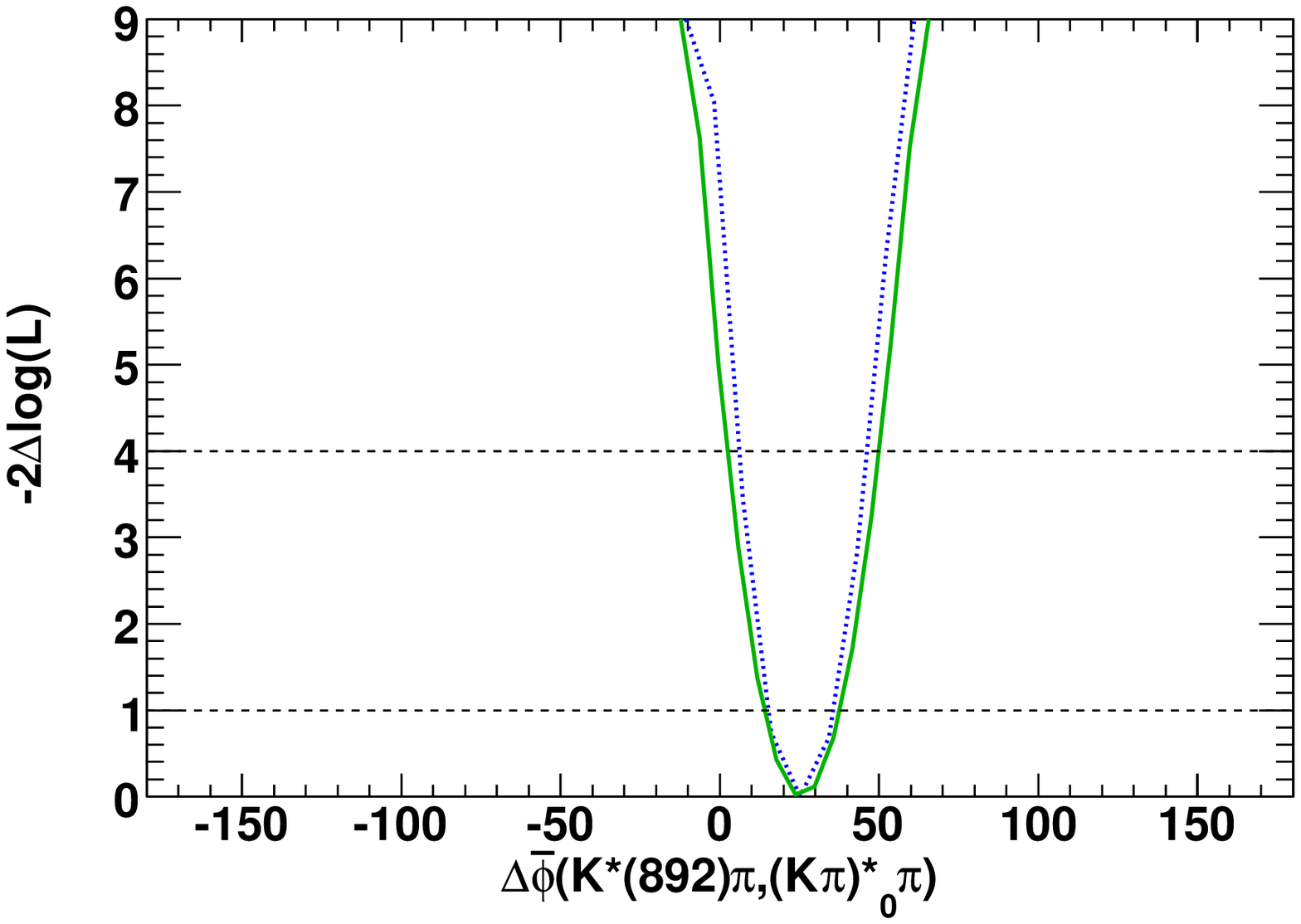}
\includegraphics[width=5.9cm,keepaspectratio]{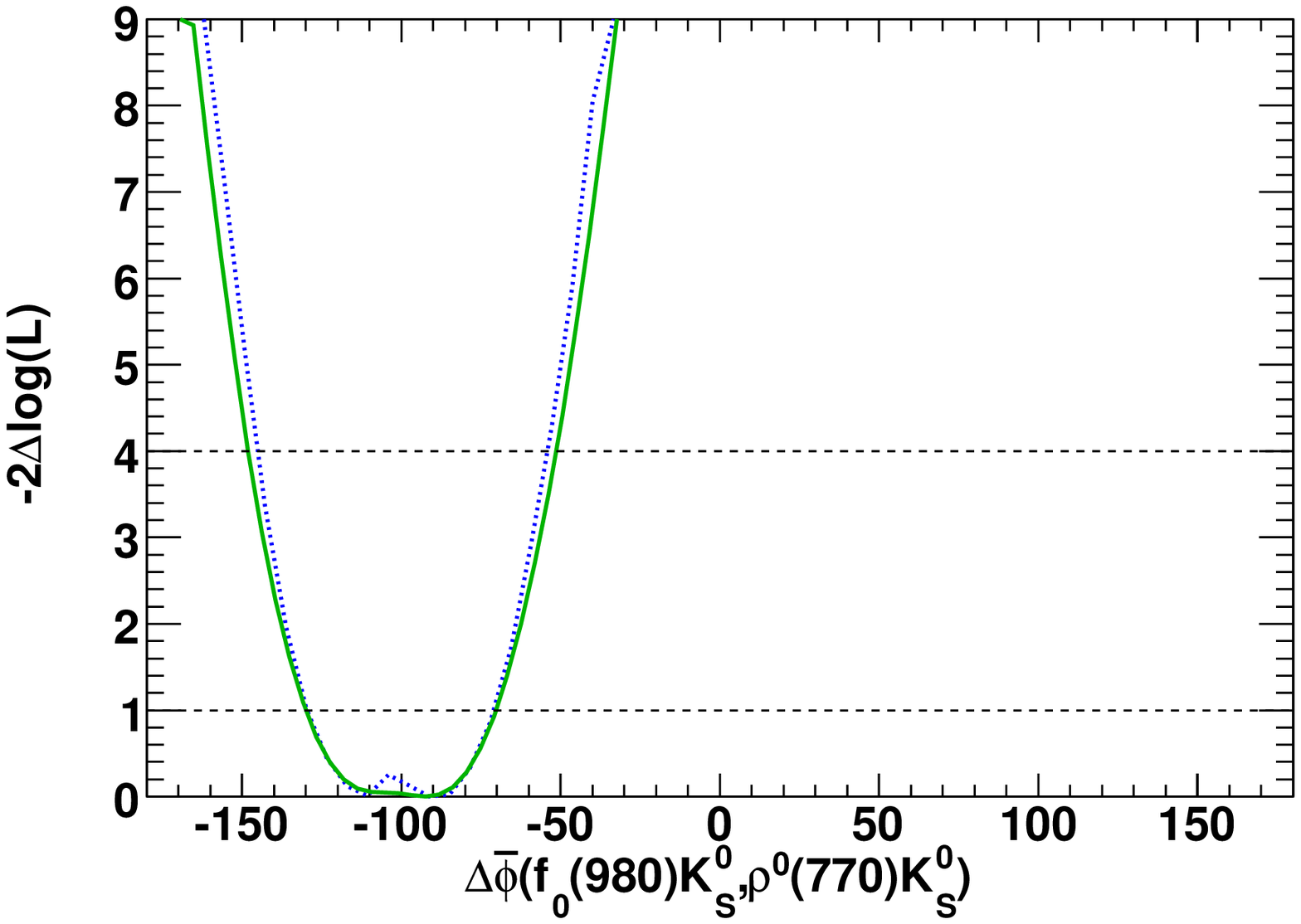}
\includegraphics[width=5.9cm,keepaspectratio]{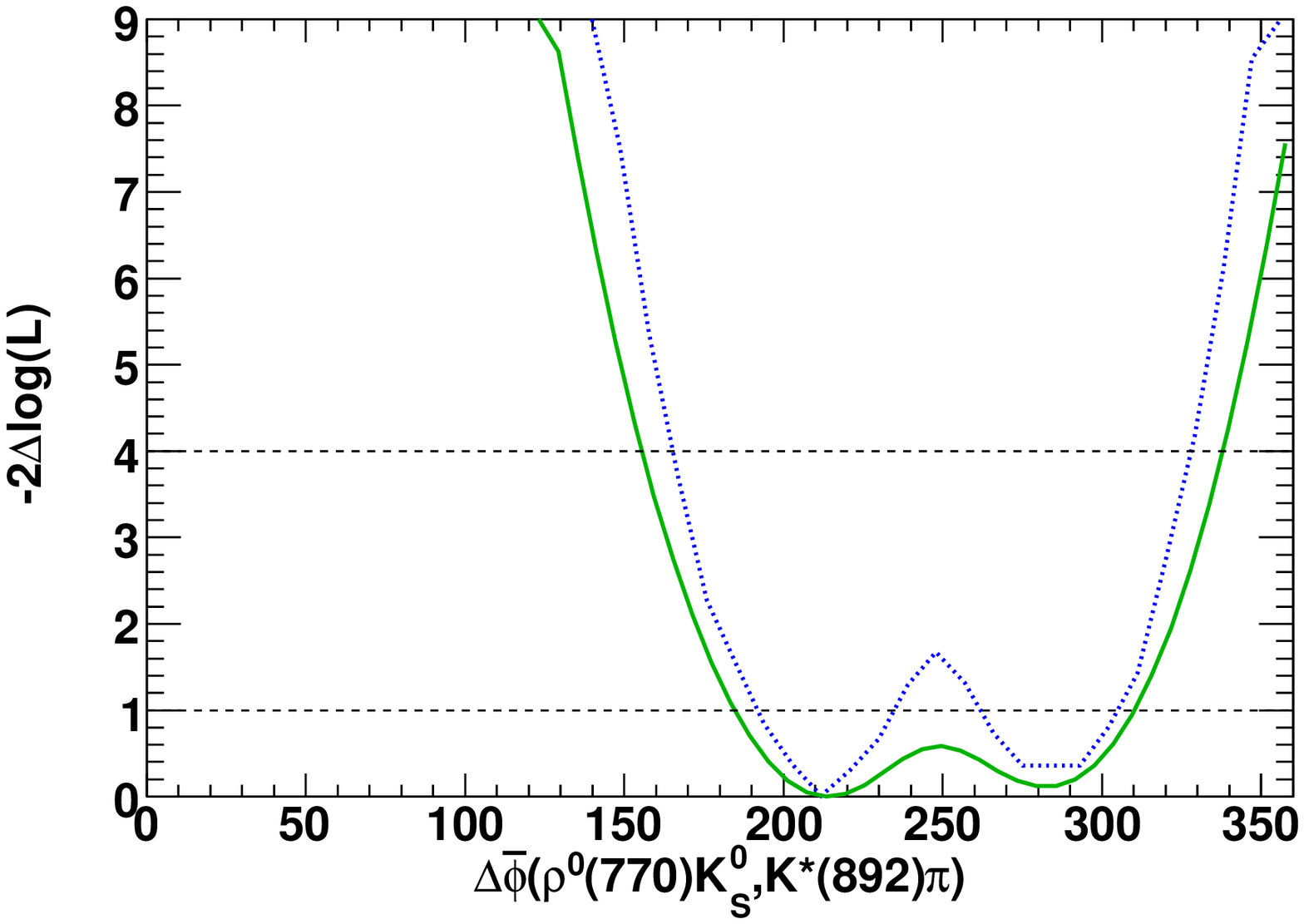}
\caption{\label{fig:LikeScans_InterPhases_2}{
Statistical (dashed line) and total (solid line) scans of $-2\Delta\logL$
as a function of the  phase differences
$\phi(K^{*}(892)\pi,(K\pi)^*_0\pi)$ (left),
$\phi(f_0(980)K^0_S,\rho^0(770)K^0_S)$ (middle), and
$\phi(\rho^0(770)K^0_S,K^{*}(892)\pi)$ (right).
The top (bottom) row shows $\Bz$ ($\Bzb$) candidates.
Horizontal dotted lines mark the
one and two standard deviation levels.}}
\end{center}
\end{figure*}

\begin{figure*}[htbp]
\begin{center}
\includegraphics[width=5.9cm,keepaspectratio]{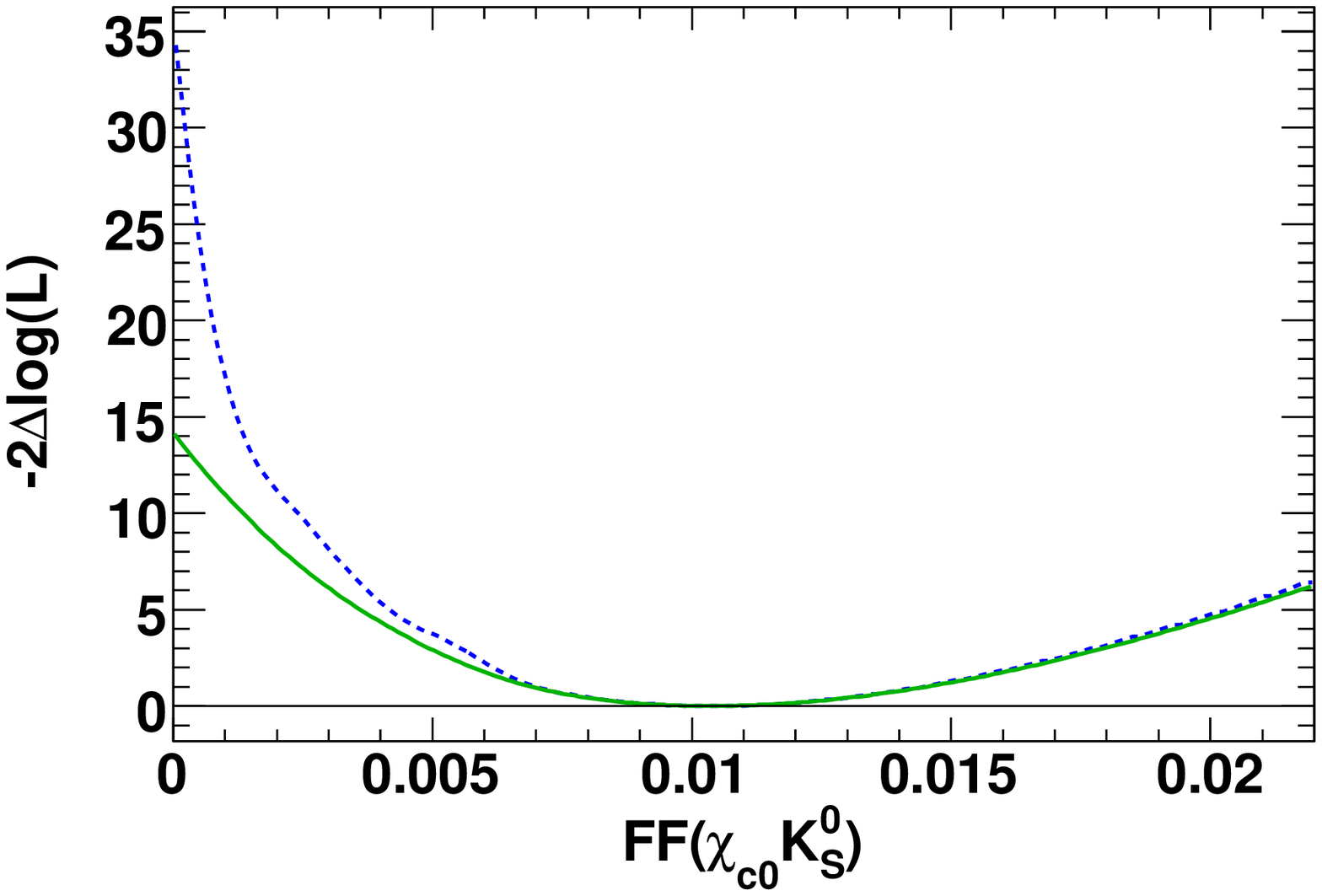}
\includegraphics[width=5.9cm,keepaspectratio]{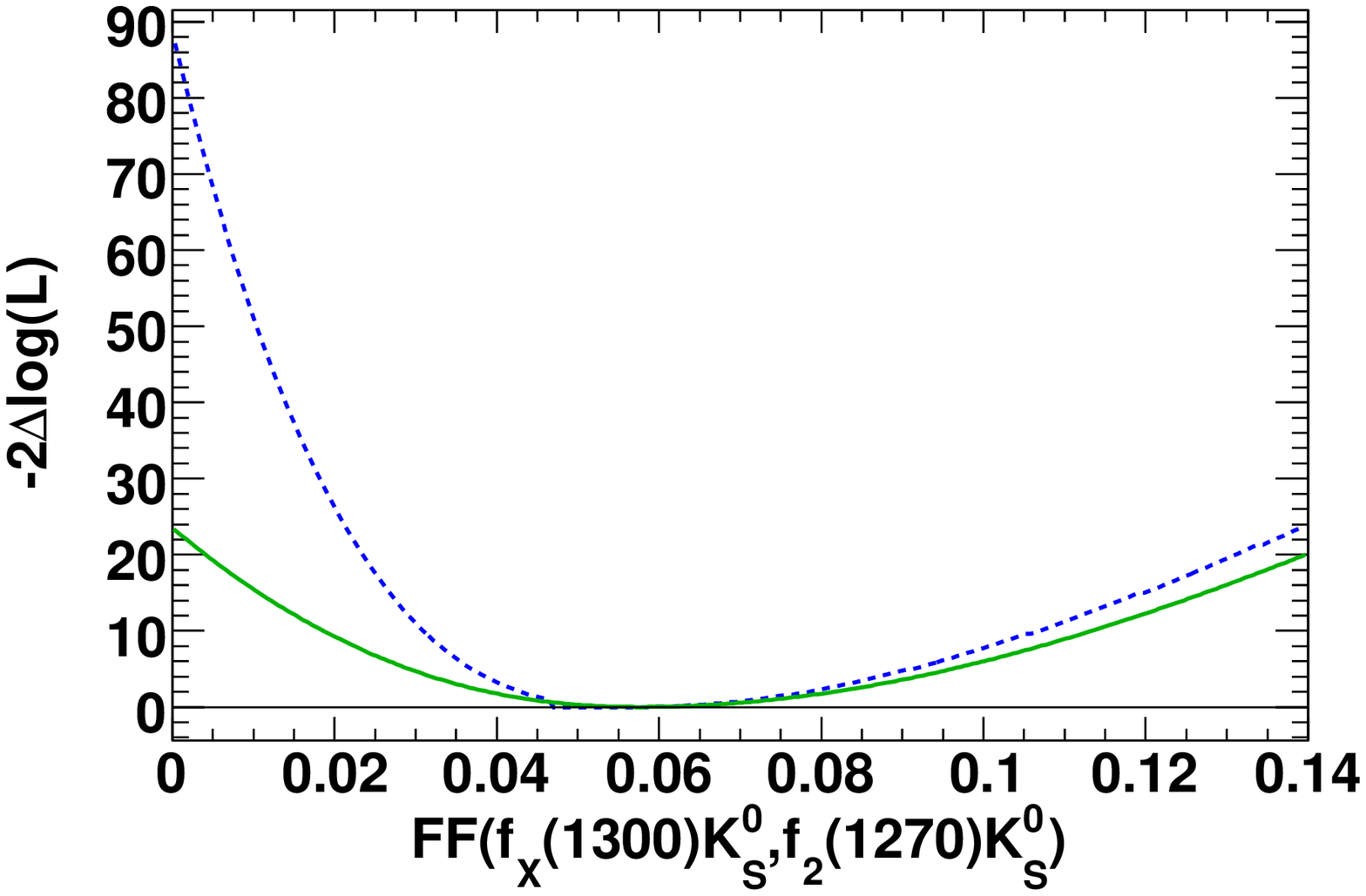}
\includegraphics[width=5.9cm,keepaspectratio]{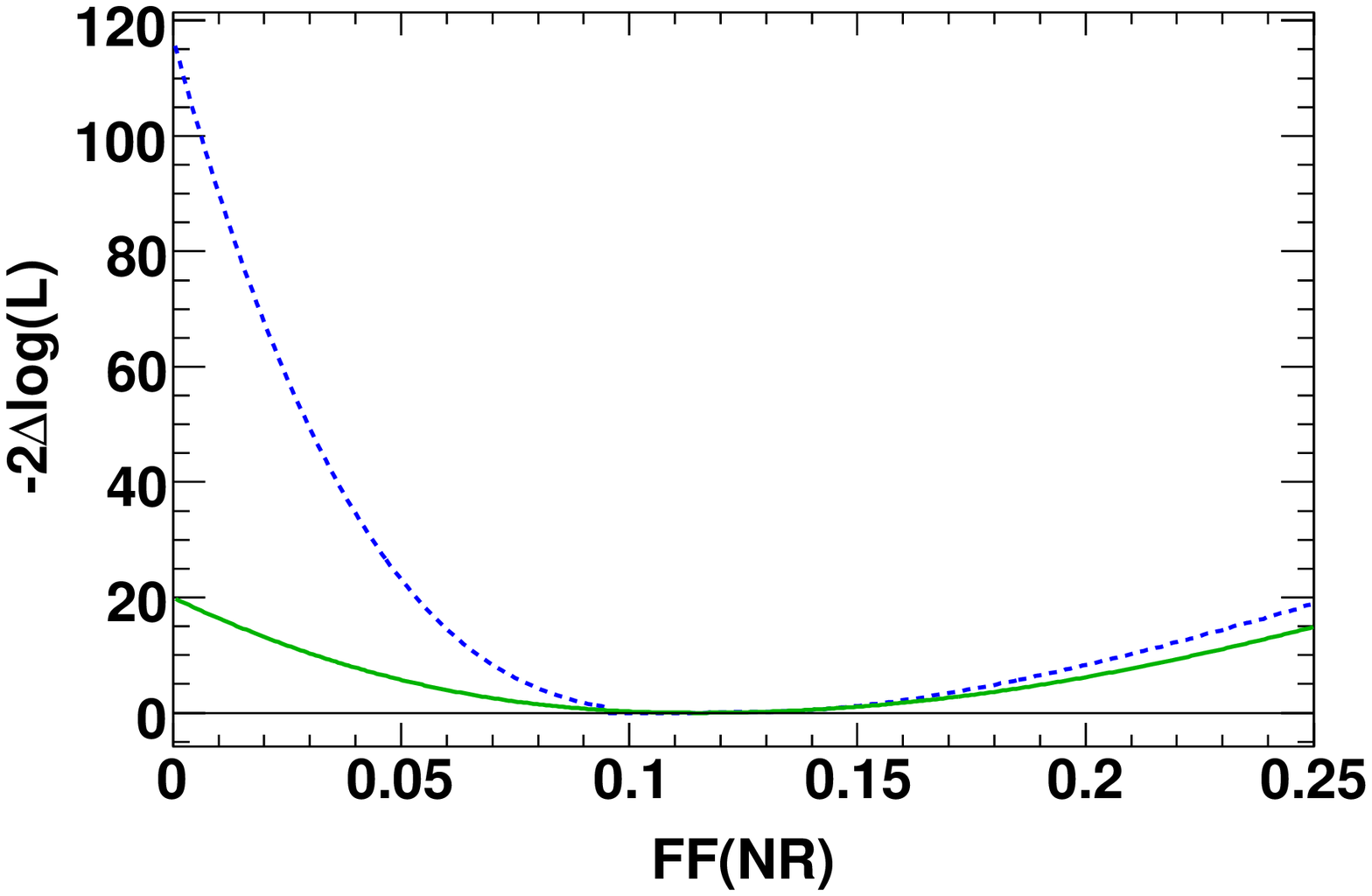}
\caption{\label{fig:fXf2chi0_significance}{
Statistical (dashed line) and total (solid line) scans in terms of $-2\Delta\logL$
as a function of the fit fractions of the  
$\chi_{c0}(1P)\KS$ component (left), the sum of fit fractions of the
$\fII\KS$ and $\fX\KS$ components (center), and the flat phase space NR component (right).
These scans are used to extract the probability of null values of these fit fractions.
}}
\end{center}
\end{figure*}

\begin{figure*}[htbp]
\begin{center}
\includegraphics[width=7.5cm,keepaspectratio]{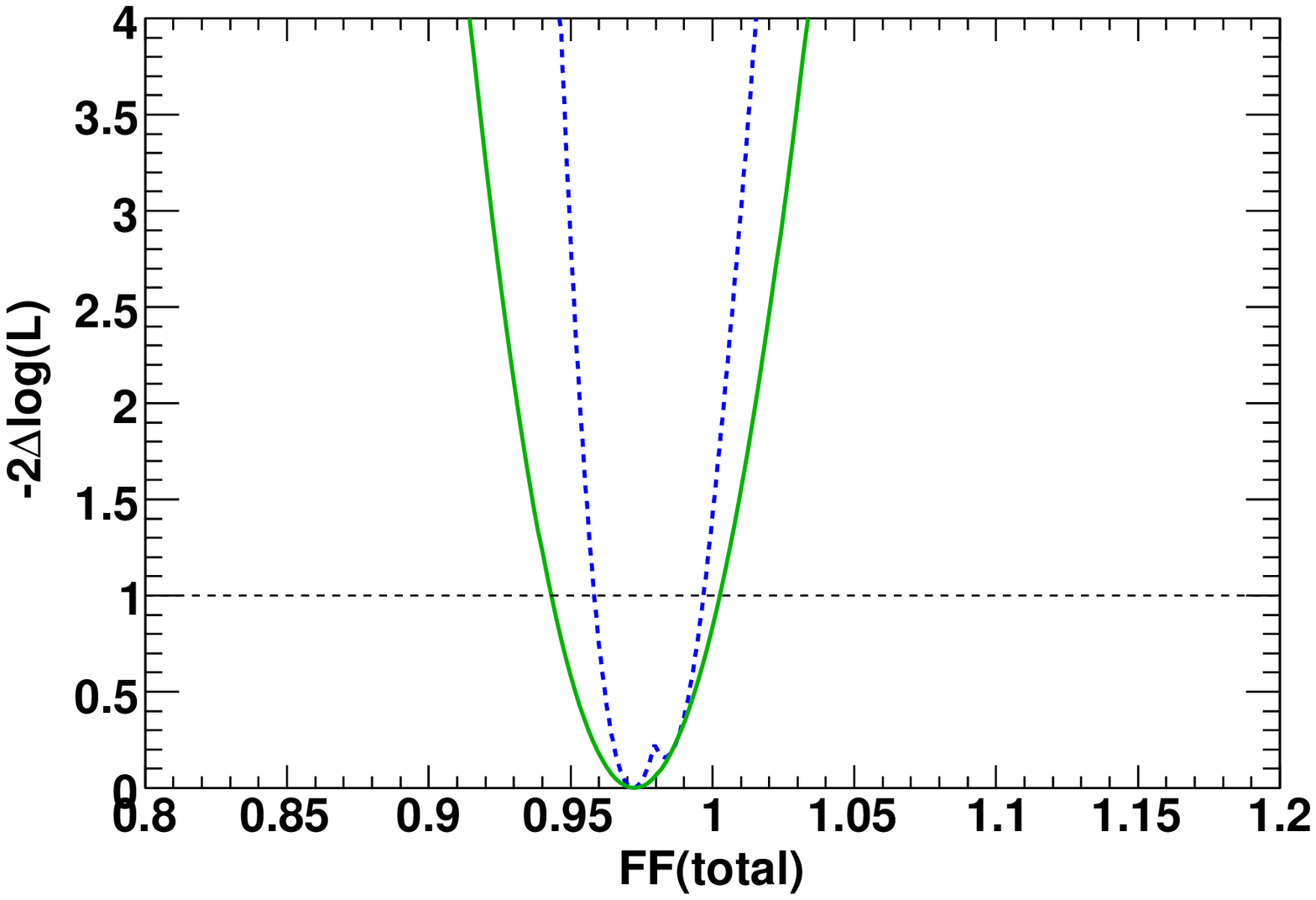}
\includegraphics[width=7.5cm,keepaspectratio]{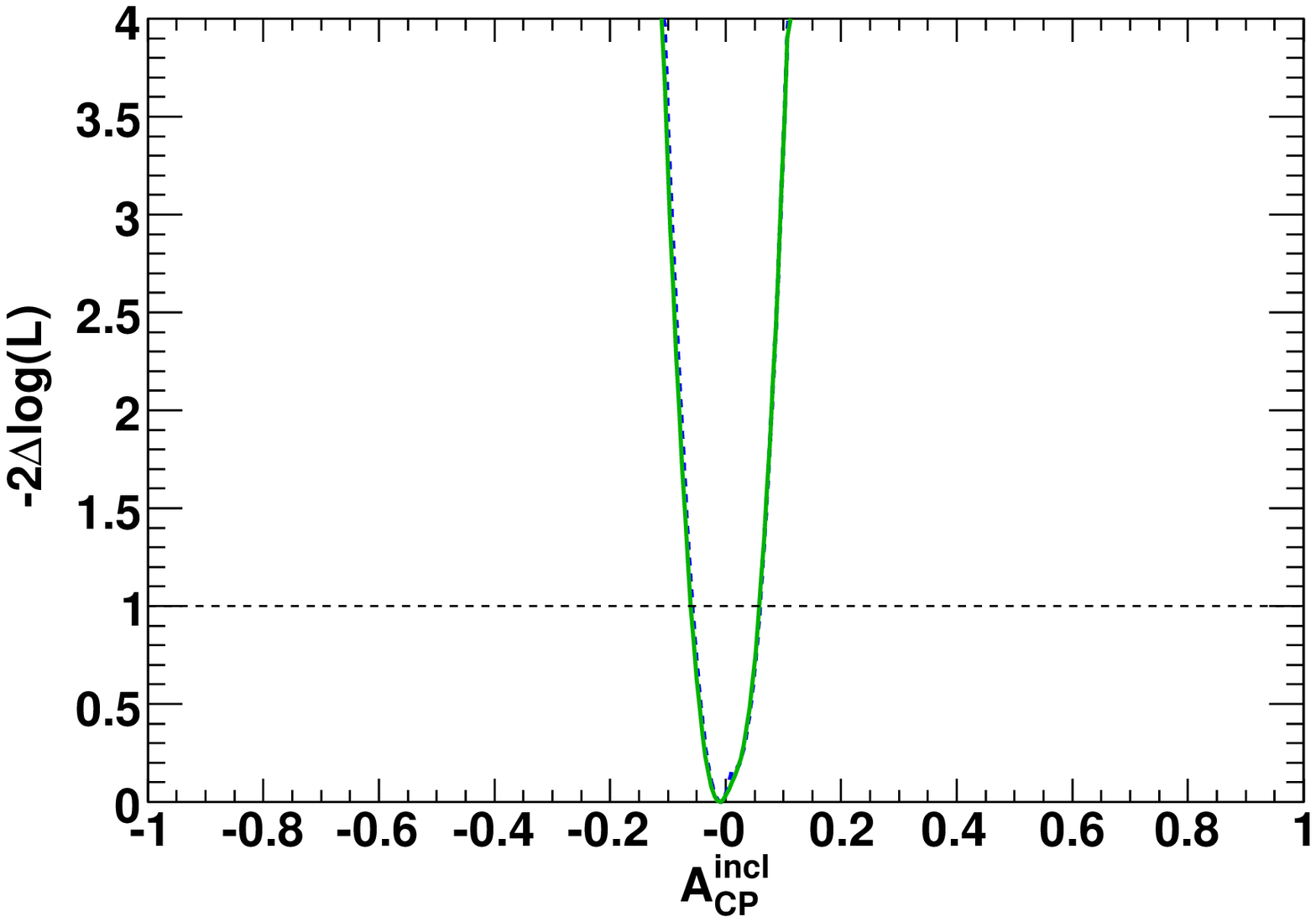}
\caption{\label{fig:Total_ACP_FF}{
Statistical (dashed line) and total (solid line) scans of $-2\Delta\logL$
as a function of the total fit fraction (left) and the inclusive direct \CP-asymmetry $A_{\CP}^{incl}$ (right).
A horizontal dotted line marks the one standard deviation level.}}
\end{center}
\end{figure*}

\begin{figure*}[htbp]
\begin{center}
\includegraphics[width=7.5cm,keepaspectratio]{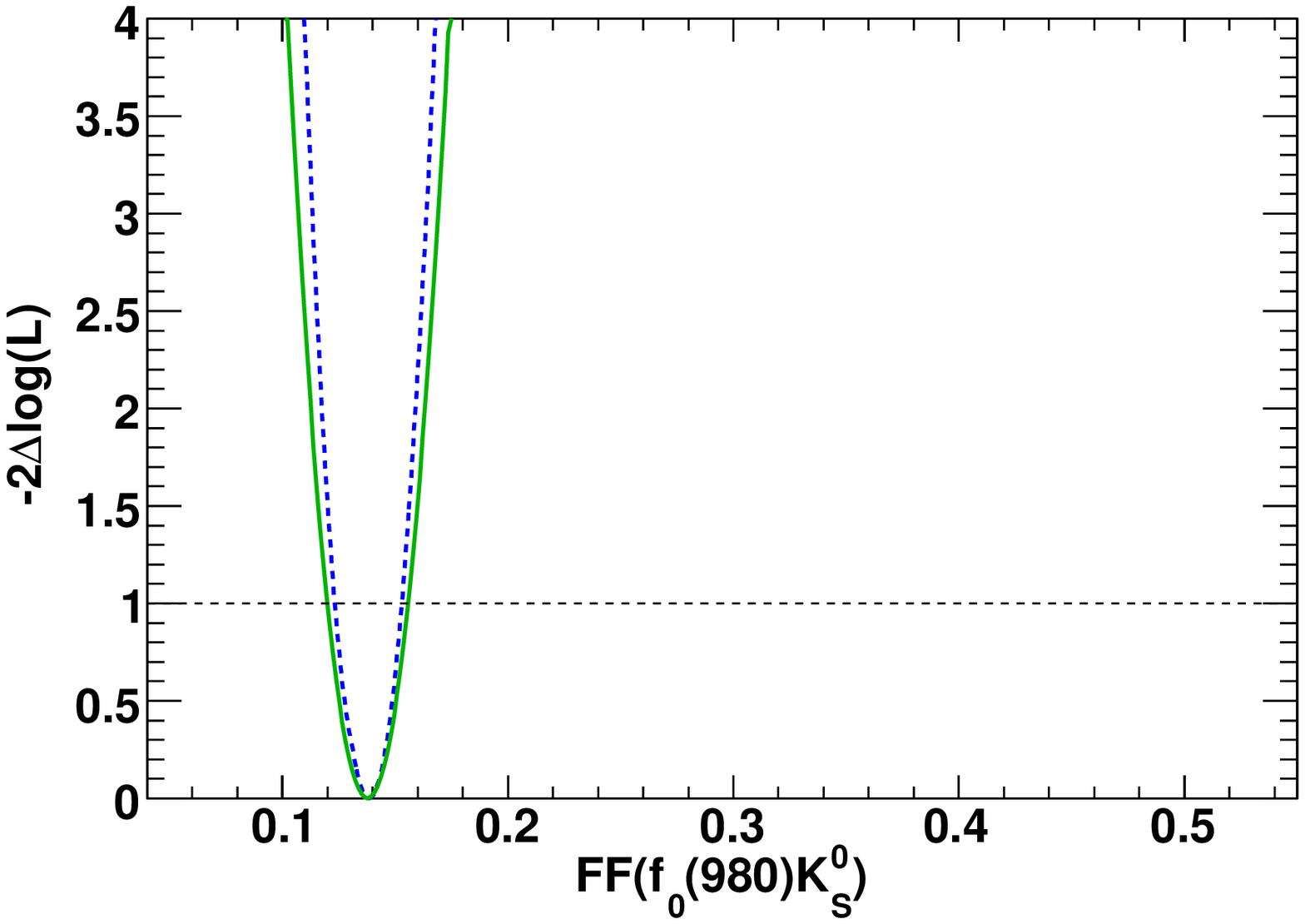}
\includegraphics[width=7.5cm,keepaspectratio]{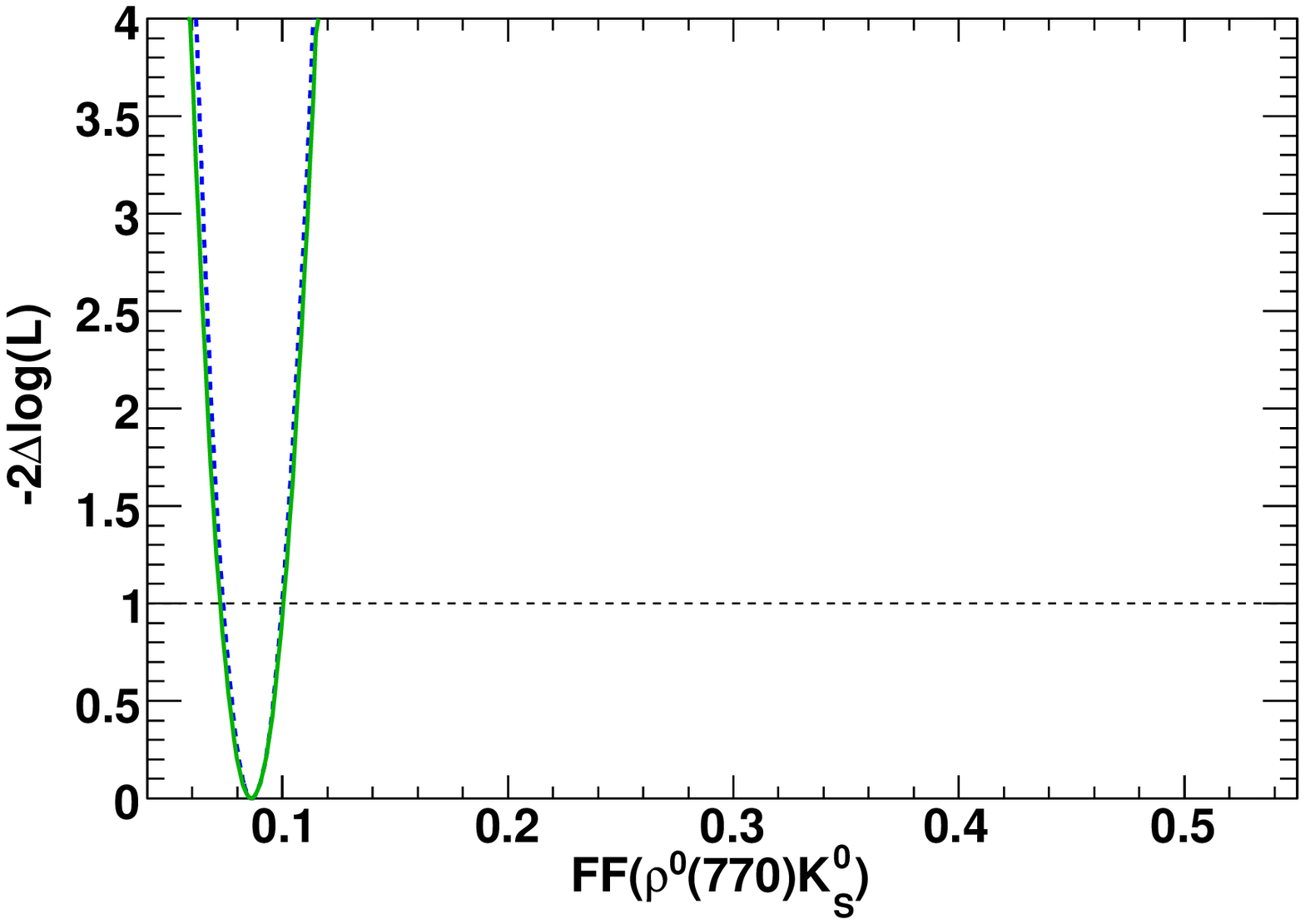}
\includegraphics[width=7.5cm,keepaspectratio]{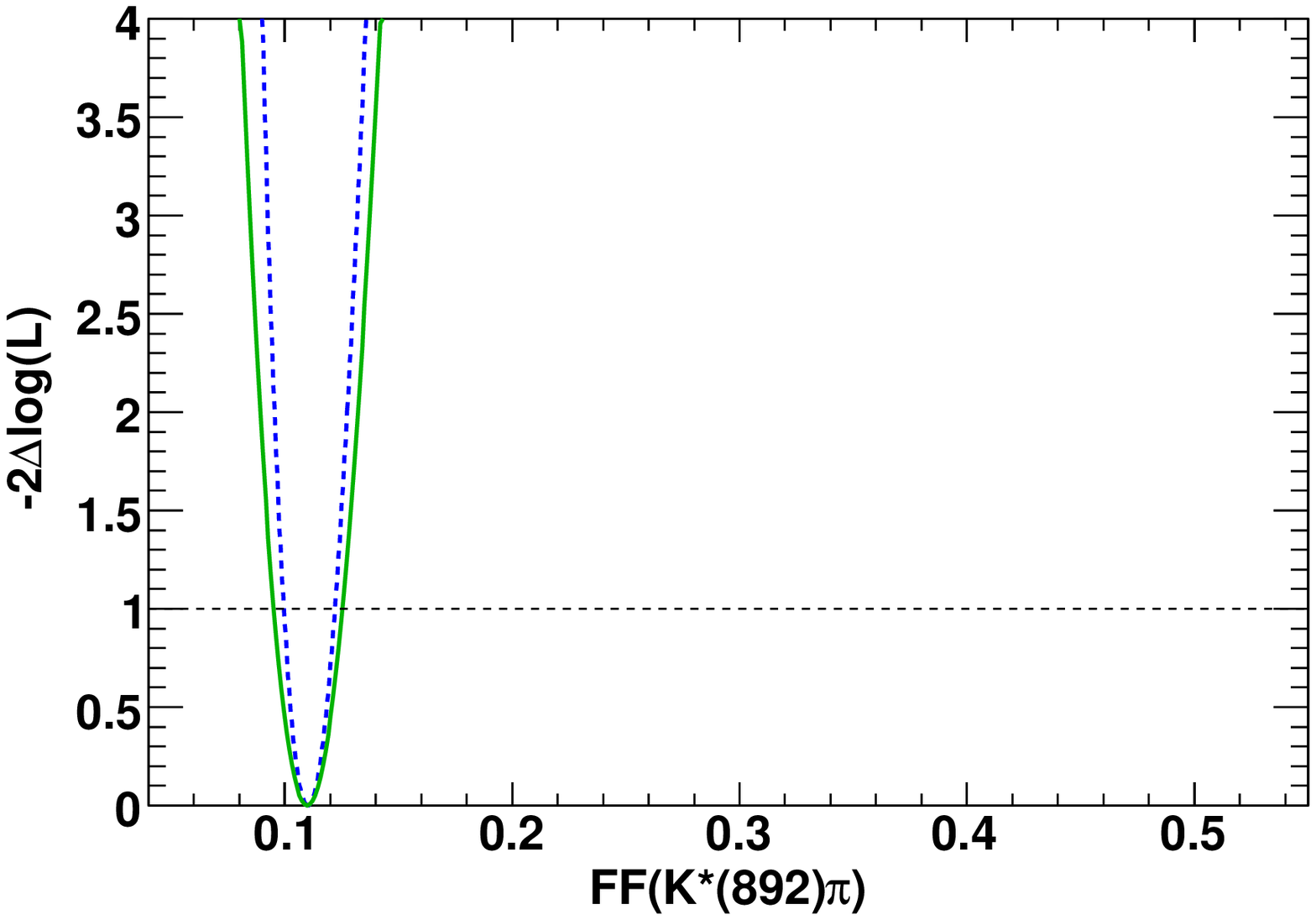}
\includegraphics[width=7.5cm,keepaspectratio]{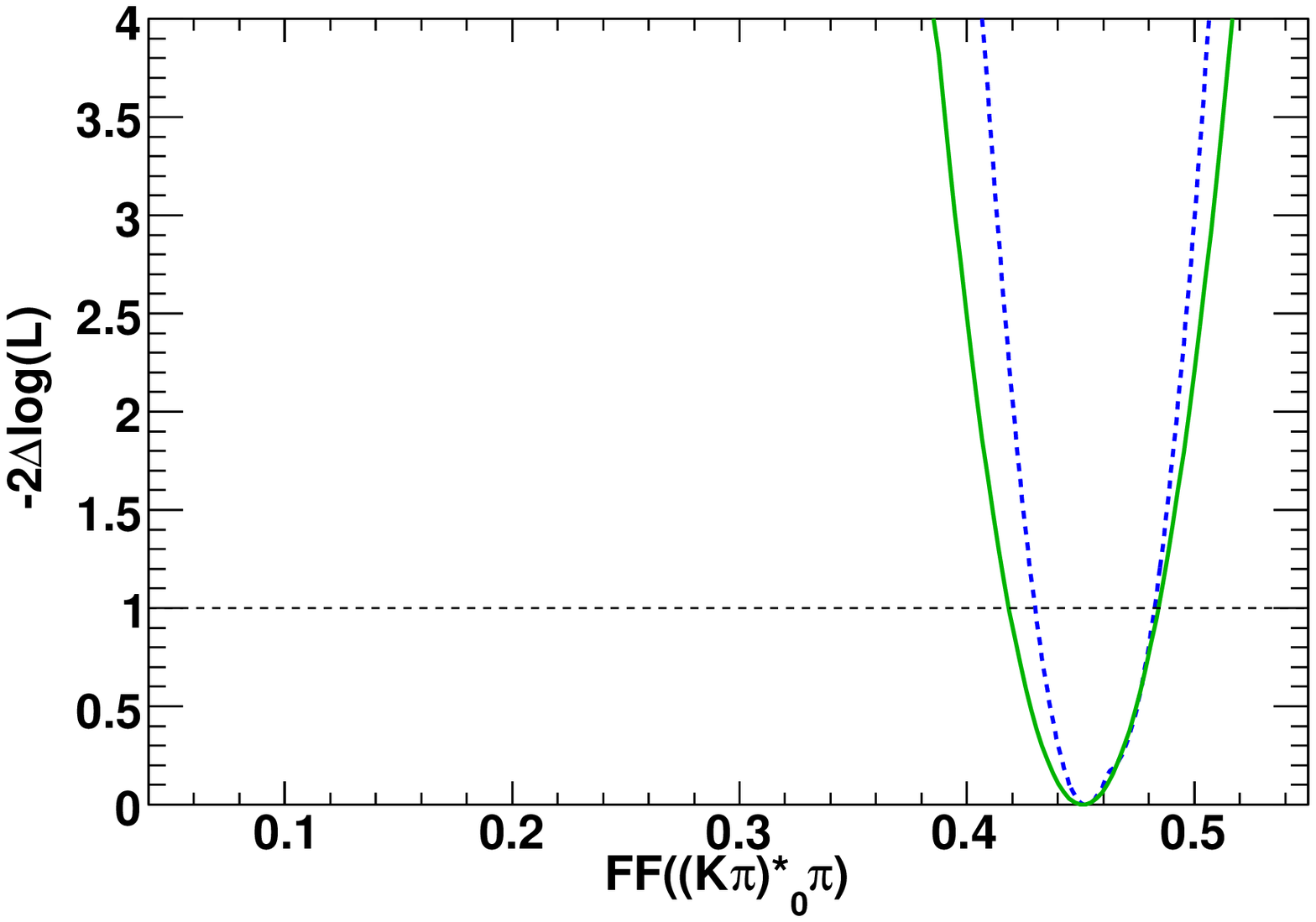}
\caption{\label{fig:LikeScans_FF_1}{
Statistical (dashed line) and total (solid line) scans of $-2\Delta\logL$
as a function of the fit fractions 
$FF(f_0(980)K^0_S)$ (top left), $FF(\rho^0(770)K^0_S)$ (top right), $FF(K^{*\pm}(892)\pi^{\mp})$ (bottom left),
and $FF((K\pi)^{*\pm}_0\pi^{\mp})$ (bottom right).
A horizontal dotted line marks the one standard deviation level.}}
\end{center}
\end{figure*}

\begin{figure*}[htbp]
\begin{center}
\includegraphics[width=7.5cm,keepaspectratio]{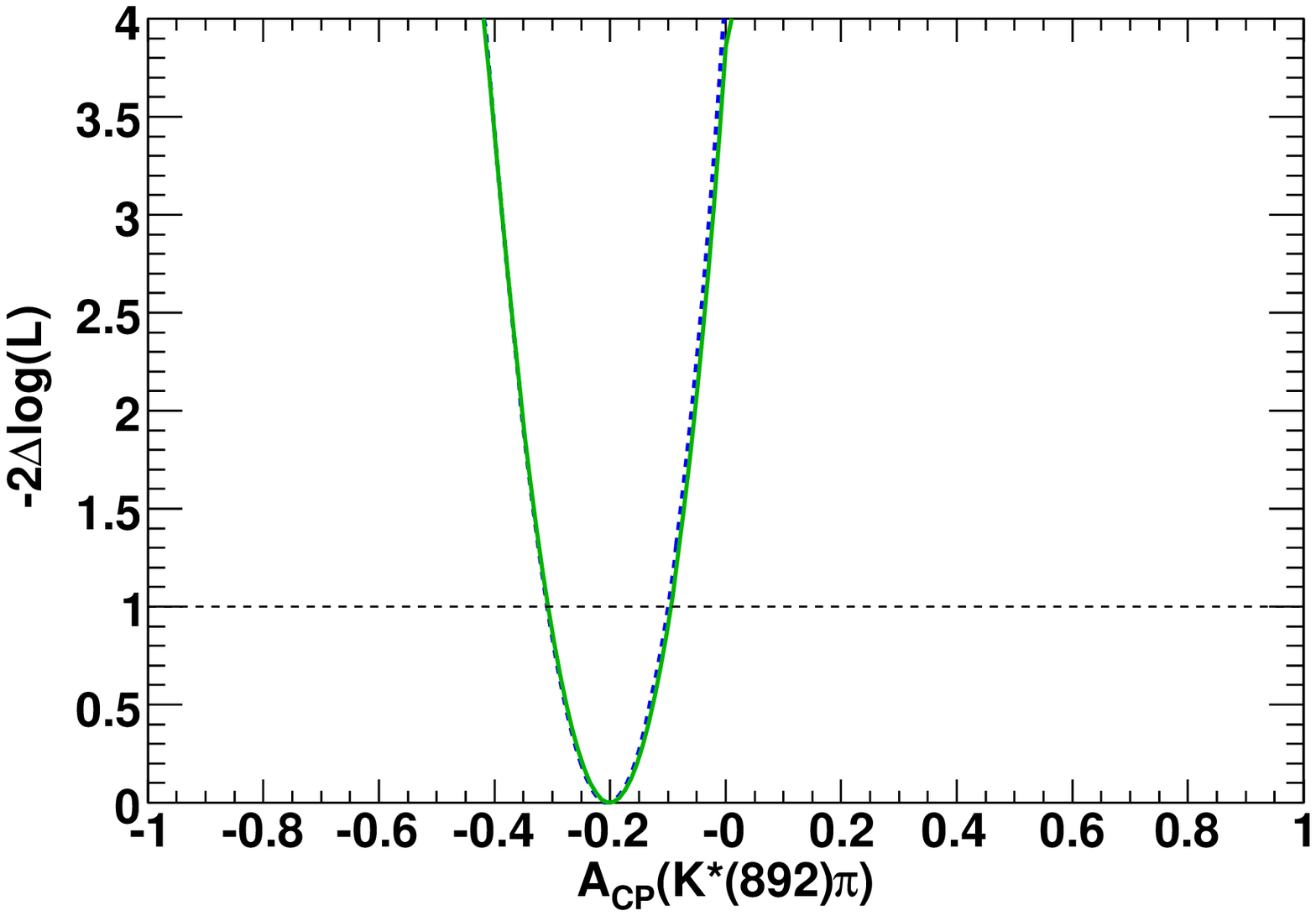}
\includegraphics[width=7.5cm,keepaspectratio]{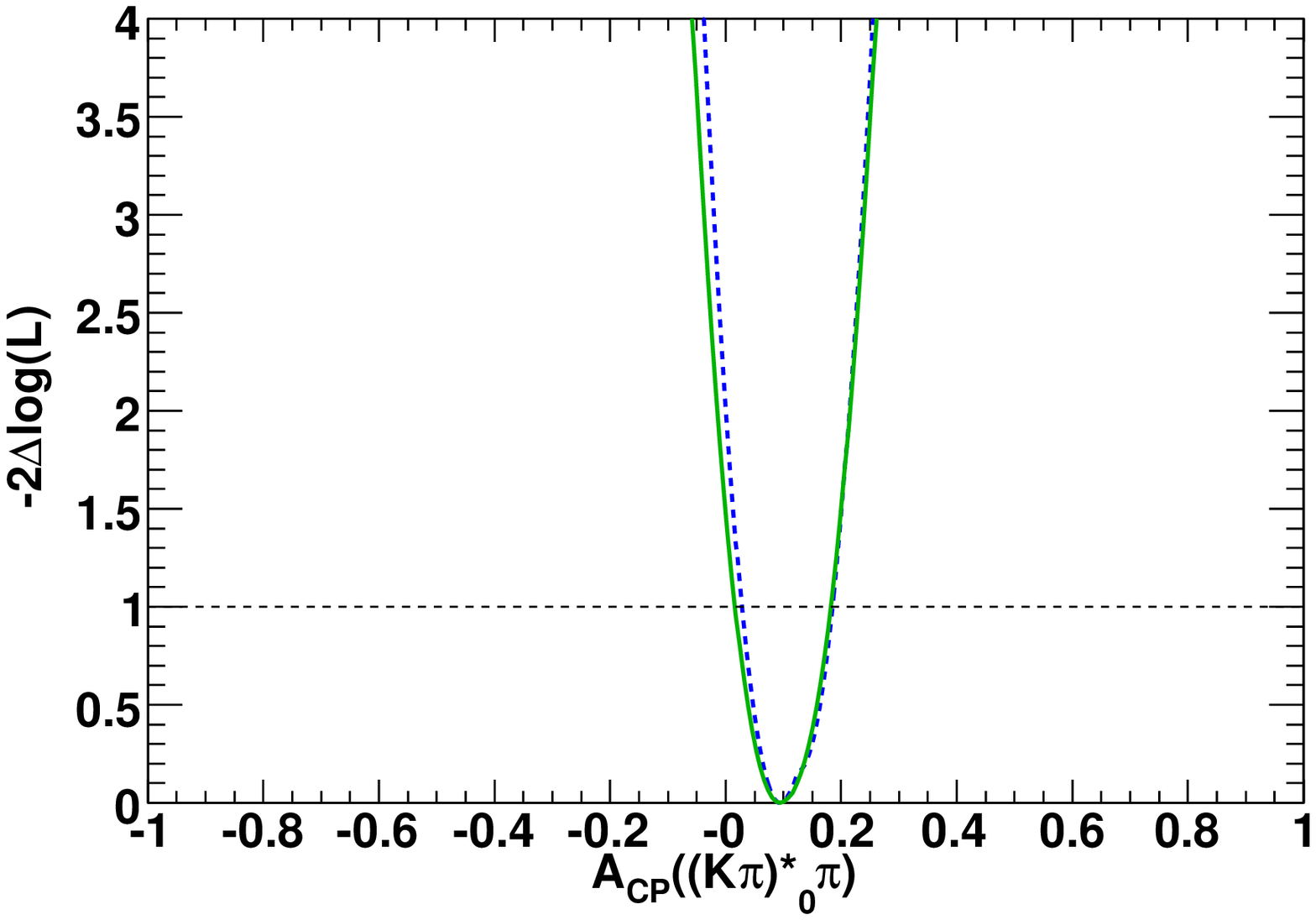}
\caption{\label{fig:LikeScans_ACP_1}{
Statistical (dashed line) and total (solid line) scans of $-2\Delta\logL$
as a function of the direct \CP asymmetries 
$\ACP(K^{*\pm}(892)\pi^{\mp})$ (left) and 
$\ACP((K\pi)^{*\pm}_0\pi^{\mp})$ (right). 
A horizontal dotted line marks the one standard deviation level.}}
\end{center}
\end{figure*}

To extract the statistical uncertainties on the Q2B parameters we perform
likelihood scans, not relying on any assumption about the shape of the
likelihood function.
Since the Q2B parameters are not directly used in the fit, we instead must
perform the scan fixing one or two parameters among the signal model
magnitudes and phases.  These are chosen in such a way that the resulting
likelihood curve can be trivially interpreted in terms of the Q2B parameter
of interest. In each case the chosen parameters are fixed at several consecutive
values, for each of which the fit to the data is repeated.
The error on the Q2B parameter is determined by the points, or the contour,
where the $-2\logL$ function changes by one unit with respect to its
minimum value.
Systematic uncertainties are discussed in Sec.~\ref{sec:Systematics}.
Results of the likelihood scans in terms of $-2\Delta\logL$ are shown in
Fig.~\ref{fig:2DLikeScans_f0rho_CBeta} to~\ref{fig:LikeScans_ACP_1}.

The measurements of time-dependent \CP-violation in the $f_0(980)K^0_S$
and $\rho^0(770)K^0_S$ modes are presented as two-dimensional likelihood scans in the $(\beta_{\rm eff},C)$
plane, shown in Fig.~\ref{fig:2DLikeScans_f0rho_CBeta}.
The scans are displayed as confidence level contours after two-dimensional
convolution with the covariance matrix of systematic uncertainties.
On the same figure are also displayed the one-dimensional likelihood scans of $\beta_{\rm eff}$.
For $f_0(980)K^0_S$ the two solutions lie below and above $45$ degrees and correspond very closely
to the trigonometric ambiguity between a given value of  $\beta_{\rm eff}$ and $90^{\circ} - \beta_{\rm eff}$
(mirror solutions). On the other hand, for $\rho^0(770)K^0_S$ both solutions are below $45$ degrees.
In this case the local solutions corresponding to the trigonometric ambiguities of the two observed
solutions are suppressed at $3.6$ and $2.0$ standard deviations, respectively. 

The $(\beta_{\rm eff},C)$ plane can be transformed to the more familiar
$(S,C)$ plane using Eq.~\eqref{eq:2betaeff} to~\eqref{eq:S}. The corresponding two-dimensional contours are shown 
in Fig.~\ref{fig:2DLikeScans_f0rho_CS}. While a part of the information on the phases is lost,
this representation has nonetheless the advantage
of allowing direct comparison with the measurement of $\sin{2\beta}$ and $C$ in
$b \to c\bar{c}s$ modes. For  $f_0(980)K^0_S$, the results agree with the expectation based on 
$b \to c\bar{c}s$ to $1.1\sigma$; for  $\rho^0(770)K^0_S$ the agreement is better than
$1\sigma$. 
For the measured values of $(\beta_{\rm eff},C)$ for $f_0(980)K^0_S$,
\CP conservation is excluded 
at $3.5\sigma$.
For  $\rho^0(770)K^0_S$, the measurement of  $(\beta_{\rm eff},C)$ 
is consistent with \CP conservation within $1\sigma$.

The measurement of the phase $\Delta\Phi(\KstarpI\pim)$ is 
presented as a  one-dimensional likelihood scan in Fig.~\ref{fig:LikeScans_InterPhases_1}.
For this flavor-specific mode, there is virtually no 
region in phase space that is accessible both to $\Bz$ and $\Bzb$; 
thus, sensitivity to this phase difference is limited.
Simulation shows that 
interference of the $\KstarpI\pim$ with the $f_0(980)K^0_S$ and $\rho^0(770)K^0_S$ modes
(for which \Bz and \Bzb amplitudes interfere via mixing) provides most of the 
sensitivity to $\Delta\Phi(\KstarpI\pim)$; unfortunately, the overlap in phase
space of these resonances is small.
As a consequence, 
only the $(-137,-5)^\circ$ interval is excluded at   $95\%$ confidence level.
Figure~\ref{fig:LikeScans_InterPhases_1} also shows the measurement of the similar
phase difference for the $(K\pi)^{*}_0$ component. As for $K^*(892)$, the measurement
sets no strong constraint on this phase. Only the interval $[-132,+25]^\circ$ is excluded at 
$95\%$ confidence level.

In contrast,   due to the sizable overlap in phase space between the $K\pi$ S- and P- waves of the same charge,
the relative phases $\phi((K\pi)^{*\pm}_0,K^{*\pm}(892))$ 
are measured to $\pm 13^\circ$ including systematics.
The one-dimensional scans are shown in Fig.~\ref{fig:LikeScans_InterPhases_2}. The
associated
observable $\Delta\phi((K\pi)^{*\pm}_0,K^{*\pm}(892))$ is compatible with \CP conservation.
Figure~\ref{fig:LikeScans_InterPhases_2} also shows the scans for $\phi(f_0(980)K^0_S,\rho^0(770)K^0_S)$, 
$\phi(\rho^0(770)K^0_S,\KstarpI\pim)$, and their corresponding \CP-conjugates.
It is clear from this figure and from Table V that the phases for the former are measured to a better accuracy. This is due to the larger overlap in phase space between the $\fI$ and the $\rhoI$. In both cases
the associated
observables $\Delta\phi$ are compatible with \CP conservation.

For the remaining resonant modes in the signal DP model:
$\fX\KS$, $\fII\KS$, NR, and $\chi_{c0}(1P)\KS$, we scan the
likelihood as a function of the corresponding fit fractions. These scans are shown in 
Fig.~\ref{fig:fXf2chi0_significance}. We obtain
a total (statistical and systematic) significance of $4.8$ and 
$3.8$ standard deviations for the  NR and $\chi_{c0}(1P)\KS$ components, respectively.
The significance for the sum of fit fractions of the $\fII\KS$ and $\fX\KS$ components is $4.8$ standard deviations
while their individual significances are $2.9\sigma$ and $2.4\sigma$, respectively.

The $(K\pi)^*_0$ component is modeled in our analysis by the LASS parametrization~\cite{LASS},
which consists of a NR effective range term plus a relativistic Breit-Wigner term for
the $\KstarII$ resonance. We separate from the corresponding branching fraction, quoted in
Table~\ref{tab:Final_Res_BFs}, the contribution of the $\KstarII$ resonance and find it to be
 $(29.9^{+2.3}_{-1.7} \pm 1.6 \pm 0.6 \pm 3.2)\times 10^{-6}$.
This value is corrected for the secondary branching
fraction using ${\mathcal B}(\KstarII \to K \pi)$ from Ref.~\cite{Amsler:2008zz} and the isospin relation 
${\mathcal B}(\KstarpII \to \Kz\pip)/{\mathcal B}(\KstarpII \to \Kp\piz)=2$.
The first uncertainty is statistical, the second is systematic, the third represents the DP signal model
dependence, and the fourth is due to the uncertainty on the secondary branching fraction.
In addition we calculate the total NR contribution by combining coherently the effective range
part of the LASS parametrization and the flat phase-space NR component.
We find this total NR fit fraction to be
 $22.1^{+2.8}_{-2.0} \pm 2.1 \pm 0.7\,\%$.
Note that this number accounts for the destructive interference between the two NR terms.
The corresponding branching fraction is
 $(11.07^{+2.51}_{-0.99} \pm 0.81 \pm 0.40)\times 10^{-6}$.

\begin{figure*}[htbp]
  \centerline{  \epsfxsize8.5cm\epsffile{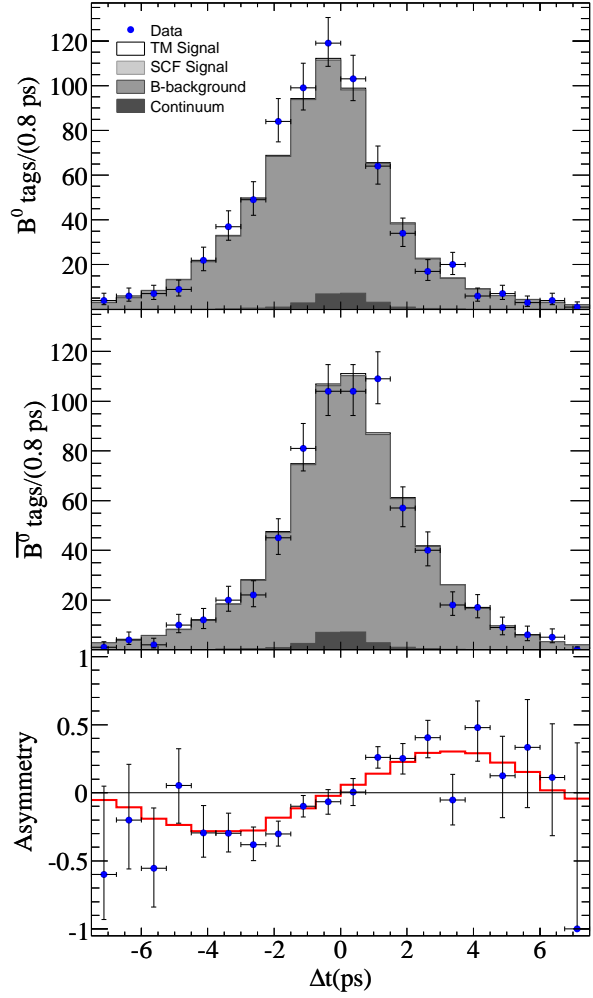}}
  \caption{\label{fig:dtAssymJpsi} Distributions of $\dt$ when the $\Bz_{\rm tag}$ is a $\Bz$ (top), $\Bzb$ (middle), and the derived $\dt$ asymmetry (bottom) 
for events in the $\jpsi \KS$ region. The solid line is the total PDF
and the points with error bars represent data.
}
\end{figure*}

As a validation of our treatment of the time-dependence, we allow
$\tau_{\Bz}$ and $\Delta m_d$ to vary in the fit. We find
$\tau_{\Bz} = 1.579 \pm 0.061 \ps$ and 
$\Delta m_d = 0.497 \pm 0.035 \ps^{-1}$
while the remaining free parameters are consistent with the nominal fit. The numbers
for $\tau_{\Bz}$ and $\Delta m_d$ are in agreement with current world averages~\cite{Barberio:2008fa}.
In addition we perform a fit floating the $S$ parameters for
$\Bz\rar J/\psi \KS$ and $\Bz\rar \psi(2S) \KS$ events.  We 
find $S=\sin(2\beta)=0.690\pm0.077$ and  $0.73\pm0.27$ for 
$J/\psi\KS$ and $\psi(2S) \KS$ respectively.  These numbers are
in agreement with the current world average~\cite{Barberio:2008fa}.
Signal enhanced distributions of $\dt$ and the $\dt$ asymmetry for events in the 
$J/\psi\KS$ region are shown in Fig.~\ref{fig:dtAssymJpsi}.
To validate the SCF modeling, we leave the average SCF fractions per tagging
category free to vary in the fit and find results that are consistent
with the MC estimation.

As a further cross-check of the results,
we performed an independent analysis and obtained compatible results~\cite{delAmoSanchez:2007zz}.
The main differences between this cross-check analysis and the one presented
here were the use of a Fisher discriminant instead of a NN, the removal of
bands in invariant mass to cut away the $\Bz \to \Dm\pip$, $\Bz \to \jpsi
\KS$ and $\Bz \to \psi(2S) \KS$ contributions, and the use of Cartesian
isobar parameters. 

\section{SYSTEMATIC STUDIES}
\label{sec:Systematics}

\begin{table*}[htbp]
\begin{center}
\caption{ \label{tab:systematics}
        Summary of systematic uncertainties on Q2B parameters. Errors on relative fractions ($\beta_{\rm eff}$ and phases) are given in percent (degrees).}
\setlength{\tabcolsep}{0.5pc}
\begin{tabular}{lcccccc}
\hline\hline
Parameter                        & DP Model     & Lineshape     & Fit Bias     & \B Background      & Other     & Total  \\
\hline\\[-9pt]                                                                                                                                     
$C(\fI\KS)$                      &  0.04        &  0.02         & $<$0.01      &  0.01              &  0.02     &  0.05  \\
$FF(\fI\KS)$                     &  0.6         &  0.69         &  0.5         &  0.07              & $<$0.01   &  1.03  \\
$\beta_{\rm eff}(\fI\KS)$            &  2.1         &  1.9          & $<$0.1       &  0.2               &  0.3      &  2.9   \\
\hline\\[-9pt]                                                                                                                          
$C(\rhoI\KS)$                    &  0.03        &  0.04         & $<$0.01      &  0.06              &  0.06     &  0.10  \\
$FF(\rhoI\KS)$                   &  0.23        &  0.31         &  0.3         &  0.09              &  0.15     &  0.52  \\
$\beta_{\rm eff}(\rhoI\KS)$          &  1.8         &  2.2          & $<$0.1       &  1.2               &  1.7      &  3.5   \\
\hline\\[-9pt]                                                                                                                          
$A_{CP}(\KstarI\pi)$             &  0.02        & $<$0.01       & $<$0.01      & $<$0.01            & $<$0.01   &  0.02  \\
$FF(\KstarI\pi)$                 &  0.8         &  0.13         & 0.4          &  0.03              &  0.43     &  1.00  \\
$\Delta\Phi(\KstarI\pi)$         &  8.1         &  2.8          & $<$0.1       &  1.4               &  3.3      &  9.3   \\
\hline\\[-9pt]                                                                                                                          
$A_{CP}((K\pi)_0^*\pi)$          &  0.02        & $<$0.01       & $<$0.01      & $<$0.01            &  0.02     &  0.03  \\
$FF((K\pi)_0^*\pi)$              &  0.90        &  0.39         &  1.8         &  0.12              &  0.33     &  2.08  \\
$\Delta\Phi((K\pi)_0^*\pi)$      &  4.4         &  2.4          & $<$0.1       &  1.3               &  3.0      &  6.0   \\
\hline\\[-9pt]                                                                                                                          
$C(\fII\KS)$                     &  0.07        &  0.04         & $<$0.01      &  0.05              &  0.06     &  0.11  \\
$FF(\fII\KS)$                    &  0.69        &  0.16         &  0.09        &  0.02              &  0.19     &  0.74  \\
\hline\\[-9pt]                                                                                                                          
$C(\fX\KS)$                      &  0.09        &  0.03         & $<$0.01      &  0.01              &  0.03     &  0.10  \\
$FF(\fX\KS)$                     &  0.87        &  0.28         &  0.14        &  0.02              &  0.17     &  0.94  \\
\hline\\[-9pt]                                                                                                                          
$C(NR)$                          &  0.04        &  0.01         & $<$0.01      &  0.01              &  0.07     &  0.08  \\
$FF(NR)$                         &  0.60        &  0.86         &  0.5         &  0.12              &  1.62     &  2.00  \\
\hline\\[-9pt]                                                                                                                          
$C(\chiczero\KS)$                &  0.05        &  0.02         & $<$0.01      &  0.01              &  0.02     &  0.06  \\
$FF(\chiczero\KS)$               &  0.09        &  0.06         &  0.04        & $<$0.01            & $<$0.01   &  0.11  \\
\hline\\[-9pt]                                                                                                                          
$A_{\CP}^{incl}$                 & $<$0.01      & $<$0.01       & $<$0.01      & $<$0.01            &  0.01     &  0.01  \\
$FF_{Tot}$                       &  1.15        &  0.55         &  2.0         &  0.08              &  0.36     &  2.40  \\
\hline\\[-9pt]                                                                                                                          
$\phi(\fI\KS,\rhoI\KS)$          &  4.4         &  2.6          & $<$0.1       &  3.4               &  4.3      &  7.5   \\
$\phi(\rhoI\KS,\KstarI\pi)$      & 12.7         &  3.0          & $<$0.1       &  3.6               &  7.3      & 15.4   \\
$\phi(\rhoI\KS,(K\pi)^*_0\pi)$   &  8.7         &  8.5          & $<$0.1       &  3.9               &  3.7      & 13.3   \\
$\phi(\KstarI\pi,(K\pi)^*_0\pi)$ &  4.7         &  0.7          & $<$0.1       &  0.3               &  4.6      &  6.6   \\
\hline\\[-9pt]

Signal Yield                     & 31.7         &  5.8          & 14.0         &  3.3               & 23.0      & 42.1   \\
\hline \hline
\end{tabular}
\vspace{1.5\baselineskip}

\vspace{-0.35cm}

\end{center}
\end{table*}

To estimate the contribution to $\BztoKspipi$ decay via other
resonances, we first
fit the data including these other decays in the fit model.
We consider possible resonances,
including  $\mbox{$\omega(782)$}$, $\rhoII$, $\mbox{$\rhoz(1700)$}$,
$f_0(1710)$, $f_2(1810)$, $\KstarpmIV$, $K^{*\pm}_2(1430)$,
$\chi_{c2}(1P)$,
and a low mass $\sigma$. A relativistic Breit--Wigner lineshape is used to parameterize these additional
resonances, with masses and widths from Ref.~\cite{Amsler:2008zz}.
As a second step we simulate high statistic samples of events, using a model based 
on the previous fits, including the additional resonances. Finally, we 
fit these simulated samples using the nominal signal model.
The systematic effect (contained in the ``DP Model'' field in 
Table~\ref{tab:systematics}) is taken from the difference observed between the generated and fitted values.
We quote this DP model
uncertainty separately from other systematics.

We vary the mass, width,
and any other parameters of all isobar fit components within their errors,
as quoted in Table~\ref{tab:model}, and assign the observed
differences in the measured amplitudes as systematic uncertainties
(``Lineshape'' in Table~\ref{tab:systematics}).

To validate the fitting tool, we perform fits on large MC samples 
of fully-reconstructed events with
the measured proportions of signal, continuum, and $B$ background events.
No significant biases are observed in these fits and therefore no corrections are applied.
The statistical uncertainties on the fit parameters are taken as systematic uncertainties
(``Fit Bias'' in Table~\ref{tab:systematics}).

Another major source of systematic uncertainty is the $B$ background model. 
The expected event yields from the background modes are varied according 
to the uncertainties in the measured or estimated branching fractions.  
Since $B$ background modes may exhibit  \CP violation, the corresponding 
parameters are varied within their uncertainties, or, if unknown, within the physical range.
As is done for the signal PDFs, we vary the $\dt$ resolution parameters and
the flavor-tagging parameters within their uncertainties and assign
the differences observed in these fits with respect to the nominal fit
as systematic errors. 
These errors are
listed as ``\B Background '' in Table~\ref{tab:systematics}.  

Other systematic effects are much less important for the measurements
of the amplitudes and  are combined in 
the ``Other'' field in Table~\ref{tab:systematics}. Details are given
below.

The parameters of the continuum PDFs are determined by the fit. No additional systematic 
uncertainties are assigned to them. An exception to this is the DP
PDF: to estimate the systematic
uncertainty from the $\mes$ sideband extrapolation, we use large 
samples of $\epem \to q\bar{q}$ MC data ($q=u,d,s,c$).
We compare the distributions of $\mprime$ and $\thetaprime$ between sidebands at
different ranges in $\mes$ and find the two such sidebands that show the
maximum discrepancy.  We assign as systematic uncertainty the effect seen
when weighting the continuum DP PDF by the ratio of these two data sets.

The uncertainties associated with $\dmd$ and $\tau$ are
estimated by varying these parameters within the uncertainties
on the world average~\cite{Amsler:2008zz}.

The signal PDFs for the $\dt$ resolution and tagging fractions 
are determined from fits to a control sample
of fully reconstructed \B decays to exclusive final states with
charm, and the uncertainties are obtained by varying the parameters
within the statistical uncertainties.

Finally, the uncertainties due to particle
identification, tracking efficiency corrections, $\KS$ reconstruction, and the calculation
of $N_{B\bar{B}}$ are $2.0\%$, $1.6\%$, $0.9\%$, and $1.1\%$, respectively.
These contribute only to the branching fraction systematic uncertainties.

The average fraction of misreconstructed signal events ($\fscfave$) predicted by the MC
simulation has been verified with fully reconstructed $\B\to D\rho$
events~\cite{Aubert:2007jn}. No significant differences between data and
the simulation were found. 
To estimate
a systematic uncertainty from 
$\fscfave$, we vary these fractions, for all tagging categories.
Tagging efficiencies, dilutions, and biases for signal events
are varied within their experimental uncertainties.
\section{SUMMARY}
\label{sec:Summary}

We have presented results from a time-dependent Dalitz plot analysis of $\BztoKspipi$ decays
obtained from a data sample of $383$ million $\FourS \to B\Bbar$ decays.
Using an amplitude analysis technique,
we measure $15$ pairs of relative magnitudes and phases for the different resonances,
taking advantage of the interference between them in the Dalitz plot.
From the measured decay amplitudes, we derive the Q2B parameters of the resonant decay modes.
Two solutions, with equivalent goodness-of-fit, were found.

Including systematic and Dalitz plot model uncertainties, the combined confidence interval 
for the measured values of
$\beta_{\rm eff}$ in $\Bz$ decays to $\fI\KS$ is
$18^\circ< \beta_{\rm eff}<76^\circ$ at $95\%$ C.L.
\CP conservation in $\Bz$ decays to $\fI\KS$ is excluded at 
$3.5\sigma$, including systematics. For $\Bz$ decays to
$\rhoI\KS$, the combined confidence interval is $-9^\circ< \beta_{\rm eff}<57^\circ$
at $95\%$ C.L.
These results are both consistent with the measurements in $b\to c\bar{c}s$ modes.

In decays to $\KstarpI \pim$, we find 
$\ACP=-0.20 \pm 0.10\pm 0.01\pm 0.02$.
For the relative phase between decay amplitudes of $\Bz \to \KstarpI \pim$ and $\Bzb \to \KstarmI \pip$,
we exclude the interval $-137^\circ<\Delta\Phi(K^{*}(892)\pi) < -5^\circ$ at $95\%$ C.L.
This last result, combined with measurements of branching ratios, direct \CP asymmetries, and relative
phases in $\KstarpI \pim$ and $\KstarzI \piz$, plus a theoretical hypothesis on the
contributions of electroweak penguins to the decay amplitudes, can be used to set non-trivial constraints
on the CKM parameters $(\rhobar,\etabar)$ by following the methods proposed 
in Refs.~\cite{Deshpande:2002be,Ciuchini:2006kv,Gronau:2006qn,Lipkin:1991st}.
\section*{ACKNOWLEDGMENTS}
\label{sec:acknowledgments}
We are grateful for the 
extraordinary contributions of our \pep2\ colleagues in
achieving the excellent luminosity and machine conditions
that have made this work possible.
The success of this project also relies critically on the 
expertise and dedication of the computing organizations that 
support \babar.
The collaborating institutions wish to thank 
SLAC for its support and the kind hospitality extended to them. 
This work is supported by the
US Department of Energy
and National Science Foundation, the
Natural Sciences and Engineering Research Council (Canada),
the Commissariat \`a l'Energie Atomique and
Institut National de Physique Nucl\'eaire et de Physique des Particules
(France), the
Bundesministerium f\"ur Bildung und Forschung and
Deutsche Forschungsgemeinschaft
(Germany), the
Istituto Nazionale di Fisica Nucleare (Italy),
the Foundation for Fundamental Research on Matter (The Netherlands),
the Research Council of Norway, the
Ministry of Education and Science of the Russian Federation, 
Ministerio de Educaci\'on y Ciencia (Spain), and the
Science and Technology Facilities Council (United Kingdom).
Individuals have received support from 
the Marie-Curie IEF program (European Union) and
the A. P. Sloan Foundation.

\section*{APPENDIX}

The full (statistical, systematic, and model dependence) correlation matrices of the isobar parameters for
solutions I and II are given in Tables~\ref{tab:cormat1} and ~\ref{tab:cormat2}, respectively.
The tables are organized in blocks for $c$, $\bar{c}$, $\arg{c}$, and $\arg{\bar{c}}$.
Here, the abbreviations $f_0$, $\rho^0$, $K^*$, $S$, $f_2$, $f_X$, $NR$, and $\chi$ represent the components $\fI\KS$, $\rhoI\KS$, $\KstarI\pi$, $(K\pi)_0^*\pi$, $\fII\KS$, $\fX\KS$, nonresonant, and $\chiczero\KS$, respectively.

\begin{table*}[htbp]
\begin{center}
\caption{
Full correlation matrix for the isobar parameters of solution I. The entries are given in percent. Since the matrix is symmetric, all elements above the diagonal are omitted. 
\label{tab:cormat1}}
\begin{tabular}{cc|rrrrrrr|rrrrrrrr}
\hline\hline
&     & \multicolumn{7}{c|}{$\left|c\right|$}                               & \multicolumn{8}{c}{$\left|\bar{c}\right|$}\\[4pt]
&        &$\rho^0\ $& $K^*\ $ & $S\ $   & $f_2\ $ & $f_X\ $ & $NR\ $  & $\chi\ $& $f_0\ $&$\rho^0\ $& $K^*\ $ & $S\ $   & $f_2\ $ & $f_X\ $ & $NR\ $  & $\chi\ $\\
\hline\\[-9pt]
\multirow{7}{*}{$\left|c\right|$} 
&$\rho^0$ & $100.0$ \\
& $K^*$   & $ 51.9$ & $100.0$ \\
& $S$     & $ 54.0$ & $ 65.0$ & $100.0$ \\
& $f_2$   & $  8.4$ & $  2.8$ & $ 21.0$ & $100.0$ \\
& $f_X$   & $ 14.9$ & $ 23.2$ & $ 32.2$ & $ 22.7$ & $100.0$ \\
& $NR$    & $  5.2$ & $ 35.0$ & $ 24.4$ & $ 12.6$ & $ 39.3$ & $100.0$ \\
& $\chi$  & $  6.4$ & $  9.9$ & $  7.8$ & $  2.0$ & $  7.4$ & $  6.1$ & $100.0$ \\
\cline{1-10}\\[-9pt]
\multirow{8}{*}{$\left|\bar{c}\right|$}
& $f_0\ $ & $ 31.3$ & $ 30.3$ & $ 39.9$ & $ 25.2$ & $ 36.7$ & $ 31.3$ & $  8.0$ & $100.0$ \\
&$\rho^0$ & $ 20.6$ & $ 48.6$ & $ 51.2$ & $  8.0$ & $ 27.7$ & $ 27.5$ & $  5.6$ & $ 17.3$ & $100.0$ \\
& $K^*$   & $ 44.7$ & $ 73.5$ & $ 56.3$ & $ -4.8$ & $ 24.9$ & $ 22.0$ & $  9.5$ & $ 22.6$ & $ 43.4$ & $100.0$ \\
& $S$     & $ 59.6$ & $ 71.9$ & $ 79.7$ & $ 21.8$ & $ 39.3$ & $ 26.9$ & $ 11.3$ & $ 35.2$ & $ 49.4$ & $ 57.7$ & $100.0$ \\
& $f_2$   & $  2.4$ & $-10.1$ & $  6.3$ & $-56.1$ & $ -1.5$ & $  3.9$ & $ -0.3$ & $ 10.7$ & $ -6.2$ & $-21.5$ & $  5.0$ & $100.0$ \\
& $f_X$   & $ 14.5$ & $ 34.1$ & $ 12.5$ & $ 16.1$ & $-23.0$ & $ 12.4$ & $  2.5$ & $ 34.5$ & $  7.3$ & $  8.3$ & $ 12.9$ & $ -6.2$ & $100.0$ \\
& $NR$    & $ 17.8$ & $ 57.6$ & $ 41.7$ & $ 12.7$ & $ 10.1$ & $ 49.7$ & $  2.4$ & $ 40.0$ & $ 32.1$ & $ 25.0$ & $ 31.7$ & $  7.5$ & $ 46.2$ & $100.0$ \\
& $\chi$  & $ 18.9$ & $ 27.0$ & $ 30.6$ & $  5.8$ & $ 11.8$ & $  9.5$ & $-84.2$ & $ 21.5$ & $ 17.8$ & $ 24.1$ & $ 27.8$ & $  0.8$ & $  8.1$ & $ 20.2$ & $100.0$ \\
\hline\\[-9pt]
\multirow{7}{*}{$\arg(c)$}
&$\rho^0$ & $-11.2$ & $ 13.3$ & $  4.0$ & $-16.1$ & $ -2.9$ & $ -2.1$ & $ -0.5$ & $ -0.2$ & $ 24.1$ & $ 16.3$ & $  3.2$ & $ -3.3$ & $  8.9$ & $  2.1$ & $  4.2$ \\
& $K^*$   & $ 25.0$ & $  8.6$ & $ -3.2$ & $ -0.2$ & $-15.7$ & $ -9.7$ & $  6.3$ & $-10.4$ & $ -3.9$ & $  5.5$ & $ 16.0$ & $  3.8$ & $  6.3$ & $ -6.5$ & $ -3.2$ \\
& $S$     & $ 33.0$ & $ 19.6$ & $  3.4$ & $ -4.7$ & $-17.3$ & $-16.5$ & $  6.2$ & $ -9.6$ & $  1.0$ & $ 18.7$ & $ 21.3$ & $ -4.2$ & $  9.6$ & $ -4.2$ & $  1.1$ \\
& $f_2$   & $ 12.1$ & $ -0.6$ & $ -9.8$ & $ -2.6$ & $-23.1$ & $-27.4$ & $  0.9$ & $-16.7$ & $ -7.2$ & $  2.2$ & $  1.1$ & $-10.6$ & $  7.2$ & $-14.1$ & $ -2.6$ \\
& $f_X$   & $ 25.0$ & $ 10.2$ & $  5.4$ & $ -0.5$ & $-11.4$ & $-11.8$ & $  1.0$ & $ -0.8$ & $  2.6$ & $  8.5$ & $ 11.8$ & $ -3.8$ & $ 15.6$ & $  2.4$ & $  0.4$ \\
& $NR$    & $ 31.6$ & $ 17.0$ & $ 39.3$ & $  1.0$ & $-27.1$ & $-31.7$ & $ -6.7$ & $ 11.3$ & $ 12.8$ & $ 14.5$ & $ 19.0$ & $  3.3$ & $ 21.5$ & $ 19.6$ & $ 14.2$ \\
& $\chi$  & $  8.6$ & $  1.8$ & $  9.8$ & $  0.6$ & $ -9.9$ & $ -8.9$ & $ -7.9$ & $  2.8$ & $  3.8$ & $  1.3$ & $  4.2$ & $  3.5$ & $  7.3$ & $  8.9$ & $ 12.4$ \\
\hline\\[-9pt]
\multirow{7}{*}{$\arg(\bar{c})$}
& $f_0\ $ & $ 32.2$ & $ 11.7$ & $ 18.9$ & $  3.5$ & $-20.3$ & $-26.2$ & $ -1.6$ & $ -3.6$ & $ -6.9$ & $  7.3$ & $ 18.2$ & $  1.8$ & $ 20.3$ & $ -7.1$ & $  4.3$ \\
&$\rho^0$ & $ 14.5$ & $ 18.0$ & $ 14.6$ & $-17.3$ & $-13.4$ & $-21.0$ & $ -0.7$ & $ -8.7$ & $ 14.3$ & $ 19.8$ & $ 13.4$ & $  1.7$ & $  7.2$ & $ -4.4$ & $  5.4$ \\
& $K^*$   & $ 17.1$ & $  7.1$ & $ 22.0$ & $  5.2$ & $-13.5$ & $-17.3$ & $ -2.1$ & $  5.0$ & $  7.2$ & $  6.5$ & $ 13.8$ & $  8.1$ & $ 12.8$ & $ 29.5$ & $  9.6$ \\
& $S$     & $ 22.5$ & $ 15.9$ & $ 25.2$ & $ -3.2$ & $-16.9$ & $-21.6$ & $ -0.5$ & $  4.2$ & $ 10.6$ & $ 17.7$ & $ 16.1$ & $  1.7$ & $ 14.1$ & $ 28.8$ & $ 10.8$ \\
& $f_2$   & $ 15.1$ & $  4.9$ & $ 15.5$ & $ -5.0$ & $-15.5$ & $-17.9$ & $ -2.1$ & $ 10.0$ & $ -2.5$ & $  3.9$ & $  2.9$ & $ 11.1$ & $ 15.7$ & $ 18.6$ & $  7.5$ \\
& $f_X$   & $  8.1$ & $  2.7$ & $ 12.3$ & $ -0.6$ & $ 16.5$ & $-20.4$ & $ -0.9$ & $ 12.2$ & $  6.1$ & $  3.4$ & $  4.8$ & $  1.4$ & $-14.6$ & $  4.7$ & $  6.5$ \\
& $NR$    & $ 15.3$ & $  4.1$ & $ 14.5$ & $ -3.0$ & $-22.6$ & $-20.8$ & $  0.8$ & $  1.7$ & $  8.2$ & $  1.8$ & $  5.2$ & $  2.6$ & $ 20.0$ & $ 15.1$ & $  3.2$ \\
& $\chi$  & $ 10.9$ & $  1.1$ & $ 12.8$ & $  0.7$ & $-13.9$ & $-18.0$ & $ -4.7$ & $  2.1$ & $  3.3$ & $  0.6$ & $  3.9$ & $  5.9$ & $  9.8$ & $ 13.4$ & $  8.2$ \\
\hline
&\multicolumn{16}{c}{ }\\[6pt]
\hline
&     & \multicolumn{7}{c|}{$\arg(c)$}                               & \multicolumn{8}{c}{$\arg(\bar{c})$}\\[4pt]

&        &$\rho^0\ $& $K^*\ $ & $S\ $   & $f_2\ $ & $f_X\ $ & $NR\ $  & $\chi\ $& $f_0\ $&$\rho^0\ $& $K^*\ $ & $S\ $   & $f_2\ $ & $f_X\ $ & $NR\ $  & $\chi\ $\\
\hline\\[-9pt] 
\multirow{7}{*}{$\arg(c)$}
&$\rho^0$ & $100.0$ \\
& $K^*$   & $ 10.4$ & $100.0$ \\
& $S$     & $ 18.2$ & $ 90.9$ & $100.0$ \\
& $f_2$   & $ 19.6$ & $ 54.1$ & $ 61.8$ & $100.0$ \\
& $f_X$   & $ 25.5$ & $ 49.3$ & $ 56.9$ & $ 58.1$ & $100.0$ \\
& $NR$    & $ 24.3$ & $ 17.2$ & $ 29.9$ & $ 31.6$ & $ 47.8$ & $100.0$ \\
& $\chi$  & $  5.0$ & $  6.7$ & $  7.9$ & $ 10.2$ & $ 17.6$ & $ 30.8$ & $100.0$ \\
\cline{1-10}\\[-9pt]
\multirow{7}{*}{$\arg(\bar{c})$}
& $f_0\ $ & $ 18.0$ & $ 34.3$ & $ 42.0$ & $ 39.8$ & $ 52.9$ & $ 55.6$ & $ 23.8$ & $100.0$ \\
&$\rho^0$ & $ 55.3$ & $ 22.2$ & $ 32.4$ & $ 25.7$ & $ 36.6$ & $ 42.2$ & $ 17.4$ & $ 58.8$ & $100.0$ \\
& $K^*$   & $  4.0$ & $ 21.5$ & $ 28.0$ & $ 23.2$ & $ 36.1$ & $ 53.9$ & $ 31.3$ & $ 46.8$ & $ 33.5$ & $100.0$ \\
& $S$     & $  9.6$ & $ 23.7$ & $ 35.1$ & $ 27.8$ & $ 41.2$ & $ 60.7$ & $ 33.3$ & $ 53.4$ & $ 42.7$ & $ 90.9$ & $100.0$ \\
& $f_2$   & $  5.5$ & $  6.4$ & $ 12.4$ & $  1.5$ & $ 29.3$ & $ 46.4$ & $ 23.5$ & $ 44.1$ & $ 36.7$ & $ 56.7$ & $ 60.8$ & $100.0$ \\
& $f_X$   & $  1.7$ & $  0.0$ & $  5.4$ & $ 13.8$ & $ 15.5$ & $ 36.4$ & $ 19.5$ & $ 22.2$ & $ 22.5$ & $ 42.1$ & $ 44.8$ & $ 39.4$ & $100.0$ \\
& $NR$    & $  7.2$ & $ 19.2$ & $ 27.5$ & $ 28.9$ & $ 42.3$ & $ 55.5$ & $ 32.9$ & $ 47.3$ & $ 37.9$ & $ 63.2$ & $ 72.5$ & $ 48.1$ & $ 48.4$ & $100.0$ \\
& $\chi$  & $  4.1$ & $  8.9$ & $ 13.3$ & $ 15.5$ & $ 27.1$ & $ 43.3$ & $ 35.9$ & $ 38.0$ & $ 26.9$ & $ 55.9$ & $ 58.9$ & $ 40.0$ & $ 33.6$ & $ 52.1$ & $100.0$ \\
\hline\hline
\end{tabular}
\end{center}
\end{table*}

\begin{table*}[htbp]
\begin{center}
\caption{
Full correlation matrix for the isobar parameters of solution II. The entries are given in percent. Since the matrix is symmetric, all elements above the diagonal are omitted. 
\label{tab:cormat2}}
\begin{tabular}{cc|rrrrrrr|rrrrrrrr}
\hline\hline
&     & \multicolumn{7}{c|}{$\left|c\right|$}                               & \multicolumn{8}{c}{$\left|\bar{c}\right|$}\\[4pt]
&        &$\rho^0\ $& $K^*\ $ & $S\ $   & $f_2\ $ & $f_X\ $ & $NR\ $  & $\chi\ $& $f_0\ $&$\rho^0\ $& $K^*\ $ & $S\ $   & $f_2\ $ & $f_X\ $ & $NR\ $  & $\chi\ $\\
\hline\\[-9pt]
\multirow{7}{*}{$\left|c\right|$}
&$\rho^0$ & $100.0$ \\
& $K^*$   & $ 46.9$ & $100.0$ \\
& $S$     & $ 49.1$ & $ 68.2$ & $100.0$ \\
& $f_2$   & $  8.7$ & $  7.7$ & $ 25.4$ & $100.0$ \\
& $f_X$   & $ 16.8$ & $ 40.3$ & $ 38.5$ & $ 26.6$ & $100.0$ \\
& $NR$    & $ -8.4$ & $ 30.2$ & $ 21.2$ & $  9.4$ & $ 49.9$ & $100.0$ \\
& $\chi$  & $  5.5$ & $ 11.7$ & $  9.3$ & $  3.4$ & $ 12.1$ & $  9.1$ & $100.0$ \\
\cline{1-10}\\[-9pt]
\multirow{8}{*}{$\left|\bar{c}\right|$}
& $f_0\ $ & $ 29.2$ & $ 42.1$ & $ 50.2$ & $ 31.5$ & $ 57.9$ & $ 34.1$ & $ 10.0$ & $100.0$ \\
&$\rho^0$ & $ 61.5$ & $ 68.1$ & $ 40.4$ & $  6.9$ & $ 20.6$ & $  6.4$ & $  6.0$ & $ 31.6$ & $100.0$ \\
& $K^*$   & $ 39.8$ & $ 75.7$ & $ 59.8$ & $  0.3$ & $ 33.1$ & $ 25.3$ & $ 10.9$ & $ 33.2$ & $ 36.3$ & $100.0$ \\
& $S$     & $ 50.6$ & $ 75.2$ & $ 83.2$ & $ 25.4$ & $ 49.9$ & $ 33.4$ & $ 13.1$ & $ 51.6$ & $ 46.0$ & $ 61.4$ & $100.0$ \\
& $f_2$   & $  0.8$ & $ -6.1$ & $  9.6$ & $-53.9$ & $  6.0$ & $ 13.3$ & $  0.2$ & $ 14.7$ & $  5.3$ & $-18.5$ & $ 10.4$ & $100.0$ \\
& $f_X$   & $ 10.0$ & $ -3.3$ & $ -0.9$ & $-10.6$ & $-68.7$ & $-17.8$ & $ -5.2$ & $-18.4$ & $  6.3$ & $ -4.0$ & $ -4.9$ & $  2.2$ & $100.0$ \\
& $NR$    & $ 23.1$ & $ 68.8$ & $ 44.7$ & $ 13.5$ & $ 39.3$ & $ 34.4$ & $  5.8$ & $ 45.6$ & $ 58.3$ & $ 32.8$ & $ 45.4$ & $ 14.7$ & $-13.8$ & $100.0$ \\
& $\chi$  & $ 22.3$ & $ 33.5$ & $ 37.8$ & $  9.8$ & $ 19.3$ & $  9.9$ & $-79.2$ & $ 31.3$ & $ 20.7$ & $ 30.2$ & $ 36.1$ & $  3.3$ & $ -2.6$ & $ 23.3$ & $100.0$ \\
\hline\\[-9pt]
\multirow{7}{*}{$\arg(c)$}
&$\rho^0$ & $-23.1$ & $ 13.7$ & $  5.5$ & $-11.4$ & $  8.0$ & $  5.2$ & $  0.0$ & $  9.0$ & $-11.9$ & $ 14.5$ & $  6.3$ & $ -0.2$ & $  0.3$ & $  3.8$ & $  6.9$ \\
& $K^*$   & $ 30.6$ & $  2.0$ & $ -2.2$ & $ -6.3$ & $-16.1$ & $-28.1$ & $ -0.0$ & $-15.2$ & $ 14.3$ & $ -1.4$ & $  6.1$ & $  2.0$ & $ 19.4$ & $-10.3$ & $  0.5$ \\
& $S$     & $ 38.1$ & $  8.9$ & $  1.8$ & $-10.1$ & $-17.9$ & $-39.5$ & $ -0.1$ & $-15.8$ & $ 17.4$ & $  9.4$ & $  7.7$ & $ -8.2$ & $ 19.7$ & $-12.1$ & $  3.7$ \\
& $f_2$   & $ 18.1$ & $-10.0$ & $-13.7$ & $ -7.4$ & $-15.4$ & $-41.3$ & $ -2.4$ & $-18.6$ & $  1.0$ & $ -6.2$ & $-10.2$ & $-12.7$ & $ 10.7$ & $-21.6$ & $ -2.7$ \\
& $f_X$   & $ 26.2$ & $ -7.8$ & $-12.2$ & $ -5.9$ & $ -7.7$ & $-35.9$ & $ -1.7$ & $-14.5$ & $  7.8$ & $ -5.7$ & $ -8.8$ & $ -9.9$ & $ 12.2$ & $-15.2$ & $ -3.6$ \\
& $NR$    & $ 32.4$ & $ -0.4$ & $ 21.4$ & $  0.5$ & $-29.5$ & $-65.2$ & $-10.4$ & $ -4.2$ & $ 12.0$ & $  0.2$ & $  0.4$ & $ -6.4$ & $ 21.2$ & $ -8.1$ & $ 10.2$ \\
& $\chi$  & $ 15.4$ & $ -2.2$ & $  0.2$ & $ -1.6$ & $ -9.9$ & $-18.3$ & $ -4.9$ & $ -5.6$ & $  5.6$ & $ -3.0$ & $ -0.2$ & $ -0.8$ & $  9.2$ & $ -5.6$ & $  4.0$ \\
\hline\\[-9pt]
\multirow{7}{*}{$\arg(\bar{c})$}
& $f_0\ $ & $ 30.1$ & $ -8.0$ & $ -2.3$ & $ -0.9$ & $-13.2$ & $-43.0$ & $ -2.8$ & $-16.7$ & $ 12.1$ & $ -7.2$ & $ -5.5$ & $ -4.9$ & $ 10.4$ & $-18.7$ & $ -1.6$ \\
&$\rho^0$ & $  7.6$ & $ 11.4$ & $  5.8$ & $ -7.5$ & $ -1.8$ & $-24.7$ & $  0.6$ & $ -7.5$ & $  4.1$ & $ 15.1$ & $  5.5$ & $-12.6$ & $  1.3$ & $ -7.0$ & $  4.0$ \\
& $K^*$   & $ 27.0$ & $  0.8$ & $  7.6$ & $  5.6$ & $  2.8$ & $-27.8$ & $  0.6$ & $ -2.0$ & $  9.1$ & $  1.5$ & $  7.1$ & $  3.2$ & $ -6.9$ & $ 13.9$ & $  4.1$ \\
& $S$     & $ 32.6$ & $  8.0$ & $  8.4$ & $ -1.1$ & $  0.6$ & $-31.3$ & $  2.1$ & $ -4.1$ & $ 12.6$ & $ 12.1$ & $  7.6$ & $ -5.6$ & $ -4.4$ & $ 12.2$ & $  4.7$ \\
& $f_2$   & $ 18.7$ & $  1.7$ & $  6.6$ & $ 10.1$ & $  9.8$ & $-22.9$ & $  0.7$ & $  7.6$ & $  8.6$ & $  3.5$ & $  0.6$ & $ -5.6$ & $-21.6$ & $  9.3$ & $  4.6$ \\
& $f_X$   & $ 21.9$ & $  1.8$ & $  4.4$ & $  9.6$ & $ -0.7$ & $-30.2$ & $  0.1$ & $ -5.0$ & $  8.1$ & $  2.8$ & $  4.0$ & $-17.3$ & $  1.0$ & $ -6.6$ & $ -0.2$ \\
& $NR$    & $ 27.7$ & $ -1.9$ & $ -3.0$ & $  3.9$ & $ -0.5$ & $-30.7$ & $  2.8$ & $-13.3$ & $  7.8$ & $ -1.5$ & $ -1.2$ & $-13.8$ & $ -7.2$ & $ -3.7$ & $ -5.0$ \\
& $\chi$  & $ 19.7$ & $ -5.0$ & $ -0.5$ & $  2.3$ & $ -4.4$ & $-27.6$ & $  2.7$ & $ -6.1$ & $  6.2$ & $ -4.1$ & $ -2.5$ & $ -1.6$ & $ -0.2$ & $ -0.1$ & $ -2.9$ \\
\hline
&\multicolumn{16}{c}{ }\\[6pt]
\hline
&     & \multicolumn{7}{c|}{$\arg(c)$}                               & \multicolumn{8}{c}{$\arg(\bar{c})$}\\[4pt]
&        &$\rho^0\ $& $K^*\ $ & $S\ $   & $f_2\ $ & $f_X\ $ & $NR\ $  & $\chi\ $& $f_0\ $&$\rho^0\ $& $K^*\ $ & $S\ $   & $f_2\ $ & $f_X\ $ & $NR\ $  & $\chi\ $\\
\hline\\[-9pt]
\multirow{7}{*}{$\arg(c)$} 
&$\rho^0$ & $100.0$ \\
& $K^*$   & $  2.9$ & $100.0$ \\
& $S$     & $  7.4$ & $ 90.6$ & $100.0$ \\
& $f_2$   & $  9.9$ & $ 56.6$ & $ 65.5$ & $100.0$ \\
& $f_X$   & $  5.9$ & $ 57.0$ & $ 64.4$ & $ 69.5$ & $100.0$ \\
& $NR$    & $ 10.1$ & $ 37.0$ & $ 50.3$ & $ 44.4$ & $ 46.6$ & $100.0$ \\
& $\chi$  & $  2.6$ & $ 39.3$ & $ 40.3$ & $ 29.1$ & $ 31.3$ & $ 28.6$ & $100.0$ \\
\cline{1-10}\\[-9pt]
\multirow{7}{*}{$\arg(\bar{c})$}
& $f_0\ $ & $ -0.6$ & $ 45.8$ & $ 53.5$ & $ 47.1$ & $ 61.0$ & $ 51.9$ & $ 27.4$ & $100.0$ \\
&$\rho^0$ & $ 41.3$ & $ 29.5$ & $ 39.2$ & $ 31.2$ & $ 39.1$ & $ 33.0$ & $ 16.5$ & $ 54.9$ & $100.0$ \\
& $K^*$   & $-11.6$ & $ 35.2$ & $ 39.7$ & $ 30.4$ & $ 42.7$ & $ 30.0$ & $ 17.6$ & $ 56.0$ & $ 32.9$ & $100.0$ \\
& $S$     & $ -8.7$ & $ 38.8$ & $ 47.7$ & $ 36.1$ & $ 49.1$ & $ 33.7$ & $ 19.5$ & $ 62.4$ & $ 41.1$ & $ 91.1$ & $100.0$ \\
& $f_2$   & $ -5.4$ & $ 12.2$ & $ 17.9$ & $  7.0$ & $ 28.5$ & $ 27.2$ & $  9.9$ & $ 52.8$ & $ 42.0$ & $ 59.3$ & $ 61.6$ & $100.0$ \\
& $f_X$   & $ -7.0$ & $ 23.2$ & $ 28.6$ & $ 28.0$ & $ 34.4$ & $ 29.9$ & $ 15.4$ & $ 34.6$ & $ 30.2$ & $ 43.3$ & $ 47.1$ & $ 41.5$ & $100.0$ \\
& $NR$    & $ -9.0$ & $ 41.4$ & $ 47.9$ & $ 44.2$ & $ 59.5$ & $ 30.9$ & $ 25.2$ & $ 68.9$ & $ 48.1$ & $ 68.6$ & $ 77.6$ & $ 54.6$ & $ 55.8$ & $100.0$ \\
& $\chi$  & $ -7.3$ & $ 29.3$ & $ 33.3$ & $ 28.8$ & $ 38.8$ & $ 29.9$ & $  8.8$ & $ 47.1$ & $ 26.9$ & $ 54.7$ & $ 58.0$ & $ 38.6$ & $ 35.8$ & $ 54.6$ & $100.0$ \\
\hline\hline
\end{tabular}
\end{center}
\end{table*}

\clearpage
\bibliography{prd}

\begin{thebibliography}{47}
\expandafter\ifx\csname natexlab\endcsname\relax\def\natexlab#1{#1}\fi
\expandafter\ifx\csname bibnamefont\endcsname\relax
  \def\bibnamefont#1{#1}\fi
\expandafter\ifx\csname bibfnamefont\endcsname\relax
  \def\bibfnamefont#1{#1}\fi
\expandafter\ifx\csname citenamefont\endcsname\relax
  \def\citenamefont#1{#1}\fi
\expandafter\ifx\csname url\endcsname\relax
  \def\url#1{\texttt{#1}}\fi
\expandafter\ifx\csname urlprefix\endcsname\relax\def\urlprefix{URL }\fi
\providecommand{\bibinfo}[2]{#2}
\providecommand{\eprint}[2][]{\url{#2}}

\bibitem[{\citenamefont{Cabibbo}(1963)}]{Cabibbo:1963yz}
\bibinfo{author}{\bibfnamefont{N.}~\bibnamefont{Cabibbo}},
  \bibinfo{journal}{Phys. Rev. Lett.} \textbf{\bibinfo{volume}{10}},
  \bibinfo{pages}{531} (\bibinfo{year}{1963}).

\bibitem[{\citenamefont{Kobayashi and Maskawa}(1973)}]{Kobayashi:1973fv}
\bibinfo{author}{\bibfnamefont{M.}~\bibnamefont{Kobayashi}} \bibnamefont{and}
  \bibinfo{author}{\bibfnamefont{T.}~\bibnamefont{Maskawa}},
  \bibinfo{journal}{Prog. Theor. Phys.} \textbf{\bibinfo{volume}{49}},
  \bibinfo{pages}{652} (\bibinfo{year}{1973}).

\bibitem[{\citenamefont{Barberio et~al.}(2008)}]{Barberio:2008fa}
\bibinfo{author}{\bibfnamefont{E.}~\bibnamefont{Barberio}} \bibnamefont{et~al.}
  (\bibinfo{collaboration}{Heavy Flavor Averaging Group - HFAG})
  (\bibinfo{year}{2008}), \eprint{arXiv:0808.1297 [hep-ex]}.

\bibitem[{\citenamefont{Grossman et~al.}(2003)\citenamefont{Grossman, Ligeti,
  Nir, and Quinn}}]{Grossman:2003qp}
\bibinfo{author}{\bibfnamefont{Y.}~\bibnamefont{Grossman}},
  \bibinfo{author}{\bibfnamefont{Z.}~\bibnamefont{Ligeti}},
  \bibinfo{author}{\bibfnamefont{Y.}~\bibnamefont{Nir}}, \bibnamefont{and}
  \bibinfo{author}{\bibfnamefont{H.}~\bibnamefont{Quinn}},
  \bibinfo{journal}{Phys. Rev.} \textbf{\bibinfo{volume}{D68}},
  \bibinfo{pages}{015004} (\bibinfo{year}{2003}).

\bibitem[{\citenamefont{Gronau et~al.}(2004{\natexlab{a}})\citenamefont{Gronau,
  Grossman, and Rosner}}]{Gronau:2003kx}
\bibinfo{author}{\bibfnamefont{M.}~\bibnamefont{Gronau}},
  \bibinfo{author}{\bibfnamefont{Y.}~\bibnamefont{Grossman}}, \bibnamefont{and}
  \bibinfo{author}{\bibfnamefont{J.~L.} \bibnamefont{Rosner}},
  \bibinfo{journal}{Phys. Lett.} \textbf{\bibinfo{volume}{B579}},
  \bibinfo{pages}{331} (\bibinfo{year}{2004}{\natexlab{a}}).

\bibitem[{\citenamefont{Gronau et~al.}(2004{\natexlab{b}})\citenamefont{Gronau,
  Rosner, and Zupan}}]{Gronau:2004hp}
\bibinfo{author}{\bibfnamefont{M.}~\bibnamefont{Gronau}},
  \bibinfo{author}{\bibfnamefont{J.~L.} \bibnamefont{Rosner}},
  \bibnamefont{and} \bibinfo{author}{\bibfnamefont{J.}~\bibnamefont{Zupan}},
  \bibinfo{journal}{Phys. Lett.} \textbf{\bibinfo{volume}{B596}},
  \bibinfo{pages}{107} (\bibinfo{year}{2004}{\natexlab{b}}).

\bibitem[{\citenamefont{Cheng et~al.}(2005{\natexlab{a}})\citenamefont{Cheng,
  Chua, and Soni}}]{Cheng:2005bg}
\bibinfo{author}{\bibfnamefont{H.-Y.} \bibnamefont{Cheng}},
  \bibinfo{author}{\bibfnamefont{C.-K.} \bibnamefont{Chua}}, \bibnamefont{and}
  \bibinfo{author}{\bibfnamefont{A.}~\bibnamefont{Soni}},
  \bibinfo{journal}{Phys. Rev.} \textbf{\bibinfo{volume}{D72}},
  \bibinfo{pages}{014006} (\bibinfo{year}{2005}{\natexlab{a}}).

\bibitem[{\citenamefont{Gronau and Rosner}(2005)}]{Gronau:2005gz}
\bibinfo{author}{\bibfnamefont{M.}~\bibnamefont{Gronau}} \bibnamefont{and}
  \bibinfo{author}{\bibfnamefont{J.~L.} \bibnamefont{Rosner}},
  \bibinfo{journal}{Phys. Rev.} \textbf{\bibinfo{volume}{D71}},
  \bibinfo{pages}{074019} (\bibinfo{year}{2005}).

\bibitem[{\citenamefont{Beneke}(2005)}]{Beneke:2005pu}
\bibinfo{author}{\bibfnamefont{M.}~\bibnamefont{Beneke}},
  \bibinfo{journal}{Phys. Lett.} \textbf{\bibinfo{volume}{B620}},
  \bibinfo{pages}{143} (\bibinfo{year}{2005}).

\bibitem[{\citenamefont{Engelhard et~al.}(2005)\citenamefont{Engelhard, Nir,
  and Raz}}]{Engelhard:2005hu}
\bibinfo{author}{\bibfnamefont{G.}~\bibnamefont{Engelhard}},
  \bibinfo{author}{\bibfnamefont{Y.}~\bibnamefont{Nir}}, \bibnamefont{and}
  \bibinfo{author}{\bibfnamefont{G.}~\bibnamefont{Raz}},
  \bibinfo{journal}{Phys. Rev.} \textbf{\bibinfo{volume}{D72}},
  \bibinfo{pages}{075013} (\bibinfo{year}{2005}).

\bibitem[{\citenamefont{Cheng et~al.}(2005{\natexlab{b}})\citenamefont{Cheng,
  Chua, and Soni}}]{Cheng:2005ug}
\bibinfo{author}{\bibfnamefont{H.-Y.} \bibnamefont{Cheng}},
  \bibinfo{author}{\bibfnamefont{C.-K.} \bibnamefont{Chua}}, \bibnamefont{and}
  \bibinfo{author}{\bibfnamefont{A.}~\bibnamefont{Soni}},
  \bibinfo{journal}{Phys. Rev.} \textbf{\bibinfo{volume}{D72}},
  \bibinfo{pages}{094003} (\bibinfo{year}{2005}{\natexlab{b}}).

\bibitem[{\citenamefont{Williamson and Zupan}(2006)}]{Williamson:2006hb}
\bibinfo{author}{\bibfnamefont{A.~R.} \bibnamefont{Williamson}}
  \bibnamefont{and} \bibinfo{author}{\bibfnamefont{J.}~\bibnamefont{Zupan}},
  \bibinfo{journal}{Phys. Rev.} \textbf{\bibinfo{volume}{D74}},
  \bibinfo{pages}{014003} (\bibinfo{year}{2006}).

\bibitem[{\citenamefont{Gershon and Hazumi}(2004)}]{Gershon:2004tk}
\bibinfo{author}{\bibfnamefont{T.}~\bibnamefont{Gershon}} \bibnamefont{and}
  \bibinfo{author}{\bibfnamefont{M.}~\bibnamefont{Hazumi}},
  \bibinfo{journal}{Phys. Lett.} \textbf{\bibinfo{volume}{B596}},
  \bibinfo{pages}{163} (\bibinfo{year}{2004}).

\bibitem[{\citenamefont{Garmash et~al.}(2004)}]{Garmash:2003er}
\bibinfo{author}{\bibfnamefont{A.}~\bibnamefont{Garmash}} \bibnamefont{et~al.}
  (\bibinfo{collaboration}{Belle}), \bibinfo{journal}{Phys. Rev.}
  \textbf{\bibinfo{volume}{D69}}, \bibinfo{pages}{012001}
  (\bibinfo{year}{2004}).

\bibitem[{\citenamefont{Aubert et~al.}(2007{\natexlab{a}})}]{Aubert:2007sd}
\bibinfo{author}{\bibfnamefont{B.}~\bibnamefont{Aubert}} \bibnamefont{et~al.}
  (\bibinfo{collaboration}{\babar}), \bibinfo{journal}{Phys. Rev. Lett.}
  \textbf{\bibinfo{volume}{99}}, \bibinfo{pages}{161802}
  (\bibinfo{year}{2007}{\natexlab{a}}).

\bibitem[{\citenamefont{Wolfenstein}(1983)}]{Wolfenstein:1983yz}
\bibinfo{author}{\bibfnamefont{L.}~\bibnamefont{Wolfenstein}},
  \bibinfo{journal}{Phys. Rev. Lett.} \textbf{\bibinfo{volume}{51}},
  \bibinfo{pages}{1945} (\bibinfo{year}{1983}).

\bibitem[{\citenamefont{Buras et~al.}(1994)\citenamefont{Buras, Lautenbacher,
  and Ostermaier}}]{Buras:1994ec}
\bibinfo{author}{\bibfnamefont{A.~J.} \bibnamefont{Buras}},
  \bibinfo{author}{\bibfnamefont{M.~E.} \bibnamefont{Lautenbacher}},
  \bibnamefont{and}
  \bibinfo{author}{\bibfnamefont{G.}~\bibnamefont{Ostermaier}},
  \bibinfo{journal}{Phys. Rev.} \textbf{\bibinfo{volume}{D50}},
  \bibinfo{pages}{3433} (\bibinfo{year}{1994}).

\bibitem[{\citenamefont{Deshpande et~al.}(2003)\citenamefont{Deshpande, Sinha,
  and Sinha}}]{Deshpande:2002be}
\bibinfo{author}{\bibfnamefont{N.~G.} \bibnamefont{Deshpande}},
  \bibinfo{author}{\bibfnamefont{N.}~\bibnamefont{Sinha}}, \bibnamefont{and}
  \bibinfo{author}{\bibfnamefont{R.}~\bibnamefont{Sinha}},
  \bibinfo{journal}{Phys. Rev. Lett.} \textbf{\bibinfo{volume}{90}},
  \bibinfo{pages}{061802} (\bibinfo{year}{2003}).

\bibitem[{\citenamefont{Ciuchini et~al.}(2006)\citenamefont{Ciuchini, Pierini,
  and Silvestrini}}]{Ciuchini:2006kv}
\bibinfo{author}{\bibfnamefont{M.}~\bibnamefont{Ciuchini}},
  \bibinfo{author}{\bibfnamefont{M.}~\bibnamefont{Pierini}}, \bibnamefont{and}
  \bibinfo{author}{\bibfnamefont{L.}~\bibnamefont{Silvestrini}},
  \bibinfo{journal}{Phys. Rev.} \textbf{\bibinfo{volume}{D74}},
  \bibinfo{pages}{051301} (\bibinfo{year}{2006}).

\bibitem[{\citenamefont{Gronau et~al.}(2007)\citenamefont{Gronau, Pirjol, Soni,
  and Zupan}}]{Gronau:2006qn}
\bibinfo{author}{\bibfnamefont{M.}~\bibnamefont{Gronau}},
  \bibinfo{author}{\bibfnamefont{D.}~\bibnamefont{Pirjol}},
  \bibinfo{author}{\bibfnamefont{A.}~\bibnamefont{Soni}}, \bibnamefont{and}
  \bibinfo{author}{\bibfnamefont{J.}~\bibnamefont{Zupan}},
  \bibinfo{journal}{Phys. Rev.} \textbf{\bibinfo{volume}{D75}},
  \bibinfo{pages}{014002} (\bibinfo{year}{2007}).

\bibitem[{\citenamefont{Lipkin et~al.}(1991)\citenamefont{Lipkin, Nir, Quinn,
  and Snyder}}]{Lipkin:1991st}
\bibinfo{author}{\bibfnamefont{H.~J.} \bibnamefont{Lipkin}},
  \bibinfo{author}{\bibfnamefont{Y.}~\bibnamefont{Nir}},
  \bibinfo{author}{\bibfnamefont{H.~R.} \bibnamefont{Quinn}}, \bibnamefont{and}
  \bibinfo{author}{\bibfnamefont{A.}~\bibnamefont{Snyder}},
  \bibinfo{journal}{Phys. Rev.} \textbf{\bibinfo{volume}{D44}},
  \bibinfo{pages}{1454} (\bibinfo{year}{1991}).

\bibitem[{\citenamefont{Dalseno et~al.}(2009)}]{BelleKspipi}
\bibinfo{author}{\bibfnamefont{J.}~\bibnamefont{Dalseno}} \bibnamefont{et~al.}
  (\bibinfo{collaboration}{Belle}), \bibinfo{journal}{Phys. Rev.}
  \textbf{\bibinfo{volume}{D79}}, \bibinfo{pages}{072004}
  (\bibinfo{year}{2009}).

\bibitem[{\citenamefont{Aubert et~al.}(2006)}]{Aubert:2005wb}
\bibinfo{author}{\bibfnamefont{B.}~\bibnamefont{Aubert}} \bibnamefont{et~al.}
  (\bibinfo{collaboration}{\babar}), \bibinfo{journal}{Phys. Rev.}
  \textbf{\bibinfo{volume}{D73}}, \bibinfo{pages}{031101}
  (\bibinfo{year}{2006}).

\bibitem[{\citenamefont{Garmash et~al.}(2007)}]{Garmash:2006fh}
\bibinfo{author}{\bibfnamefont{A.}~\bibnamefont{Garmash}} \bibnamefont{et~al.}
  (\bibinfo{collaboration}{Belle}), \bibinfo{journal}{Phys. Rev.}
  \textbf{\bibinfo{volume}{D75}}, \bibinfo{pages}{012006}
  (\bibinfo{year}{2007}).

\bibitem[{\citenamefont{Chang et~al.}(2004)}]{Chang:2004um}
\bibinfo{author}{\bibfnamefont{P.}~\bibnamefont{Chang}} \bibnamefont{et~al.}
  (\bibinfo{collaboration}{Belle}), \bibinfo{journal}{Phys. Lett.}
  \textbf{\bibinfo{volume}{B599}}, \bibinfo{pages}{148} (\bibinfo{year}{2004}).

\bibitem[{\citenamefont{Aubert et~al.}(2008{\natexlab{a}})}]{Aubert:2007bs}
\bibinfo{author}{\bibfnamefont{B.}~\bibnamefont{Aubert}} \bibnamefont{et~al.}
  (\bibinfo{collaboration}{\babar}), \bibinfo{journal}{Phys. Rev.}
  \textbf{\bibinfo{volume}{D78}}, \bibinfo{pages}{052005}
  (\bibinfo{year}{2008}{\natexlab{a}}).

\bibitem[{\citenamefont{Aubert et~al.}(2008{\natexlab{b}})}]{Aubert:2008bj}
\bibinfo{author}{\bibfnamefont{B.}~\bibnamefont{Aubert}} \bibnamefont{et~al.}
  (\bibinfo{collaboration}{\babar}), \bibinfo{journal}{Phys. Rev.}
  \textbf{\bibinfo{volume}{D78}}, \bibinfo{pages}{012004}
  (\bibinfo{year}{2008}{\natexlab{b}}).

\bibitem[{\citenamefont{Garmash et~al.}(2006)}]{Garmash:2005rv}
\bibinfo{author}{\bibfnamefont{A.}~\bibnamefont{Garmash}} \bibnamefont{et~al.}
  (\bibinfo{collaboration}{Belle}), \bibinfo{journal}{Phys. Rev. Lett.}
  \textbf{\bibinfo{volume}{96}}, \bibinfo{pages}{251803}
  (\bibinfo{year}{2006}).

\bibitem[{\citenamefont{Chang et~al.}(2008)\citenamefont{Chang, Li, and
  Yang}}]{Chang:2008tf}
\bibinfo{author}{\bibfnamefont{Q.}~\bibnamefont{Chang}},
  \bibinfo{author}{\bibfnamefont{X.-Q.} \bibnamefont{Li}}, \bibnamefont{and}
  \bibinfo{author}{\bibfnamefont{Y.-D.} \bibnamefont{Yang}},
  \bibinfo{journal}{JHEP} \textbf{\bibinfo{volume}{0809}}, \bibinfo{pages}{038}
  (\bibinfo{year}{2008}).

\bibitem[{\citenamefont{Blatt and Weisskopf}(1952)}]{blatt-weisskopf}
\bibinfo{author}{\bibfnamefont{J.}~\bibnamefont{Blatt}} \bibnamefont{and}
  \bibinfo{author}{\bibfnamefont{V.~E.} \bibnamefont{Weisskopf}},
  \emph{\bibinfo{title}{Theoretical Nuclear Physics}} (\bibinfo{publisher}{J.
  Wiley (New York)}, \bibinfo{year}{1952}).

\bibitem[{\citenamefont{Amsler et~al.}(2008)}]{Amsler:2008zz}
\bibinfo{author}{\bibfnamefont{C.}~\bibnamefont{Amsler}} \bibnamefont{et~al.}
  (\bibinfo{collaboration}{Particle Data Group}), \bibinfo{journal}{Phys.
  Lett.} \textbf{\bibinfo{volume}{B667}}, \bibinfo{pages}{1}
  (\bibinfo{year}{2008}).

\bibitem[{\citenamefont{Flatte}(1976)}]{Flatte}
\bibinfo{author}{\bibfnamefont{S.~M.} \bibnamefont{Flatte}},
  \bibinfo{journal}{Phys. Lett.} \textbf{\bibinfo{volume}{B63}},
  \bibinfo{pages}{224} (\bibinfo{year}{1976}).

\bibitem[{\citenamefont{Gounaris and Sakurai}(1968)}]{Gounaris:1968mw}
\bibinfo{author}{\bibfnamefont{G.~J.} \bibnamefont{Gounaris}} \bibnamefont{and}
  \bibinfo{author}{\bibfnamefont{J.~J.} \bibnamefont{Sakurai}},
  \bibinfo{journal}{Phys. Rev. Lett.} \textbf{\bibinfo{volume}{21}},
  \bibinfo{pages}{244} (\bibinfo{year}{1968}).

\bibitem[{\citenamefont{Aston et~al.}(1988)}]{LASS}
\bibinfo{author}{\bibfnamefont{D.}~\bibnamefont{Aston}} \bibnamefont{et~al.}
  (\bibinfo{collaboration}{LASS}), \bibinfo{journal}{Nucl. Phys.}
  \textbf{\bibinfo{volume}{B296}}, \bibinfo{pages}{493} (\bibinfo{year}{1988}).

\bibitem[{\citenamefont{Bugg}(2003)}]{Bugg:2003kj}
\bibinfo{author}{\bibfnamefont{D.~V.} \bibnamefont{Bugg}},
  \bibinfo{journal}{Phys. Lett.} \textbf{\bibinfo{volume}{B572}},
  \bibinfo{pages}{1} (\bibinfo{year}{2003}).

\bibitem[{\citenamefont{Ablikim et~al.}(2005)}]{valFlatte}
\bibinfo{author}{\bibfnamefont{M.}~\bibnamefont{Ablikim}} \bibnamefont{et~al.}
  (\bibinfo{collaboration}{BES}), \bibinfo{journal}{Phys. Lett.}
  \textbf{\bibinfo{volume}{B607}}, \bibinfo{pages}{243} (\bibinfo{year}{2005}).

\bibitem[{\citenamefont{Aubert et~al.}(2005{\natexlab{a}})}]{Aubert:2005sk}
\bibinfo{author}{\bibfnamefont{B.}~\bibnamefont{Aubert}} \bibnamefont{et~al.}
  (\bibinfo{collaboration}{\babar}), \bibinfo{journal}{Phys. Rev.}
  \textbf{\bibinfo{volume}{D72}}, \bibinfo{pages}{052002}
  (\bibinfo{year}{2005}{\natexlab{a}}).

\bibitem[{\citenamefont{Aubert et~al.}(2002)}]{babar}
\bibinfo{author}{\bibfnamefont{B.}~\bibnamefont{Aubert}} \bibnamefont{et~al.}
  (\bibinfo{collaboration}{\babar}), \bibinfo{journal}{Nucl. Instrum. Meth.}
  \textbf{\bibinfo{volume}{A479}}, \bibinfo{pages}{1} (\bibinfo{year}{2002}).

\bibitem[{\citenamefont{Gay et~al.}(1995)\citenamefont{Gay, Michel, Proriol,
  and Deschamps}}]{Gay:1995sm}
\bibinfo{author}{\bibfnamefont{P.}~\bibnamefont{Gay}},
  \bibinfo{author}{\bibfnamefont{B.}~\bibnamefont{Michel}},
  \bibinfo{author}{\bibfnamefont{J.}~\bibnamefont{Proriol}}, \bibnamefont{and}
  \bibinfo{author}{\bibfnamefont{O.}~\bibnamefont{Deschamps}}
  (\bibinfo{year}{1995}), \bibinfo{note}{prepared for 4th International
  Workshop on Software Engineering and Artificial Intelligence for High-energy
  and Nuclear Physics (AIHENP 95), Pisa, Italy, 3-8 April 1995}.

\bibitem[{\citenamefont{Aubert et~al.}(2005{\natexlab{b}})}]{BabarS2b}
\bibinfo{author}{\bibfnamefont{B.}~\bibnamefont{Aubert}} \bibnamefont{et~al.}
  (\bibinfo{collaboration}{\babar}), \bibinfo{journal}{Phys. Rev. Lett.}
  \textbf{\bibinfo{volume}{94}}, \bibinfo{pages}{161803}
  (\bibinfo{year}{2005}{\natexlab{b}}).

\bibitem[{\citenamefont{Gardner and Tandean}(2004)}]{Gardner:2003su}
\bibinfo{author}{\bibfnamefont{S.}~\bibnamefont{Gardner}} \bibnamefont{and}
  \bibinfo{author}{\bibfnamefont{J.}~\bibnamefont{Tandean}},
  \bibinfo{journal}{Phys. Rev.} \textbf{\bibinfo{volume}{D69}},
  \bibinfo{pages}{034011} (\bibinfo{year}{2004}).

\bibitem[{\citenamefont{Aubert et~al.}(2007{\natexlab{b}})}]{Aubert:2007jn}
\bibinfo{author}{\bibfnamefont{B.}~\bibnamefont{Aubert}} \bibnamefont{et~al.}
  (\bibinfo{collaboration}{\babar}), \bibinfo{journal}{Phys. Rev.}
  \textbf{\bibinfo{volume}{D76}}, \bibinfo{pages}{012004}
  (\bibinfo{year}{2007}{\natexlab{b}}).

\bibitem[{\citenamefont{Oreglia}(1980)}]{Oreglia:1980cs}
\bibinfo{author}{\bibfnamefont{M.}~\bibnamefont{Oreglia}}, Ph.D. thesis,
  \bibinfo{school}{SLAC-R-0236} (\bibinfo{year}{1980}),
  \bibinfo{note}{{A}ppendix D}.

\bibitem[{\citenamefont{Gaiser}(1982)}]{Gaiser:1982yw}
\bibinfo{author}{\bibfnamefont{J.}~\bibnamefont{Gaiser}}, Ph.D. thesis,
  \bibinfo{school}{SLAC-R-0255} (\bibinfo{year}{1982}),
  \bibinfo{note}{{A}ppendix F}.

\bibitem[{\citenamefont{Skwarnicki}(1986)}]{Skwarnicki:1986xj}
\bibinfo{author}{\bibfnamefont{T.}~\bibnamefont{Skwarnicki}}, Ph.D. thesis,
  \bibinfo{school}{{{DESY-F31-86-02}}} (\bibinfo{year}{1986}),
  \bibinfo{note}{{A}ppendix E}.

\bibitem[{\citenamefont{Albrecht et~al.}(1990)}]{argus}
\bibinfo{author}{\bibfnamefont{H.}~\bibnamefont{Albrecht}} \bibnamefont{et~al.}
  (\bibinfo{collaboration}{ARGUS}), \bibinfo{journal}{Z. Phys.}
  \textbf{\bibinfo{volume}{C48}}, \bibinfo{pages}{543} (\bibinfo{year}{1990}).

\bibitem[{\citenamefont{del Amo~Sanchez}(2007)}]{delAmoSanchez:2007zz}
\bibinfo{author}{\bibfnamefont{P.}~\bibnamefont{del Amo~Sanchez}}, Ph.D.
  thesis, \bibinfo{school}{BABAR-THESIS-07-007} (\bibinfo{year}{2007}).

\end{thebibliography}
\bibliographystyle{apsrev}

\end{document}